\pdfoutput=1
\documentclass{article}
\usepackage[margin=1in]{geometry}

\usepackage{hyperref}

\usepackage{amsmath,amssymb,amsthm}
\usepackage[T1]{fontenc}
\usepackage[utf8]{inputenc}

\usepackage{array,authblk,enumitem,multirow,pifont,textcomp,url,xcolor,xspace}

\usepackage{graphicx}
\usepackage{subfigure}
\usepackage[ruled,vlined,linesnumbered]{algorithm2e}
\SetAlFnt{\small}

\usepackage[sort,numbers]{natbib}

\newtheorem{definition}{Definition}

\newtheorem{theorem}{Theorem}

\theoremstyle{definition}
\newtheorem{example}{Example}

\begin{document}

\title{Efficient Sampling Algorithms for Approximate Temporal Motif Counting (Extended Version)\footnote{Jingjing Wang and Yanhao Wang contributed equally to this research.
This research was partially supported by 
the China Scholarships Council grant 201906130131,
the NSFC grant 61632009,
and the grant MOE2017-T2-1-141 from the Singapore Ministry of Education.
}}

\author[1]{Jingjing Wang\thanks{wangjingjing2019@hnu.edu.cn}}
\author[2]{Yanhao Wang\thanks{yanhao90@comp.nus.edu.sg}}
\author[1]{Wenjun Jiang\thanks{jiangwenjun@hnu.edu.cn}}
\author[3,4]{Yuchen Li\thanks{yuchenli@smu.edu.sg}}
\author[2]{Kian-Lee Tan\thanks{tankl@comp.nus.edu.sg}}
\affil[1]{Hunan University}
\affil[2]{National University of Singapore}
\affil[3]{Singapore Management University}
\affil[4]{Alibaba-Zhejiang University Joint Research Institute of Frontier Technologies}

\maketitle

\begin{abstract}
  A great variety of complex systems ranging from user interactions in communication networks
  to transactions in financial markets can be modeled as \emph{temporal graphs},
  which consist of a set of vertices and a series of timestamped and directed edges. 
  \emph{Temporal motifs} in temporal graphs are generalized from
  subgraph patterns in static graphs which take into account
  edge orderings and durations in addition to structures.
  Counting the number of occurrences of temporal motifs
  is a fundamental problem for temporal network analysis.
  However, existing methods either cannot support temporal motifs
  or suffer from performance issues. In this paper, we focus on
  approximate temporal motif counting via random sampling.
  We first propose a generic edge sampling (ES)
  algorithm for estimating the number of instances of any temporal motif.
  Furthermore, we devise an improved EWS algorithm that hybridizes
  edge sampling with wedge sampling for counting temporal motifs with $3$ vertices
  and $3$ edges. We provide comprehensive analyses of the theoretical bounds
  and complexities of our proposed algorithms. Finally, we conduct
  extensive experiments on several real-world datasets, and the results
  show that our ES and EWS algorithms have higher efficiency,
  better accuracy, and greater scalability than
  the state-of-the-art sampling method for temporal motif counting.
\end{abstract}

\section{Introduction}\label{sec:intro}

Graphs are one of the most fundamental data structures
that are widely used for modeling complex systems across diverse domains
from bioinformatics~\cite{DBLP:journals/bioinformatics/Przulj07},
to neuroscience~\cite{DBLP:journals/ploscb/VarshneyCPHC11},
to social sciences~\cite{FAUST2010221}.
Modern graph datasets increasingly incorporate temporal information
to describe the dynamics of relations over time.
Such graphs are referred to as
\emph{temporal graphs}~\cite{HOLME201297}
and typically represented by a set of vertices and
a sequence of timestamped and directed edges between vertices called \emph{temporal edges}.
For example, a communication
network~\cite{DBLP:conf/cikm/ZhaoTHOJL10,DBLP:conf/sigmod/GurukarRR15,DBLP:journals/pvldb/WangFLT17,DBLP:journals/tois/WangLFT18,DBLP:conf/edbt/WangLT19}
can be denoted by a temporal graph where each person
is a vertex and each message sent from one person to another is a temporal edge.
Similarly, computer networks and financial transactions
can also be modeled as temporal graphs.
Due to the ubiquitousness of temporal graphs,
they have attracted much attention~\cite{DBLP:conf/sigmod/GurukarRR15,DBLP:conf/wsdm/ParanjapeBL17,DBLP:conf/cikm/ZhaoTHOJL10,DBLP:conf/icde/LiSQYD18,DBLP:conf/cikm/NamakiWSLG17,DBLP:conf/cikm/GalimbertiBBCG18,DBLP:journals/pvldb/ShaLHT17,DBLP:journals/pvldb/GuoLST17}
recently.

\begin{figure}
  \centering
  \includegraphics[width=0.24\textwidth]{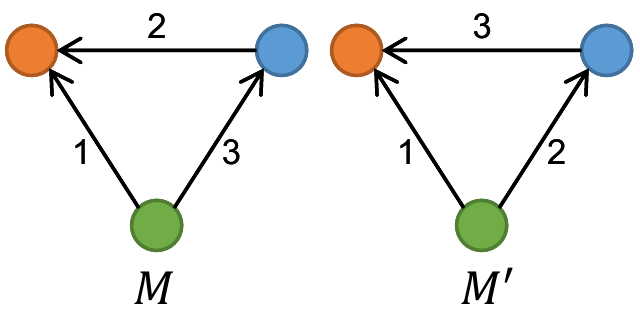}
  \vspace{-1em}
  \caption{Examples for temporal motifs}
  \label{fig:example0}
\end{figure}

One fundamental problem in temporal graphs with wide real-world
applications such as network characterization~\cite{DBLP:conf/wsdm/ParanjapeBL17},
structure prediction~\cite{DBLP:conf/wsdm/LiuBC19},
and fraud detection~\cite{DBLP:journals/pvldb/KumarC18},
is to count the number of occurrences of small (connected) subgraph patterns
(i.e., \emph{motifs}~\cite{Milo824}).
To capture the temporal dynamics in network analysis,
the notion of \emph{motif}~\cite{DBLP:conf/wsdm/ParanjapeBL17,DBLP:conf/edbt/KosyfakiMPT19,Kovanen_2011,DBLP:conf/wsdm/LiuBC19}
in temporal graphs is more general than its counterpart
in static graphs. It takes into account not only the subgraph structure
(i.e., \emph{subgraph isomorphism}~\cite{DBLP:journals/jacm/Ullmann76,DBLP:conf/sigmod/GuoL0HXT20})
but also the temporal information including edge ordering and motif duration.
As an illustrative example, $M$ and $M'$ in Figure~\ref{fig:example0}
are different temporal motifs. 
Though $M$ and $M'$ have exactly the same structure, they are different in the ordering of edges.
Consequently, although there has been a considerable amount of work
on subgraph counting in static graphs~\cite{DBLP:journals/tkdd/WangLRTZG14,DBLP:conf/www/JhaSP15,DBLP:conf/icde/WangLTZ16,DBLP:conf/wsdm/0002C00P17,DBLP:conf/www/PinarSV17,DBLP:journals/tkde/WangZZLCLTTG18,DBLP:conf/sdm/KoldaPS13,DBLP:conf/icdm/TurkogluT17,DBLP:conf/www/TurkT19,DBLP:conf/cikm/EtemadiLT16},
they cannot be used for temporal motif counting directly.

Generally, it is a challenging task to count temporal motifs.
Firstly, the problem is at least as hard as subgraph counting in static graphs,
whose time complexity increases exponentially with the number of edges in the query subgraph.
Secondly, it becomes even more computationally difficult
because the temporal information is considered.
For example, counting the number of instances of $k$-stars is simple
in static graphs; however, counting temporal $k$-stars is proven to be
NP-hard~\cite{DBLP:conf/wsdm/LiuBC19} due to the combinatorial nature of edge ordering.
Thirdly, temporal graphs are a kind of \emph{multigraph} that is permitted to have
multiple edges between the same two vertices at different timestamps.
As a result, there may exist many different instances of a temporal motif
within the same set of vertices, which leads to more challenges
for counting problems.
There have been a few methods for exact temporal motif
counting~\cite{DBLP:conf/wsdm/ParanjapeBL17}
or enumeration~\cite{DBLP:conf/bigdataconf/MackeyPFCC18,DBLP:journals/pvldb/KumarC18}.
However, they suffer from efficiency issues and often
cannot scale well in massive temporal graphs with hundreds of millions
of edges~\cite{DBLP:conf/wsdm/LiuBC19}.

In many scenarios, it is not necessary to count motifs exactly, and
finding an approximate number is sufficient for practical use.
A recent work~\cite{DBLP:conf/wsdm/LiuBC19} has proposed a
sampling method for approximate temporal motif counting.
It partitions a temporal graph into equal-time intervals,
utilizes an exact algorithm~\cite{DBLP:conf/bigdataconf/MackeyPFCC18}
to count the number of motif instances in a subset of intervals,
and computes an estimate from the per-interval counts.
However, this method still cannot achieve satisfactory
performance in massive datasets.
On the one hand, it fails to provide an accurate estimate
when the sampling rate and length of intervals are small.
On the other hand, its efficiency does not significantly improve upon that of exact methods
when the sampling rate and length of intervals are too large.

\vspace{1mm}
\noindent\textbf{Our Contributions:}
In this paper, we propose more efficient and accurate sampling algorithms
for approximate temporal motif counting.
First of all, we propose a generic Edge Sampling (ES) algorithm
to estimate the number of instances of any $k$-vertex $l$-edge temporal motif
in a temporal graph. The basic idea of our ES algorithm is to first uniformly
draw a set of random edges from the temporal graph, then exactly count
the number of \emph{local} motif instances that contain each sampled edge
by enumerating them, and finally compute the global motif count from local counts.
The ES algorithm exploits the \textsc{BackTracking} (BT) algorithm~\cite{DBLP:journals/jacm/Ullmann76,DBLP:conf/bigdataconf/MackeyPFCC18} for subgraph isomorphism
to enumerate local motif instances. We devise simple heuristics to determine
the matching order of a motif for the BT algorithm to reduce the search space.

Furthermore, temporal motifs with $3$ vertices and $3$ edges
(i.e., triadic patterns) are one of the most important classes of motifs,
whose distribution is an indicator to characterize
temporal networks~\cite{DBLP:conf/sdm/KoldaPS13,DBLP:conf/wsdm/ParanjapeBL17,DBLP:journals/snam/UzupyteW20,FAUST2010221}.
Therefore, we propose an improved Edge-Wedge Sampling (EWS) algorithm
that combines \emph{edge sampling} with
\emph{wedge sampling}~\cite{DBLP:conf/sdm/KoldaPS13,DBLP:conf/icdm/TurkogluT17}
specialized for counting any $3$-vertex $3$-edge temporal motif.
Instead of enumerating all instances containing a sampled edge,
the EWS algorithm generates a sample of \emph{temporal wedges}
(i.e., $3$-vertex $2$-edge motifs)
and estimates the number of local instances by counting how many
edges can match the query motif together with each sampled temporal wedge.
In this way, EWS avoids the computationally intensive enumeration
and greatly improves the efficiency upon ES.
Moreover, we analyze the theoretical bounds and complexities of both ES and EWS.

Finally, we test our algorithms on several real-world
datasets. The experimental results confirm the efficiency and
effectiveness of our algorithms: ES and EWS can provide
estimates with relative errors less than $1\%$ and $2\%$ in $37.5$ and $2.3$
seconds on a temporal graph with over $100$M edges, respectively.
In addition, they run up to $10.3$ and $48.5$ times faster than
the state-of-the-art sampling method while having lower estimation errors.

\vspace{1mm}
\noindent\textbf{Organization:}
The remainder of this paper is organized as follows.
Section~\ref{sec:related:work} reviews the related work.
Section~\ref{sec:def} introduces the background and
formulation of \emph{temporal motif counting}.
Section~\ref{sec:alg} presents the ES and EWS algorithms
for temporal motif counting and analyzes them theoretically.
Section~\ref{sec:exp} describes the setup and results of the experiments.
Finally, Section~\ref{sec:conclusion} provides some concluding remarks.

\section{Related Work}\label{sec:related:work}

\noindent\textbf{Random Sampling for Motif Counting:}
In recent years, there have been great efforts to (approximately) count
the number of occurrences of a motif in a large graph via random sampling.
First of all, many sampling methods
such as subgraph sampling~\cite{DBLP:conf/kdd/TsourakakisKMF09},
edge sampling~\cite{DBLP:conf/kdd/AhmedDNK14,DBLP:conf/kdd/LimK15,DBLP:journals/pvldb/WangQSZTG17},
color sampling~\cite{DBLP:journals/ipl/PaghT12},
neighborhood sampling~\cite{DBLP:journals/pvldb/PavanTTW13},
wedge sampling~\cite{DBLP:conf/kdd/JhaSP13,DBLP:conf/sdm/KoldaPS13,DBLP:conf/icdm/TurkogluT17,DBLP:conf/www/TurkT19},
and reservoir sampling~\cite{DBLP:conf/kdd/StefaniERU16},
were proposed for approximate triangle counting
(see~\cite{DBLP:journals/tkde/WuYL16} for an experimental analysis).
Moreover, sampling methods were also used for estimating more complex motifs,
e.g., 4-vertex motifs~\cite{DBLP:conf/www/JhaSP15,DBLP:conf/kdd/Sanei-MehriST18},
5-vertex motifs~\cite{DBLP:conf/www/PinarSV17,DBLP:journals/tkdd/WangLRTZG14,DBLP:conf/icde/WangLTZ16,DBLP:journals/tkde/WangZZLCLTTG18}, motifs with 6 or more vertices~\cite{DBLP:conf/wsdm/0002C00P17},
and $k$-cliques~\cite{DBLP:conf/www/JainS17}.
However, all above methods were proposed for static graphs and
did not consider the temporal information and ordering of edges.
Thus, they could not be applied to temporal motif counting directly.

\vspace{1mm}
\noindent\textbf{Motifs in Temporal Networks:}
Prior studies have considered different types of \emph{temporal network motifs}.
Viard et al.~\cite{DBLP:conf/asunam/ViardLM15,DBLP:journals/tcs/ViardLM16}
and Himmel et al.~\cite{DBLP:conf/asunam/HimmelMNS16} extended the notion of \emph{maximal clique}
to temporal networks and proposed efficient algorithms for maximal clique enumeration.
Li et al.~\cite{DBLP:conf/icde/LiSQYD18} proposed the notion of
\emph{$(\theta,\tau)$-persistent $k$-core}
to capture the persistence of a community in temporal networks.
However, these notions of temporal motifs were different from ours
since they did not take \emph{edge ordering} into account.
Zhao et al.~\cite{DBLP:conf/cikm/ZhaoTHOJL10} and Gurukar et al.~\cite{DBLP:conf/sigmod/GurukarRR15}
studied the \emph{communication motifs}, which are frequent subgraphs to characterize
the patterns of information propagation in social networks.
Kovanen et al.~\cite{Kovanen_2011} and Kosyfaki et al.~\cite{DBLP:conf/edbt/KosyfakiMPT19}
defined the \emph{flow motifs} to model flow transfer among a set of vertices within a time window
in temporal networks. Although both definitions accounted for edge ordering,
they were more restrictive than ours because the former assumed any two adjacent edges must occur within a fixed time span while the latter assumed edges in a motif must be consecutive
events for a vertex~\cite{DBLP:conf/wsdm/ParanjapeBL17}.

\vspace{1mm}
\noindent\textbf{Temporal Motif Counting \& Enumeration:}
There have been several existing studies on counting and enumerating temporal
motifs. Paranjape et al.~\cite{DBLP:conf/wsdm/ParanjapeBL17} first formally defined
the notion of \emph{temporal motifs}
we use in this paper. They proposed exact algorithms for counting temporal motifs
based on subgraph enumeration in static graphs and timestamp-based pruning.
Kumar and Calders~\cite{DBLP:journals/pvldb/KumarC18} proposed an efficient algorithm called 2SCENT to
enumerate all simple temporal cycles in a directed interaction network.
Although 2SCENT was shown to be effective for cycles,
it could not be used for enumerating temporal motifs of any other type.
Mackey et al.~\cite{DBLP:conf/bigdataconf/MackeyPFCC18} proposed an efficient \textsc{BackTracking}
algorithm for temporal subgraph isomorphism. The algorithm could count temporal motifs
exactly by enumerating all of them.
Liu et al.~\cite{DBLP:conf/wsdm/LiuBC19} proposed an interval-based sampling framework
for counting temporal motifs. To the best of our knowledge, this is the only
existing work on approximate temporal motif counting via sampling.
In this paper, we present two improved sampling algorithms for temporal motif counting
and compare them with the algorithms
in~\cite{DBLP:conf/wsdm/ParanjapeBL17,DBLP:conf/bigdataconf/MackeyPFCC18,DBLP:journals/pvldb/KumarC18,DBLP:conf/wsdm/LiuBC19}
for evaluation.

\section{Preliminaries}\label{sec:def}

In this section, we formally define \emph{temporal graphs}, \emph{temporal motifs},
and the problem of \emph{temporal motif counting} on a temporal graph.
Here, we follow the definition of \emph{temporal motifs} in~\cite{DBLP:conf/wsdm/ParanjapeBL17,DBLP:conf/wsdm/LiuBC19,DBLP:conf/bigdataconf/MackeyPFCC18}
for its simplicity and generality. Other types of temporal motifs have been discussed
in Section~\ref{sec:related:work}.

\vspace{1mm}
\noindent\textbf{Temporal Graph:}
A \emph{temporal graph} $T=(V_T,E_T)$ is defined by
a set $V_T$ of $n$ vertices and a sequence $E_T$ of $m$ temporal edges
among vertices in $V_T$.
Each temporal edge $e=(u,v,t)$ where $u,v \in V_T$ and $t \in \mathbb{R}^{+}$
is a timestamped directed edge from $u$ to $v$ at time $t$.
There may be many temporal edges from $u$ to $v$ at different timestamps
(e.g., a user can comment on the posts of another user many times on \emph{Reddit}).
For ease of presentation, we assume the timestamp $t$ of each temporal edge $e$ is unique
so that the temporal edges in $E_T$ are strictly ordered.
Note that our algorithms can also handle the case when timestamps
are non-unique by using any consistent rule to break ties.

\vspace{1mm}
\noindent\textbf{Temporal Motif:}
We formalize the notion of \emph{temporal motifs}~\cite{DBLP:conf/wsdm/ParanjapeBL17,DBLP:conf/wsdm/LiuBC19}
in the following definition.
\begin{definition}[Temporal Motif]
  A temporal motif $M=(V_M,$ $E_M,\sigma)$ consists of a (connected) graph 
  with a set of $k$ vertices $V_M$ and a set of $l$ edges $E_M$,
  and an ordering $\sigma$ on the edges in $E_M$.
\end{definition}

Intuitively, a temporal motif $M$ can be represented as an ordered sequence of edges
$\langle e^{\prime}_1=(u^{\prime}_1,v^{\prime}_1),\ldots,e^{\prime}_l=(u^{\prime}_l,v^{\prime}_l) \rangle$.
Given a temporal motif $M$ as a template pattern,
we aim to count how many times this pattern appears in a temporal graph $T$.
Furthermore, we only consider the instances where the pattern is formed within a short time span.
For example, an instance formed in an hour is more interesting than
one formed accidentally in one year on a communication network~\cite{DBLP:conf/cikm/ZhaoTHOJL10,DBLP:conf/sigmod/GurukarRR15,DBLP:conf/wsdm/ParanjapeBL17}.
Therefore, given a temporal graph $T$ and a temporal motif $M$,
our goal is to find a sequence of edges $S \subseteq E_T$ such that
(1) $S$ exactly matches (i.e., \emph{is isomorphic to}) $M$,
(2) $S$ is in the same order as specified by $\sigma$,
and (3) all edges in $S$ occur within a time span of at most $\delta$.
We call such an edge sequence $S$ as
a \emph{$\delta$-instance}~\cite{DBLP:conf/wsdm/ParanjapeBL17,DBLP:conf/wsdm/LiuBC19} of $M$
and the difference between $t_l$ and $t_1$ as the \emph{duration} $\Delta(S)$ of instance $S$.
The formal definition is given in the following.
\begin{definition}[Motif $\delta$-instance]
  A sequence of $l$ edges $S=\langle (w_1,x_1,t_1),\ldots,(w_l,$ $x_l,t_l) \rangle$
  ($t_1 < \ldots < t_l$) from a temporal graph $T$ is a $\delta$-instance of
  a temporal motif
  $M=\langle (u^{\prime}_1,v^{\prime}_1),\ldots,(u^{\prime}_l,v^{\prime}_l) \rangle$ if
  (1) there exists a bijection $f$ between the vertex sets of $S$ and $M$
  such that $f(w_i)=u^{\prime}_i$ and $f(x_i)=v^{\prime}_i$ for $i = 1,\ldots,l$;
  and (2) the duration $\Delta(S)$ is at most $\delta$, i.e., $t_l - t_1 \leq \delta$.
\end{definition}

\begin{figure}
  \centering
  \includegraphics[width=0.5\textwidth]{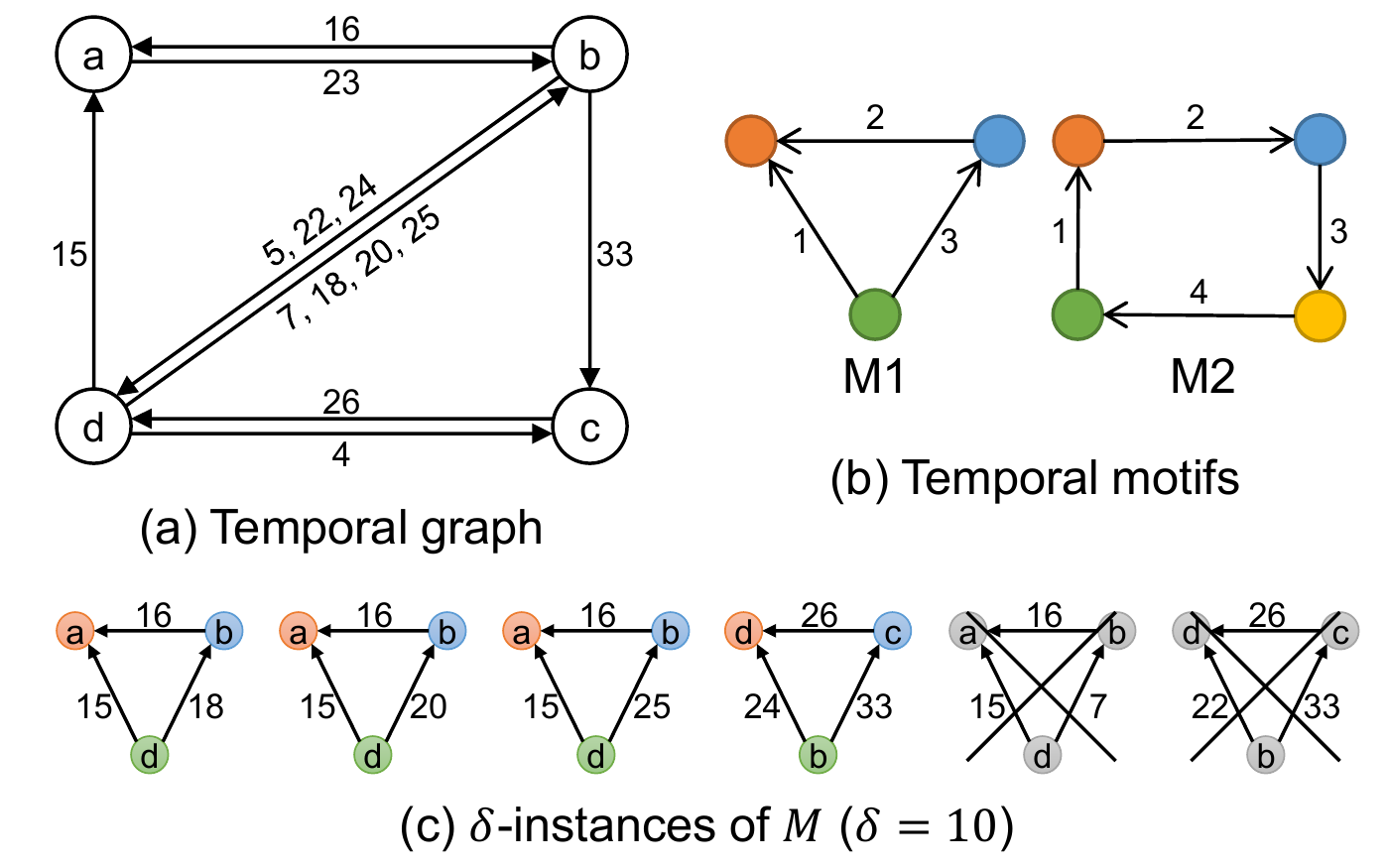}
  \vspace{-1em}
  \caption{Example for temporal graph and motifs}
  \label{fig:example}
\end{figure}
\begin{example}
  In Figure~\ref{fig:example}(a), we illustrate a temporal graph
  with $4$ vertices and $13$ temporal edges.
  Let us consider the problem of finding all $\delta$-instances ($\delta=10$)
  of temporal motif $M1$ in Figure~\ref{fig:example}(b).
  As shown in Figure~\ref{fig:example}(c), there are $4$ valid $10$-instances of $M1$ found.
  These instances can match $M1$ in terms of both structure and edge ordering
  and their durations are within $10$. In addition, we also give $2$ invalid instances of $M1$,
  which are isomorphic to $M1$ but violate either the edge ordering or duration constraint.
\end{example}

\noindent\textbf{Temporal Motif Counting:}
According to the above notions, we present the \emph{temporal motif counting}
problem studied in this paper.

\begin{definition}[Temporal Motif Counting]\label{def:count}
  For a temporal graph $T$, a temporal motif $M$, and a time span $\delta$,
  the temporal motif counting problem returns the number $C_M$
  of $\delta$-instances of $M$ appeared in $T$.
\end{definition}

The temporal motif counting problem has proven to be
NP-hard for very simple motifs, e.g.~$k$-stars~\cite{DBLP:conf/wsdm/LiuBC19},
because the edge ordering is taken into account.
According to previous results~\cite{DBLP:conf/wsdm/LiuBC19},
although there is a simple polynomial algorithm to
count the number of $k$-stars on a static graph,
it is NP-hard to exactly count the number of temporal $k$-stars.
Typically, counting temporal motifs exactly on massive graphs
with millions or even billions of edges is a computationally intensive
task~\cite{DBLP:conf/wsdm/ParanjapeBL17,DBLP:conf/wsdm/LiuBC19}.
Therefore, we focus on designing efficient and scalable sampling algorithms
for estimating the number of temporal motifs approximately in Section~\ref{sec:alg}.
The frequently used notations are summarized in Table~\ref{tab:freq}.

\begin{table}[t]
  \small
  \centering
  \caption{Frequently used notations}\label{tab:freq}
  \begin{tabular}{c l}
    \hline
    \textbf{Symbol} & \textbf{Description} \\
    \hline
    $T$                   & Temporal graph \\
    $V_T,E_T$             & Set of vertices and edges in $T$ \\
    $n,m$                 & Number of vertices and edges in $T$ \\
    $M$                   & Temporal motif \\
    $V_M,E_M$             & Set of vertices and edges in $M$ \\
    $k,l$                 & Number of vertices and edges in $M$ \\
    $\delta$              & Maximum time span of a motif instance \\
    $S$                   & Motif $\delta$-instance \\
    $C_M$                 & Number of $\delta$-instances of $M$ in $T$ \\
    $\widehat{C}_M$       & Unbiased estimator of $C_M$ \\
    $p$                   & Probability of edge sampling \\
    $\widehat{E}_T$		    & Set of sampled edges from $E_T$\\
    $\eta(e)$             & Number of $\delta$-instances of $M$ containing edge $e$ \\
    $\eta_j(e)$           & Number of $\delta$-instances of $M$ when $e$ is mapped to $e_j^\prime$ \\
    $q$                   & Probability of wedge sampling \\
    $W$                   & Temporal wedge \\
    $\eta(W)$             & Number of $\delta$-instances of $M$ containing $W$ \\
    $W_j^\prime$          & Temporal wedge pattern for $M$ when $e$ is mapped to $e_j^\prime$ \\
    $\widehat{\mathcal{W}}_j(e)$ & Set of sampled $\delta$-instances of $W_j^\prime$  \\
    $\widehat{\eta}_j(e)$ & Unbiased estimator of $\eta_j(e)$ \\
    \hline
  \end{tabular}
\end{table}

\section{Our Algorithms}\label{sec:alg}

In this section, we present our proposed algorithms for
approximate temporal motif counting in detail.
We first describe our generic Edge Sampling (ES) algorithm in Section~\ref{subsec:alg:1}.
Then, we introduce our improved EWS algorithm specific for counting
$3$-vertex $3$-edge temporal motifs in Section~\ref{subsec:alg:2}.
In addition, we theoretically analyze the expected values and variances
of the estimates returned by both algorithms.
Finally, we discuss the streaming implementation of our algorithms
in Section~\ref{subsec:alg:discuss}.

\subsection{The Generic Edge Sampling Algorithm}\label{subsec:alg:1}

The Edge Sampling (ES) algorithm is motivated by
an exact subgraph counting algorithm called
\emph{edge iterator}~\cite{DBLP:journals/tkde/WuYL16}.
Given a temporal graph $T$, a temporal motif $M$, and a time span $\delta$,
we use $\eta(e)$ to denote the number of local $\delta$-instances of $M$ containing an edge $e$.
To count all $\delta$-instances of $M$ in $T$ exactly,
we can simply count $\eta(e)$ for each $e \in E_T$ and then sum them up.
In this way, each instance is counted $l$ times
and the total number of instances is equal to the sum divided by $l$,
i.e.,~$C_M=\frac{1}{l}\sum_{e \in E_T}\eta(e)$.

Based on the above idea, we propose the ES algorithm for estimating $C_M$:
For each edge $e \in E_T$, we randomly sample it and compute $\eta(e)$
with fixed probability $p$. Then, we acquire an unbiased estimator
$\widehat{C}_M$ of $C_M$ by adding up $\eta(e)$ for each sampled edge $e$
and scaling the sum by a factor of $\frac{1}{pl}$, i.e.,
$\widehat{C}_M=\frac{1}{pl}\sum_{e \in \widehat{E}_T}\eta(e)$
where $\widehat{E}_T$ is the set of sampled edges.

Now the remaining problem becomes how to compute $\eta(e)$ for an edge $e$.
The ES algorithm adopts the well-known \textsc{BackTracking}
algorithm~\cite{DBLP:journals/jacm/Ullmann76,DBLP:conf/bigdataconf/MackeyPFCC18}
to enumerate all $\delta$-instances that contain an edge $e$ for computing $\eta(e)$.
Specifically, the \textsc{BackTracking} algorithm runs $l$ times for each edge $e$;
in the $j$\textsuperscript{th} run, it first maps edge $e$ to the $j$\textsuperscript{th}
edge $e_j^\prime$ of $M$ and then uses a tree search
to find all different combinations of the remaining $l-1$ edges that can form
$\delta$-instances of $M$ with edge $e$.
Let $\eta_j(e)$ be the number of $\delta$-instances of $M$ where $e$ is mapped
to $e_j^\prime$. It is obvious that $\eta(e)$ is equal
to the sum of $\eta_j(e)$ for $j=1,\ldots,l$,
i.e., $\eta(e)=\sum_{j=1}^{l}\eta_j(e)$.

\begin{algorithm}
  \caption{Edge Sampling}\label{alg:es}
  \KwIn{Temporal graph $T$, temporal motif $M$, time span $\delta$, edge sampling probability $p$.}
  \KwOut{Estimator $\widehat{C}_M$ of the number of $\delta$-instances of $M$ in $T$}
  Initialize $\widehat{E}_T \gets \varnothing$\;\label{ln:es:sample:s}
  \ForEach{$e\in E_T$}
  {
    Toss a biased coin with success probability $p$\;
    \If{success}
    {
      $\widehat{E}_T \gets \widehat{E}_T \cup \{e\}$\;\label{ln:es:sample:t}
    }
  }
  \ForEach{$e=(u,v,t) \in \widehat{E}_T$\label{ln:es:count:s}}
  {
    Set $\eta(e) \gets 0$\;
    \For{$j \in 1,\ldots,l$}
    {
      Generate an initial instance $S_j^{(1)}$ by mapping $e$ to $e_j^\prime$\;
      \label{ln:es:init}
      Run \textsc{BackTracking} on $E_T[t-\delta,t+\delta]$ starting from $S_j^{(1)}$
      to find the set $\mathcal{S}_j(e) = \{S_j(e) \,:\, S_j(e)$
      is a $\delta$-instance of $M$ where $e$ is mapped to $e_j^\prime\}$\;
      \label{ln:es:bt}
      Set $\eta_j(e) \gets |\mathcal{S}_j(e)|$ and $\eta(e) \gets \eta(e) + \eta_j(e)$\;
      \label{ln:es:count:t}
    }
  }
  \Return{$\widehat{C}_M \gets \frac{1}{pl}\sum_{e \in \widehat{E}_T}\eta(e)$}\;
  \label{ln:es:estimate}
\end{algorithm}

We depict the procedure of our ES algorithm in Algorithm~\ref{alg:es}.
The first step of ES is to generate a random sample $\widehat{E}_T$ of edges
from the edge set $E_T$ where the probability of adding any edge is $p$
(Lines~\ref{ln:es:sample:s}--\ref{ln:es:sample:t}).
Then, in the second step (Lines~\ref{ln:es:count:s}--\ref{ln:es:count:t}),
it counts the number $\eta(e)$ of local $\delta$-instances of $M$ for each sampled edge $e$
by running the \textsc{BackTracking} algorithm
to enumerate each instance $S_j(e)$ that is a $\delta$-instance of $M$
and maps $e$ to $e_j^\prime$ for $j=1,\ldots,l$.
Note that \textsc{BackTracking} (BT) runs on a subset $E_T[t-\delta,t+\delta]$
of $E_T$ which consists of all edges with timestamps from $t-\delta$ to $t+\delta$
for edge $e=(u,v,t)$ since it is safe to ignore any other edge
due to the duration constraint.
Here, we omit the detailed procedure of the BT algorithm
because it generally follows an existing algorithm for subgraph
isomorphism in temporal graphs~\cite{DBLP:conf/bigdataconf/MackeyPFCC18}.
The main difference between our algorithm
and the one in~\cite{DBLP:conf/bigdataconf/MackeyPFCC18} lies
in the matching order, which will be discussed later. 
After counting $\eta(e)$ for each sampled edge $e$,
it finally returns an estimate $\widehat{C}_M$ of $C_M$
(Line~\ref{ln:es:estimate}).

\vspace{1mm}
\noindent\textbf{Matching Order for \textsc{BackTracking}:}
Now we discuss how to determine the matching order of a temporal motif.
The BT algorithm in~\cite{DBLP:conf/bigdataconf/MackeyPFCC18}
adopts a time-first matching order:
it always matches the edges of $M$ in order of
$\langle e_1^\prime,\ldots,e_l^\prime \rangle$.
The advantage of this matching order is that it best exploits
the temporal information for search space pruning. For a partial instance
$S^{(j)}=\langle (w_1,x_1,t_1),\ldots,(w_j,$ $x_j,t_j) \rangle$ after
$e_j^\prime$ is mapped, the search space for mapping $e_{j+1}^\prime$
is restricted to $E_T[t_j,t_1+\delta]$.
However, the time-first matching order may not work well in the
ES algorithm. First, it does not consider the \emph{connectivity} of
the matching order: If $e_{j+1}^\prime$ is not connected with any prior edge,
it has to be mapped to all edges in $E_T[t_j,t_1+\delta]$,
which may lead to a large number of redundant partial matchings.
Second, the time-first order is violated by Line~\ref{ln:es:init} of Algorithm~\ref{alg:es}
when $j>1$ since it first maps $e$ to $e_{j}^\prime$.

In order to overcome the above two drawbacks, we propose two heuristics to determine the matching
order of a given motif $M$ for reducing the search space,
and generate $l$ matching orders for $M$, in each of which $e_{j}^\prime$ ($j=1,\ldots,l$)
is placed first:
(1) \emph{enforcing connectivity}: For each $i=2,\ldots,l $, the $i$\textsuperscript{th}
edge in the matching order must be adjacent to at least one prior edge that has been matched;
(2) \emph{boundary edge first}: If there are multiple unmatched edges that satisfy
the \emph{connectivity} constraint, the boundary edge (i.e., the first or last unmatched
edge in the ordering $\sigma$ of $M$) will be matched first.
The first rule can avoid redundant partial matchings and
the second rule can restrict the temporal range of tree search,
both of which are effective for search space pruning.

\begin{example}
  We consider how to decide the matching orders of $M2$ (i.e., $4$-simple temporal cycle)
  in Figure~\ref{fig:example}(b). When $e_1^\prime$ is placed first, we can select
  $e_2^\prime$ or $e_4^\prime$ as the second edge according to the
  \emph{enforcing connectivity} rule; and $e_4^\prime$ is selected
  according to the \emph{boundary edge first} rule. Then, either $e_2^\prime$ or
  $e_3^\prime$ can be selected as the next edge since they both satisfy two rules.
  Therefore, either $\langle e_1^\prime,e_4^\prime,e_2^\prime,e_3^\prime \rangle$
  or $\langle e_1^\prime,e_4^\prime,e_3^\prime,e_2^\prime \rangle$ is a valid matching order.
  Accordingly, $\langle e_2^\prime,e_1^\prime,e_4^\prime,e_3^\prime \rangle$,
  $\langle e_3^\prime,e_4^\prime,e_1^\prime,e_2^\prime \rangle$,
  and $\langle e_4^\prime,e_1^\prime,e_2^\prime,e_3^\prime \rangle$
  are valid matching orders when $e_2^\prime$, $e_3^\prime$,
  and $e_4^\prime$ in $M2$ are placed first, respectively.
\end{example}

\begin{figure}
  \centering
  \includegraphics[width=0.6\textwidth]{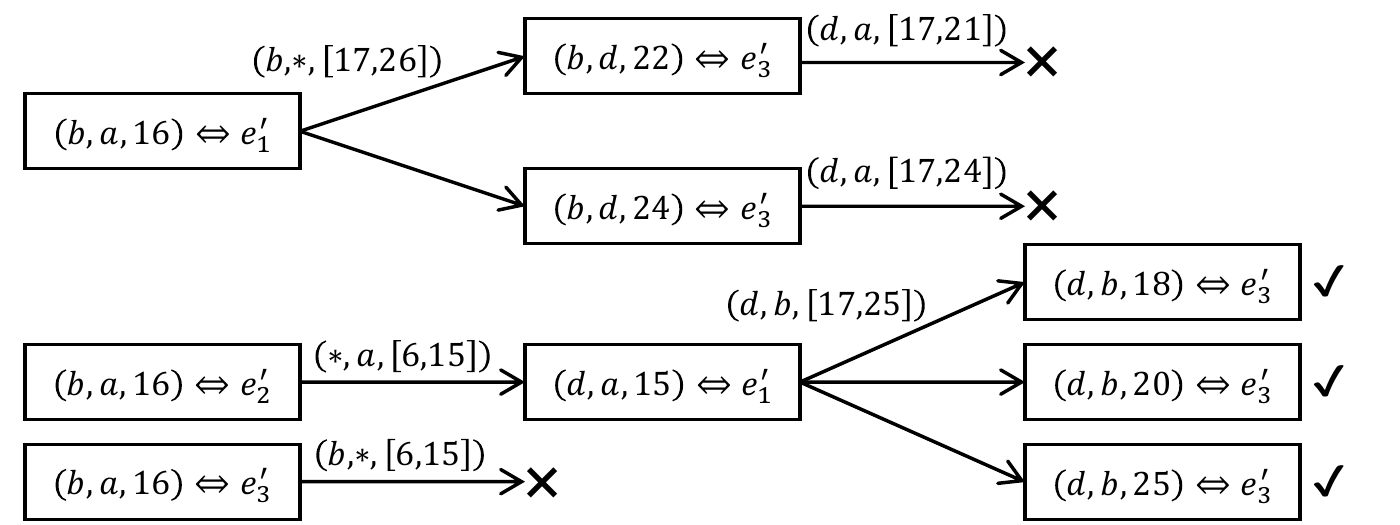}
  \vspace{-1em}
  \caption{Example of enumerating $\delta$-instances ($\delta=10$) of motif $M1$
  for edge $(b,a,16)$ in Figure~\ref{fig:example} using \textsc{Backtracking}}
  \label{fig:backtracking}
\end{figure}

\begin{example}
  In Figure~\ref{fig:backtracking}, we show how to use \textsc{Backtracking}
  to enumerate $\delta$-instances of $M1$ ($\delta=10$) for $e=(b,a,16)$
  in Figure~\ref{fig:example}. There are three tree search procedures in each of which
  $e$ is mapped to $e_1^\prime$, $e_2^\prime$, and $e_3^\prime$, respectively.
  The condition of each mapping step is given in form of $(v_s,v_t,[t_s,t_t])$
  where $v_s$ and $v_t$ are the starting and ending vertices and $[t_s,t_t]$
  is the range of timestamps.
  Here, ``$*$'' means that it can be mapped to an arbitrary unmapped vertex.
  Moreover, we use `$\checkmark$' and `\ding{53}' to denote a successful matching
  and a failed partial matching, respectively. We find three $10$-instances
  of $M1$ and thus $\eta(e)=3$. When we run ES with $p=0.25$ and
  $\widehat{E}_T=\{(d,c,4),(b,a,16),(b,d,24)\}$, since the numbers of $10$-instances
  containing $(d,c,4)$ and $(b,d,24)$ are respectively $0$ and $1$,
  we can compute $\widehat{C}_M=\frac{3+0+1}{0.25 \times 3} \approx 5.33$.
\end{example}

\noindent\textbf{Theoretical Analysis:}
Next, we analyze the estimate $\widehat{C}_M$ returned by Algorithm~\ref{alg:es}
theoretically.
We first prove that $\widehat{C}_M$ is an unbiased estimator of $C_M$ in Theorem~\ref{thm:es:exp}.
The variance of $\widehat{C}_M$ is given in Theorem~\ref{thm:es:var}.

\begin{theorem}\label{thm:es:exp}
  The expected value $\mathbb{E}[\widehat{C}_M]$ of $\widehat{C}_M$
  returned by Algorithm~\ref{alg:es} is $C_M$.
\end{theorem}
\begin{proof}
  Here, we consider the edges in $E_T$ are indexed by $[1,m]$
  and use an indicator $\omega_i$ to denote whether the $i$\textsuperscript{th}
  edge $e_i$ is sampled, i.e.,
  \begin{equation*}
  \omega_i = \begin{cases}
              1, & e_i \in \widehat{E}_T \\
              0, & e_i \notin \widehat{E}_T
             \end{cases}
  \end{equation*}
  Then, we have
  \begin{equation}\label{Eq:n1}
    \widehat{C}_M = \frac{1}{pl}\sum_{e \in \widehat{E}_T}\eta(e)
    = \frac{1}{pl}\sum_{i=1}^{m}\omega_{i}\cdot\eta(e_i)
  \end{equation}
  Next, based on Equation~\ref{Eq:n1} and the fact that $\mathbb{E}[\omega_{i}]=p$, we have
  \begin{equation*}
    \mathbb{E}[\widehat{C}_M] = \frac{1}{pl}\sum_{i=1}^{m}\mathbb{E}[\omega_{i}]\cdot\eta(e_i)
    =\frac{1}{l}\sum_{i=1}^{m}\eta(e_i)=C_M
  \end{equation*}
  and conclude the proof.
\end{proof}

\begin{theorem}\label{thm:es:var}
  The variance $\textnormal{Val}[\widehat{C}_M]$ of $\widehat{C}_M$
  returned by Algorithm~\ref{alg:es} is at most $\frac{1-p}{p} \cdot C_M^2$.
\end{theorem}
\begin{proof}
 According to Equation~\ref{Eq:n1}, we have
 \begin{equation*}
   \textnormal{Val}[\widehat{C}_M]
   =\textnormal{Val}\Big[\sum_{i=1}^m \frac{\eta(e_i)}{pl} \cdot \omega_i \Big]
   =\sum_{i,j=1}^{m} \frac{\eta(e_i)}{pl} \cdot \frac{\eta(e_j)}{pl}
   \cdot \textnormal{Cov}(\omega_i,\omega_j)
 \end{equation*}
 Because the indicators $\omega_{i}$ and $\omega_{j}$ are independent if $i \neq j$,
 we have $\textnormal{Cov}(\omega_i,\omega_j)=0$ for any $i \neq j$.
 In addition, $\textnormal{Cov}(\omega_i,\omega_i)=\textnormal{Val}[\omega_i]=p-p^2$.
 Based on the above results, we have
 \begin{align*}
   \textnormal{Val}[\widehat{C}_M]
   & =\sum_{i=1}^{m} \frac{\eta^2(e_i)}{p^2 l^2}(p-p^2)
     =\frac{1-p}{pl^2}\sum_{i=1}^{m}\eta^2(e_i) \\
   & \leq \frac{1-p}{pl^2}\Big(\sum_{i=1}^{m}\eta(e_i)\Big)^2
     =\frac{1-p}{p} \cdot C_M^2
 \end{align*}
 and conclude the proof.
\end{proof}

Finally, we can acquire Theorem~\ref{thm:es:est} by applying Chebyshev's inequality
to the result of Theorem~\ref{thm:es:var}.
\begin{theorem}\label{thm:es:est}
  $\textnormal{Pr}[|\widehat{C}_M-C_M|\geq\varepsilon\cdot C_M]\leq\frac{1-p}{p\varepsilon^2}$
\end{theorem}
\begin{proof}
  By applying Chebyshev's inequality, we have
  $\textnormal{Pr}[|\widehat{C}_M-C_M|\geq\varepsilon\cdot C_M]
  \leq\frac{\textnormal{Val}[\widehat{C}_M]}{\varepsilon^2 C_M^2}$
  and thus prove the theorem by substituting $\textnormal{Val}[\widehat{C}_M]$
  with $\frac{1-p}{p} \cdot C_M^2$ according to Theorem~\ref{thm:es:var}.
\end{proof}
According to Theorem~\ref{thm:es:est}, we can say $\widehat{C}_M$
is an $(\varepsilon,\gamma)$-estimator of $C_M$ for parameters $\varepsilon,\gamma \in (0,1)$,
i.e., $\textnormal{Pr}[|\widehat{C}_M-C_M|<\varepsilon\cdot C_M]> 1-\gamma$,
when $p=\frac{1}{1+\gamma\varepsilon^{2}}$.

\vspace{1mm}
\noindent\textbf{Time Complexity:}
We first analyze the time complexity of computing $\eta(e)$ for an edge $e$.
For \textsc{BackTracking}, the search space of each matching step is
at most the number of (in-/out-)edges within range $[t-\delta,t]$
or $[t,t+\delta]$ connected with a vertex $v$.
Here, we use $d_{\delta}$ to denote the maximum number of (in-/out-)edges
connected with one vertex within any $\delta$-length time interval.
The time complexity of \textsc{BackTracking} is $O(d_{\delta}^{l-1})$ and
thus the time complexity of computing $\eta(e)$ is $O(l d_{\delta}^{l-1})$.
Therefore, ES provides an $(\varepsilon,\gamma)$-estimator of $C_M$
in $O(\frac{ml d_{\delta}^{l-1}}{1+\gamma\varepsilon^2})$ time.

\subsection{The Improved EWS Algorithm}\label{subsec:alg:2}

\begin{figure}
  \centering
  \includegraphics[width=0.5\textwidth]{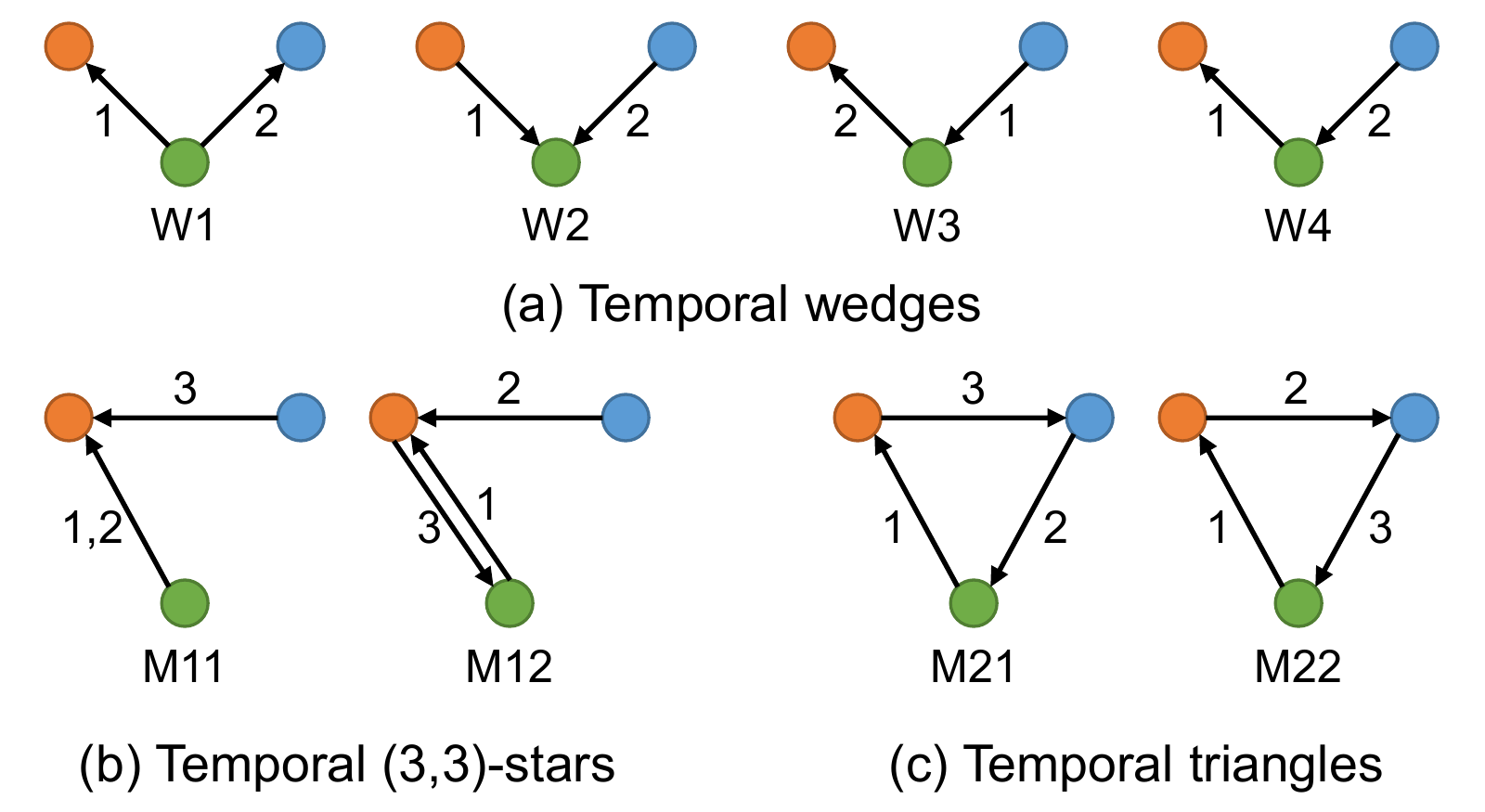}
  \vspace{-1em}
  \caption{Temporal wedges, temporal $(3,3)$-stars, and temporal triangles}
  \label{fig:wedge}
\end{figure}

The ES algorithm in Section~\ref{subsec:alg:1} is generic and able to
count any connected temporal motif. 
Nevertheless, there are still opportunities to further reduce the computational
overhead of ES when the query motif is limited to $3$-vertex $3$-edge
temporal motifs (i.e., triadic patterns),
which are one of the most important classes of motifs to characterize 
temporal networks~\cite{DBLP:conf/sdm/KoldaPS13,DBLP:conf/wsdm/ParanjapeBL17,DBLP:journals/snam/UzupyteW20,FAUST2010221}.
In this section, we propose an improved Edge-Wedge Sampling (EWS) algorithm that combines
\emph{edge sampling} with \emph{wedge sampling} for counting $3$-vertex $3$-edge temporal motifs.

Wedge sampling~\cite{DBLP:conf/sdm/KoldaPS13,DBLP:conf/icdm/TurkogluT17,DBLP:conf/www/TurkT19,DBLP:journals/tkde/WuYL16} is a widely used method for triangle counting.
Its basic idea is to draw a sample of wedges (i.e., $3$-vertex $2$-edge subgraph patterns)
uniformly from a graph and check the ratio of ``closed wedges'' (i.e., form a triangle in the graph)
 to estimate the number of triangles.
However, traditional wedge-sampling methods are proposed for undirected static graphs
and cannot be directly used on temporal graphs.
First, they consider that all wedges are isomorphic and treat them equally.
But there are four temporal wedge patterns with different edge directions and orderings
as illustrated in Figure~\ref{fig:wedge}(a).
Second, they are designed for simple graphs where one wedge can form
at most one triangle. However, since temporal graphs are multigraphs and there may exist
multiple edges between the same two vertices, one temporal wedge can participate in
more than one instance of a temporal motif.
Therefore, in the EWS algorithm, we extend \emph{wedge sampling} for temporal motif counting by addressing both issues.

The detailed procedure of EWS is presented in Algorithm~\ref{alg:ews}.
First of all, it uses the same method as ES to sample a set $\widehat{E}_T$
of edges (Line~\ref{ln:ews:sample}).
For each sampled edge $e\in \widehat{E}_T$ and $j = 1,2,3$,
it also maps $e$ to $e_j^\prime$ 
for computing $\eta_j(e)$ (Line~\ref{ln:ews:mapping}),
i.e., the number of $\delta$-instances of $M$ where $e$ is mapped to $e_j^\prime$.
But, instead of running \textsc{BackTracking} to compute $\eta_j(e)$ exactly,
it utilizes \emph{temporal wedge sampling} to estimate $\eta_j(e)$ approximately 
without fully enumeration (Lines~\ref{ln:ews:wedge:begin}--\ref{ln:ews:est:end}),
which is divided into two subroutines as discussed later.
At last, it acquires an estimate $\widehat{C}_M$ of $C_M$
from each estimate $\widehat{\eta}_j(e)$ of $\eta_j(e)$
using a similar method to ES (Line~\ref{ln:ews:estimate}).

\begin{algorithm} 
  \caption{Edge-Wedge Sampling}\label{alg:ews}
  \KwIn{Temporal graph $T$, temporal motif $M$, time span $\delta$, edge sampling probability $p$, wedge sampling probability $q$.}
  \KwOut{Estimator $\widehat{C}_M$ of the number of $\delta$-instances of $M$ in $T$}
  Generate $\widehat{E}_T$ using Line~\ref{ln:es:sample:s}--\ref{ln:es:sample:t}
  of Algorithm~\ref{alg:es}\;\label{ln:ews:sample}
  \ForEach{$e=(u,v,t) \in \widehat{E}_T$\label{ln:ews:wedge:s}}
  {
    \For{$j \gets 1,2,3$}
    {
      Map edge $e$ to $e_j^\prime$\;\label{ln:ews:mapping}
      Initialize $\widehat{\eta}_j(e) \gets 0$
      and $\widehat{\mathcal{W}}_j(e) \gets \varnothing$\;\label{ln:ews:wedge:begin}
      \uIf{$M$ is a temporal $(3,3)$-star\label{ln:ews:wedge:gb}}
      {
        Select $W_j^\prime$ including $e_j^\prime$ centered at the center of $M$\;
      }
      \ElseIf{$M$ is a temporal triangle}
      {
        Select $W_j^\prime$ including $e_j^\prime$ centered at the vertex mapped to
        the one with a lower degree in $u$ and $v$\;
      }\label{ln:ews:wedge:ge}
      $E_j(e) \gets$ all edges that form $\delta$-instances of $W_j^\prime$ with $e$\;
      \label{ln:ews:edges}
      \ForEach{$g \in E_j(e)$ \label{ln:ews:wedge:p}}
      {
        Add a $\delta$-instance $W$ of $W_j^\prime$ comprising $e$ and $g$
        to $\widehat{\mathcal{W}}_j(e)$ with probability $q$\;
      }\label{ln:ews:wedge:end}
      \ForEach{$W \in \widehat{\mathcal{W}}_j(e)$ \label{ln:ews:est:begin}}
      {
        Let $\eta(W)$ be the number of edges that form $\delta$-instances of $M$
        together with $W$\;\label{ln:ews:est:W}
        $\widehat{\eta}_j(e) \gets \widehat{\eta}_j(e) + \frac{\eta(W)}{q}$\;
      } \label{ln:ews:est:end}
    }
    \label{ln:ews:wedge:t}
  }
  \Return{$\widehat{C}_M \gets \frac{1}{3p}\sum_{e \in \widehat{E}_T}
  \sum_{j=1}^{3}\widehat{\eta}_j(e)$}\;\label{ln:ews:estimate}
\end{algorithm}

\noindent\textbf{Sample Temporal Wedges (Lines~\ref{ln:ews:wedge:begin}--\ref{ln:ews:wedge:end}):}
The first step of \emph{temporal wedge sampling} is to determine
which  temporal wedge pattern is to be matched according to the query motif $M$
and the mapping from $e$ to $e_j^\prime$.
Specifically, we categorize 3-vertex 3-edge temporal motifs into two types,
i.e., \emph{temporal $(3,3)$-stars} and \emph{temporal triangles} as shown
in Figure~\ref{fig:wedge}, based on whether they are closed.
Interested readers may refer to~\cite{DBLP:conf/wsdm/ParanjapeBL17}
for a full list of all $3$-vertex $3$-edge temporal motifs.
For a star or wedge pattern, the vertex connected with all edges
is its \emph{center}. Given that $e=(u,v,t)$ has been mapped to $e_j^\prime$,
EWS should find a temporal wedge pattern $W_j^\prime$ containing $e_j^\prime$ from $M$ for sampling.
Here, different strategies are adopted to determine $W_j^\prime$
for star and triangle motifs (Lines~\ref{ln:ews:wedge:gb}--\ref{ln:ews:wedge:ge}):
If $M$ is a temporal $(3,3)$-star, it must select  $W_j^\prime$ that
contains $e_j^\prime$ and has the same center as $M$;
If $M$ is a temporal triangle, it may use the vertex mapped to either $u$ or $v$
as the center to generate a wedge pattern. In this case, the center of $W_j^\prime$ will be
mapped to the vertex with a lower degree between $u$ and $v$ for search space reduction.
After deciding $W_j^\prime$, it enumerates all edges that form a $\delta$-instance of
$W_j^\prime$ together with $e$ as $E_j(e)$ from the adjacency list of the central vertex (Line~\ref{ln:ews:edges}).
By selecting each edge $g \in E_j(e)$ with probability $q$,
it generates a sample $\widehat{\mathcal{W}}_j(e)$ of
$\delta$-instances of $W_j^\prime$ (Lines~\ref{ln:ews:wedge:p}---\ref{ln:ews:wedge:end}).

\vspace{1mm}
\noindent\textbf{Estimate $\eta_j(e)$ (Lines~\ref{ln:ews:est:begin}--\ref{ln:ews:est:end}):}
Now, it estimates $\eta_j(e)$ from the set $\widehat{\mathcal{W}}_j(e)$ of
sampled temporal wedges. For each $W \in \widehat{\mathcal{W}}_j(e)$,
it counts the number $\eta(W)$ of $\delta$-instances of $M$ that contain $W$ (Line~\ref{ln:ews:est:W}).
Specifically, after matching $W$ with $W_j^\prime$, it can determine
the starting and ending vertices as well as the temporal range for the mapping
the third edge of $M$. For the fast computation of $\eta(W)$,
EWS maintains a hash table that uses an ordered combination
$\langle u, v\rangle$ ($u,v \in V_T$) as the key and
a sorted list of the timestamps of all edges from $u$ to $v$ as the value
on the edge set $E_T$ of $T$.
In this way, $\eta(W)$ can be computed by a hash search
followed by at most two binary searches on the sorted list.
Finally, $\eta_j(e)$ can be estimated by summing up $\eta(W)$
for each $W \in \widehat{\mathcal{W}}_j(e)$ (Line~\ref{ln:ews:est:end}),
i.e., $\widehat{\eta}_j(e)=\frac{1}{q}\sum_{W\in\widehat{\mathcal{W}}_j(e)}\eta(W)$ .

\begin{figure}
  \centering
  \includegraphics[width=0.6\textwidth]{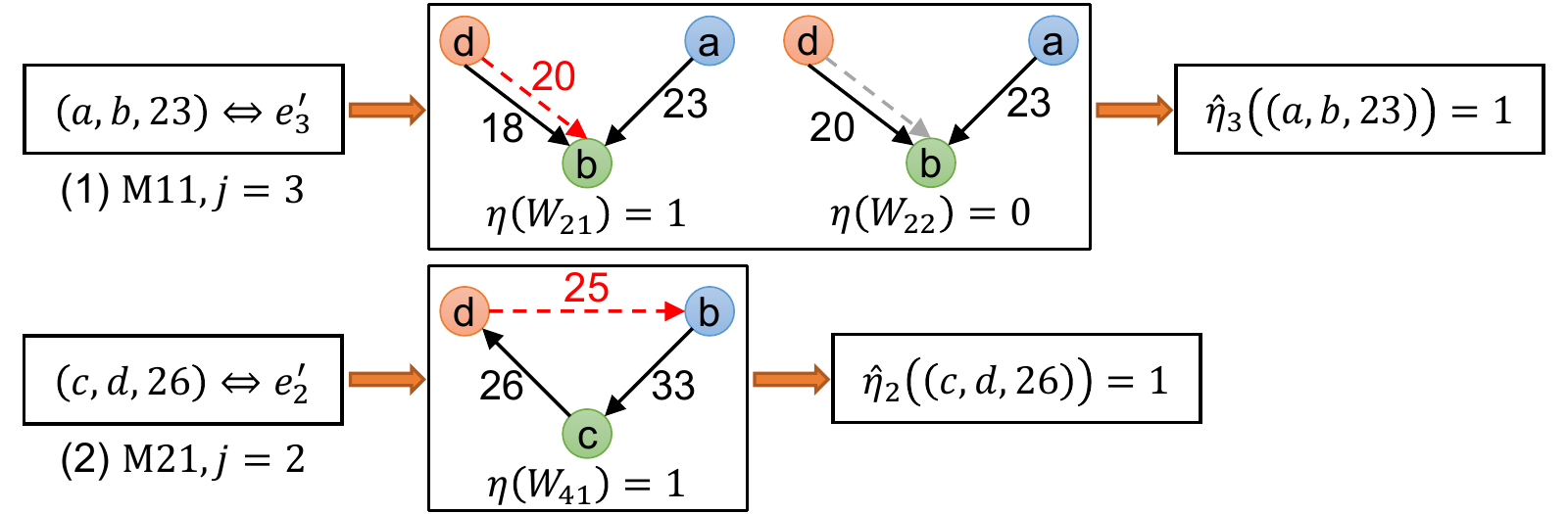}
  \vspace{-1em}
  \caption{Examples for the EWS algorithm}
  \label{fig:ews}
\end{figure}

\begin{example}
  In Figure~\ref{fig:ews}, we show how to compute $\widehat{\eta}_j(e)$
  using EWS on the temporal graph in Figure~\ref{fig:example}.
  In this example, $q$ is set to $1$, i.e., all temporal wedges found are sampled.
  When $(a,b,23)$ is mapped to $e_3^\prime$ of $M11$ in Figure~\ref{fig:wedge},
  we have $W_3^\prime = W2$ and find $2$ instances $W21$ and $W22$ of $W2$.
  Then, we acquire $\eta(W21)=1$ and $\eta(W22)=0$ and thus $\widehat{\eta}_3((a,b,23))=1$.
  For an edge $(c,d,26)$ mapped to $e_2^\prime$ of $M21$ in Figure~\ref{fig:wedge},
  $c$ is used as the central vertex since $deg(c)=3<deg(d)=5$.
  Then, we have $W_2^\prime = W4$ and there is only
  one instance $W41$ of $W4$ found. As $\eta(W41)=1$, we get
  $\widehat{\eta}_2((c,d,26))=1$ accordingly.
\end{example}

\noindent\textbf{Theoretical Analysis:}
Next, we analyze the estimate $\widehat{C}_M$ returned by Algorithm~\ref{alg:ews}
theoretically. We prove the unbiasedness and variances of $\widehat{C}_M$ in Theorem~\ref{thm:ews:exp} and Theorem~\ref{thm:ews:var}, respectively.
Detailed proofs are also provided in the technical report.

\begin{theorem}\label{thm:ews:exp}
  The expected value $\mathbb{E}[\widehat{C}_M]$ of $\widehat{C}_M$
  returned by Algorithm~\ref{alg:ews} is $C_M$.
\end{theorem}
\begin{proof}
  By applying the result of Theorem~\ref{thm:es:exp}, we only need to
  show $\mathbb{E}[\widehat{\eta}_j(e)]=\eta_j(e)$ to prove Theorem~\ref{thm:ews:exp}.
  Here, we index the edges in $E_j(e)$ by $[1,\ldots,|E_j(e)|]$ and
  use an indicator $\omega_r$ to denote whether the wedge $W_r$
  w.r.t.~the $r$\textsuperscript{th} edge in $E_j(e)$ is sampled.
  We have the following equality:
  \begin{displaymath}
    \mathbb{E}[\widehat{\eta}_j(e)]
    = \frac{1}{q}\sum_{r=1}^{|E_j(e)|}\mathbb{E}[\omega_r]\cdot\eta(W_r)
    = \sum_{r=1}^{|E_j(e)|}\eta(W_r)
    = \eta_j(e)
  \end{displaymath}
  and conclude the proof.
\end{proof}

\begin{theorem}\label{thm:ews:var}
  The variance $\textnormal{Val}[\widehat{C}_M]$ of $\widehat{C}_M$
  returned by Algorithm~\ref{alg:ews} is at most $\frac{1-pq}{pq} \cdot C_M^2$.
\end{theorem}
\begin{proof}
 Let us index the edges in $E_T$ by $[1,m]$ and the edges in
 $E_j(e_i)$ when $e_i$ is mapped to $e_j^\prime$ by $[1,m_{ij}]$
 where $m_{ij}=|E_j(e_i)|$. Similar to the proof of Theorem~\ref{thm:es:var},
 we have
 \begin{align*}
   \textnormal{Val}[\widehat{C}_M]
   & = \textnormal{Val}\Big[\frac{1}{3pq} \sum_{i=1}^{m}\sum_{j=1}^{3}\sum_{r=1}^{m_{ij}} \omega_i \cdot \omega_{ijr} \cdot \eta(W_{ijr}) \Big] \\
   & = \sum_{i=1}^{m}\sum_{j=1}^{3}\sum_{r=1}^{m_{ij}} \frac{\eta^2(W_{ijr})}{9 p^2 q^2} \cdot \textnormal{Var}[\omega_i \cdot \omega_{ijr}] \\
   & = \frac{1-pq}{9pq} \cdot \sum_{i=1}^{m}\sum_{j=1}^{3}\sum_{r=1}^{m_{ij}} \eta^2(W_{ijr})
     \leq \frac{1-pq}{pq} \cdot C_M^2
 \end{align*}
 where $\eta(W_{ijr})$ is the number of $\delta$-instances of $M$ containing
 a temporal wedge $W_{ijr}$ and $\omega_{ijr}$ is its indicator,
 the second equality holds for the independence of $\omega_i$ and $\omega_{ijr}$,
 the third equality holds because $\textnormal{Var}[\omega_i \cdot \omega_{ijr}]=pq-p^2 q^2$,
 and the last inequality holds for
 $C_M=\frac{1}{3}\sum_{i=1}^{m}\sum_{j=1}^{3}\sum_{r=1}^{m_{ij}}\eta(W_{ijr})$.
\end{proof}
According to the result of Theorem~\ref{thm:ews:var} and Chebyshev's inequality, we have
$\textnormal{Pr}[|\widehat{C}_M-C_M|\geq\varepsilon\cdot C_M]\leq\frac{1-pq}{pq\varepsilon^2}$
and $\widehat{C}_M$ is an $(\varepsilon,\gamma)$-estimator of $C_M$ for
parameters $\varepsilon,\gamma \in (0,1)$ when $pq=\frac{1}{1+\gamma\varepsilon^{2}}$.

\vspace{1mm}
\noindent\textbf{Time Complexity:}
We first analyze the time to compute $\widehat{\eta}_j(e)$.
First, $|E_j(e)|$ is bounded by the maximum number of (in-/out-)edges
connected with one vertex within any $\delta$-length time interval, i.e., $d_{\delta}$.
Second, the time to compute $\eta(W)$ using a hash table
is $O(\log{h})$ where $h$ is the maximum number of edges between any two vertices.
Therefore, the time complexity per edge in EWS is $O(d_{\delta}\log{h})$.
This is lower than $O(d_{\delta}^2)$ time per edge in ES ($k=l=3$).
Finally, EWS provides an $(\varepsilon,\gamma)$-estimator of $C_M$
in $O(\frac{m d_{\delta}\log{h}}{1+\gamma\varepsilon^2})$ time.

\subsection{Streaming Implementation}\label{subsec:alg:discuss}

To deal with a dataset that is too large to fit in memory or generated in a
streaming manner, it is possible to adapt our algorithms to a streaming setting.
Assuming that all edges are sorted in chronological order, our algorithms can determine
whether to sample an edge or not when it arrives.
Then, for each sampled edge $e=(u,v,t)$, we only need the edges
with timestamps in $[t-\delta,t+\delta]$ to compute its local count $\eta(e)$ or $\widehat\eta(e)$.
After a one-pass scan over the temporal graph stream, we can obtain an estimate of the number of a temporal motif in the stream.
Generally, our algorithms can process any temporal graph stream in one pass
by always maintaining the edges in the most recent time interval of length $2\delta$
while having the same theoretical bounds as in the batch setting.

\section{Experimental Evaluation}\label{sec:exp}

In this section, we evaluate the empirical performance of our proposed
algorithms on real-world datasets. We first introduce the experimental
setup in Section~\ref{subsec:exp:setup}. The experimental results
are presented in Section~\ref{subsec:exp:results}.

\subsection{Experimental Setup}\label{subsec:exp:setup}

\noindent\textbf{Experimental Environment:}
All experiments were conducted on a server running Ubuntu 18.04.1 LTS
with an Intel\textsuperscript{\textregistered} Xeon\textsuperscript{\textregistered}
Gold 6140 2.30GHz processor and 250GB main memory.
All datasets and our code are publicly available\footnote{\url{https://github.com/jingjing-hnu/Temporal-Motif-Counting}}.
We downloaded the
code\footnote{\url{http://snap.stanford.edu/temporal-motifs/}}$^,$\footnote{\url{https://github.com/rohit13k/CycleDetection}}$^,$\footnote{\url{https://gitlab.com/paul.liu.ubc/sampling-temporal-motifs}}
of baselines published by the authors
and followed the instructions for compilation and usage.
All algorithms were implemented in \texttt{C++11}
compiled by \texttt{GCC v7.4} with \texttt{-O3} optimizations,
and ran on a single thread.

\begin{table}
  \small
  \centering
  \caption{Statistics of datasets}
  \label{tab:Datasets}
  \begin{tabular}{|c|c|c|c|c|}
    \hline
    \textbf{Dataset} & $|V_T|$ & $|E|$ & $|E_T|$ & \textbf{Time span} \\
    \hline
    \textbf{AU} & $157,222$  &$544,621$ & $726,639$  & $7.16$ years \\
    \hline
    \textbf{SU} & $192,409$  & $854,377$& $1,108,716$ & $7.60$ years \\
    \hline
    \textbf{SO} & $2,584,164$ & $34,875,684$& $47,902,865$ & $7.60$ years \\
    \hline
    \textbf{BC} & $48,098,591$ &$86,798,226$ & $113,100,979$  & $7.08$ years \\
    \hline
    \textbf{RC} & $5,688,164$ &$329,485,956$ & $399,523,749$  & $7.44$ years \\ 
    \hline
  \end{tabular}
\end{table}

\vspace{1mm}
\noindent\textbf{Datasets:}
We used five different real-world datasets in our experiments
including AskUbuntu (AU), SuperUser (SU), StackOverflow (SO),
BitCoin (BC), and RedditComments (RC).
All datasets were downloaded from publicly available sources
like the SNAP repository~\cite{DBLP:journals/tist/LeskovecS16}.
Each dataset is a sequence of temporal edges in chronological order.
We report the statistics of these datasets in Table~\ref{tab:Datasets},
where $|V_T|$ is the number of vertices, $|E|$ is the number of (static) edges,
$|E_T|$ is the number of temporal edges, and \emph{time span}
is the overall time span of the entire dataset.

\begin{figure}
  \centering
  \includegraphics[width=0.5\textwidth]{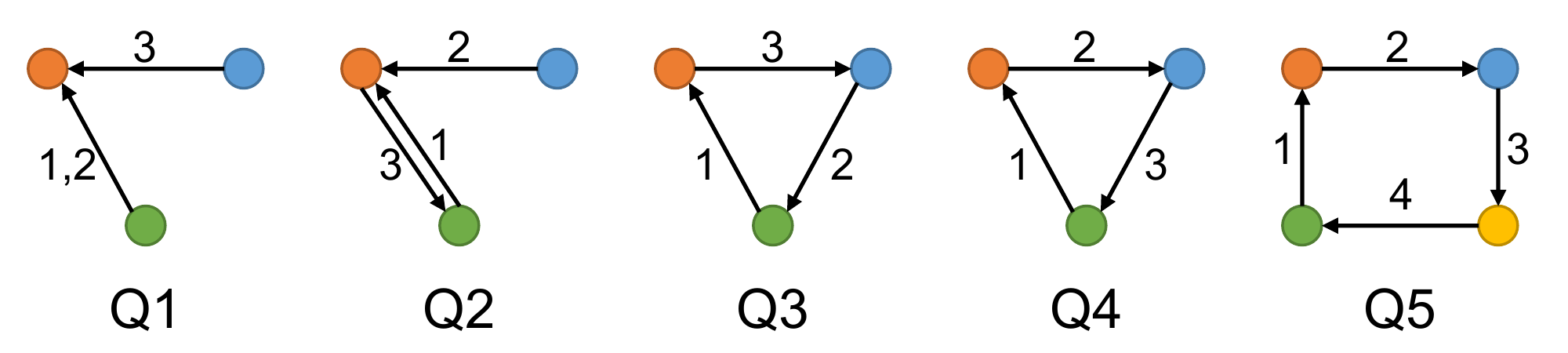}
  \vspace{-1em}
  \caption{Query motifs}
  \label{fig:queries}
\end{figure}

\vspace{1mm}
\noindent\textbf{Algorithms:}
The algorithms compared are listed as follows.
\begin{itemize}
  \item \textbf{EX}: An exact algorithm for temporal motif counting
  in~\cite{DBLP:conf/wsdm/ParanjapeBL17}. The available implementation
  is applicable only to $3$-edge motifs and cannot support motifs with
  $4$ or more edges (e.g.,~Q5 in Figure~\ref{fig:queries}).
  \item \textbf{2SCENT}: An algorithm for simple temporal cycle
  (e.g.,~Q4 and Q5 in Figure~\ref{fig:queries})
  enumeration in~\cite{DBLP:journals/pvldb/KumarC18}.
  \item \textbf{BT}: A \textsc{BackTracking} algorithm for temporal subgraph isomorphism
  in~\cite{DBLP:conf/bigdataconf/MackeyPFCC18}. It provides the exact
  count of any temporal motif by enumerating all of them.
  \item \textbf{IS-BT}: An interval-based sampling algorithm for
  temporal motif counting in~\cite{DBLP:conf/wsdm/LiuBC19}.
  BT~\cite{DBLP:conf/bigdataconf/MackeyPFCC18} is used
  as a subroutine for any motif with more than $2$ vertices.
  \item \textbf{ES}: Our generic edge sampling algorithm for temporal motif counting
  in Section~\ref{subsec:alg:1}.
  \item \textbf{EWS}: Our improved algorithm
  that combines edge sampling with wedge sampling
  for counting temporal motifs with $3$ vertices and $3$ edges
  (e.g.~Q1--Q4 in Figure~\ref{fig:queries}) in Section~\ref{subsec:alg:2}.
\end{itemize}

\noindent\textbf{Queries:}
The five query motifs we use in the experiments are listed in Figure~\ref{fig:queries}.
Since different algorithms specialize in different types of motifs, we
select a few motifs that can best represent the specializations of all algorithms.
As discussed above, an algorithm may not be applicable to some of the motifs.
In this case, the algorithm is ignored in the experiments on these motifs.

\vspace{1mm}
\noindent\textbf{Performance Measures:}
The efficiency is measured by the CPU time
(in seconds) of an algorithm to count a query motif in a temporal graph.
The accuracy of a sampling algorithm is measured by
the relative error $\frac{|\widehat{x}-x|}{x}$
where $x$ is the exact number of instances of a query motif in a temporal graph
and $\widehat{x}$ is an estimate of $x$ returned by an algorithm.
In each experiment, we run all algorithms
$10$ times and use the average CPU time and
relative errors for comparison.

\subsection{Experimental Results}
\label{subsec:exp:results}

\begin{table*}
  \small
  \setlength\tabcolsep{2 pt}
  \centering
  \caption{Running time (in seconds) and average errors ($\%$) of all algorithms on each dataset. We use ``---'' and ``\ding{53}'' to denote ``motif not supported'' and
  ``running out of memory'', respectively. For IS-BT, ES, and EWS, we show their speedup ratios over BT for comparison. We use ``*'' to mark the results of ES and EWS for $p=0.1$ instead of $p=0.01$.}
  \label{tab:results}
  \begin{tabular}{|c|c|c|c|c|c|c|c|c|c|c|}
    \hline
    \multirow{2}{*}{\textbf{Dataset}} & \multirow{2}{*}{\textbf{Motif}} & \textbf{EX} & \textbf{2SCENT} & \textbf{BT} & \multicolumn{2}{c|}{\textbf{IS-BT}} & \multicolumn{2}{c|}{\textbf{ES}} & \multicolumn{2}{c|}{\textbf{EWS}} \\ \cline{3-11}
    & & time (s) & time (s) & time (s) & error & time (s) & error & time (s) & error & time (s) \\
    \hline
    \multirow{5}{*}{\textbf{AU}}
    & Q1 & \multirow{2}{*}{1.8} & \multirow{3}{*}{\textbf{---}} & 0.758 & 4.84\% & 0.402/1.9x &\textbf{4.32\%} & 0.059/12.8x & 4.32\% & \textbf{0.027/28.1x} \\ \cline{2-2} \cline{5-11} 
    & Q2 &  &  & 1.104 & \textbf{4.16\%} & 0.434/2.5x & 4.57\% & 0.048/23.0x & 4.57\% & \textbf{0.029/38.1x} \\ \cline{2-3} \cline{5-11} 
    & Q3 & \multirow{2}{*}{2.3} &  & 0.884 & 3.97\% & 0.50/1.8x & *\textbf{3.73\%} & *0.605/1.5x & *3.73\% & *\textbf{0.183/4.8x} \\ \cline{2-2} \cline{4-11}
    & Q4 &  & \multirow{2}{*}{23.68} & 1.038 & 4.67\% & 0.492/2.1x & *\textbf{4.63\%} & *0.628/1.7x &*4.63\% & *\textbf{0.173/6x} \\ \cline{2-3} \cline{5-11} 
    & Q5 & \textbf{---} &  & 1.262 & \textbf{3.98\%} & 0.536/2.4x & *4.62\% & *\textbf{0.322/3.9x} & \multicolumn{2}{c|}{\textbf{---}}\\ \hline
    \multirow{5}{*}{\textbf{SU}}
    & Q1 & \multirow{2}{*}{3.26} & \multirow{3}{*}{\textbf{---}} & 1.499 & 3.99\% & 0.620/2.4x & \textbf{3.06\%} & 0.102/14.7x & 3.06\% & \textbf{0.052/28.8x} \\ \cline{2-2} \cline{5-11} 
    & Q2 &  &  & 1.650 & 3.23\% & 0.671/2.5x & \textbf{2.47\%} & 0.083/19.9x & 2.47\% & \textbf{0.046/35.9x} \\ \cline{2-3} \cline{5-11} 
    & Q3 & \multirow{2}{*}{4.6} &  & 1.506 & 4.85\% & 0.723/2.1x & \textbf{4.66\%} & 0.113/13.3x & 4.66\% & \textbf{0.030/50.2x} \\ \cline{2-2} \cline{4-11} 
    & Q4 &  & \multirow{2}{*}{46.0} & 1.434 & \textbf{3.79\%} & 0.725/2.0x & 4.63\% & 0.128/11.2x & 4.63\% & \textbf{0.042/34.1x} \\ \cline{2-3} \cline{5-11} 
    & Q5 & \textbf{---} &  & 1.521 & 4.55\% & 0.759/2.0x & *\textbf{4.52\%} & *\textbf{0.453/3.4x} & \multicolumn{2}{c|}{\textbf{---}}\\ \hline
    \multirow{5}{*}{\textbf{SO}}
    & Q1 & \multirow{2}{*}{169} & \multirow{3}{*}{\textbf{---}} & 105.8 & 4.82\% & 8.626/12.3x &\textbf{0.97\%} & 4.419/23.9x & 1.22\% & \textbf{1.528/69.2x} \\ \cline{2-2} \cline{5-11} 
    & Q2 &  &  & 110.7 & 4.82\% & 27.48/4.0x & \textbf{0.20\%} & 3.985/27.8x & 0.89\% & \textbf{1.514/73.1x} \\ \cline{2-3} \cline{5-11} 
    & Q3 & \multirow{2}{*}{466} &  & 107.4 & 4.30\% & 25.70/4.2x & \textbf{1.36\%} & 4.031/26.6x & 3.6\% & \textbf{1.235/87x} \\ \cline{2-2} \cline{4-11} 
    & Q4 &  & \multirow{2}{*}{243.7} & 105.5 & 4.90\% & 6.775/15.6x & \textbf{1.78\%} & 3.936/26.8x & 3.31\% & \textbf{1.153/91.5x}\\ \cline{2-3} \cline{5-11} 
    & Q5 & \textbf{---} &  & 91.83 & 4.91\% & 9.451/9.7x & \textbf{3.48\%} & \textbf{1.505/61.0x} & \multicolumn{2}{c|}{\textbf{---}}\\ \hline
    \multirow{5}{*}{\textbf{BC}}
    & Q1 & \multirow{2}{*}{8143} & \multirow{3}{*}{\textbf{---}} & 220.0 & 4.75\% & 50.02/4.4x & \textbf{0.64\%} & 59.12/3.7x & 0.67\% & \textbf{9.463/23.2x} \\ \cline{2-2} \cline{5-11} 
    & Q2 &  &  & 399.8 & 4.90\% & 125.1/3.2x & \textbf{1.11\%} & 34.74/11.5x & 1.16\% & \textbf{8.126/49.2x} \\ \cline{2-3} \cline{5-11} 
    & Q3 & \multirow{2}{*}{8116} &  & 396.8 & 3.89\% & 90.19/4.4x & \textbf{1.49\%} & 41.49/9.6x & 3.02\% & \textbf{2.121/187x} \\ \cline{2-2} \cline{4-11} 
    & Q4 &  & \multirow{2}{*}{473.7} & 473.4 & 4.93\% & 95.47/5.0x & \textbf{0.83\%} & 37.43/12.6x & 1.91\% & \textbf{2.262/209x} \\ \cline{2-3} \cline{5-11} 
    & Q5 & \textbf{---} &  & 596.4 & 4.83\% & 319.7/1.9x & \textbf{2.92\%} & \textbf{20.47/29.1x} & \multicolumn{2}{c|}{\textbf{---}}\\ \hline
    \multirow{5}{*}{\textbf{RC}}
    & Q1 & \multirow{2}{*}{2799} & \multirow{3}{*}{\textbf{---}} & 1966 & 4.76\% & 840.5/2.3x & \textbf{3.27\%} & 257.4/7.6x & 3.36\% & \textbf{31.49/62.4x} \\ \cline{2-2} \cline{5-11} 
    & Q2 &  &  & 2113 & 4.67\% & 428/4.9x & 0.63\% & 120.6/17.5x & \textbf{0.6\%} & \textbf{30.57/69.1x} \\ \cline{2-3} \cline{5-11} 
    & Q3 &  \multirow{2}{*}{\ding{53}} &  & 2069 & 4.61\% & 784.4/2.6x & 2.42\% & 76.09/27.2x & \textbf{2.27\%} & \textbf{16.17/128x} \\ \cline{2-2} \cline{4-11} 
    & Q4 &  & \multirow{2}{*}{2245} & 1897 & 4.86\% & 683/2.8x & \textbf{3.47\%} & 68.60/27.7x & 4.57\% & \textbf{15.91/119x} \\ \cline{2-3} \cline{5-11} 
    & Q5 & \textbf{---} &  & 1613 & 4.41\% & 706.6/2.3x & *\textbf{4.32\%} & *\textbf{120.3/13.4x} & \multicolumn{2}{c|}{\textbf{---}}\\ \hline
  \end{tabular}
\end{table*}

\begin{figure*}
  \centering
  \subfigure[Q2 on SU]{
    \label{subfig:q:a1}
    \includegraphics[height=0.9in]{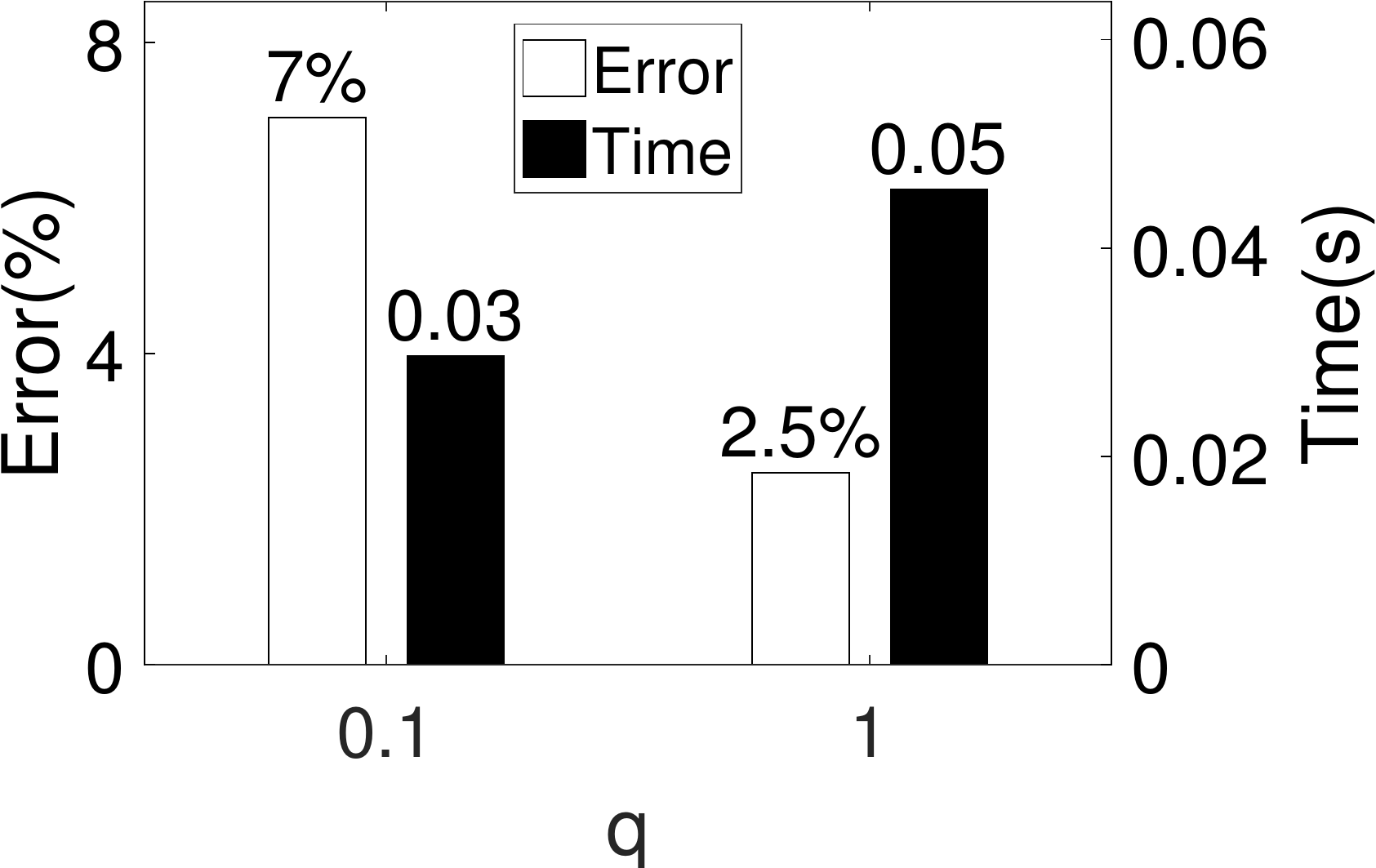}
  }
  \subfigure[Q3 on SU]{
    \label{subfig:q:a2}
    \includegraphics[height=0.9in]{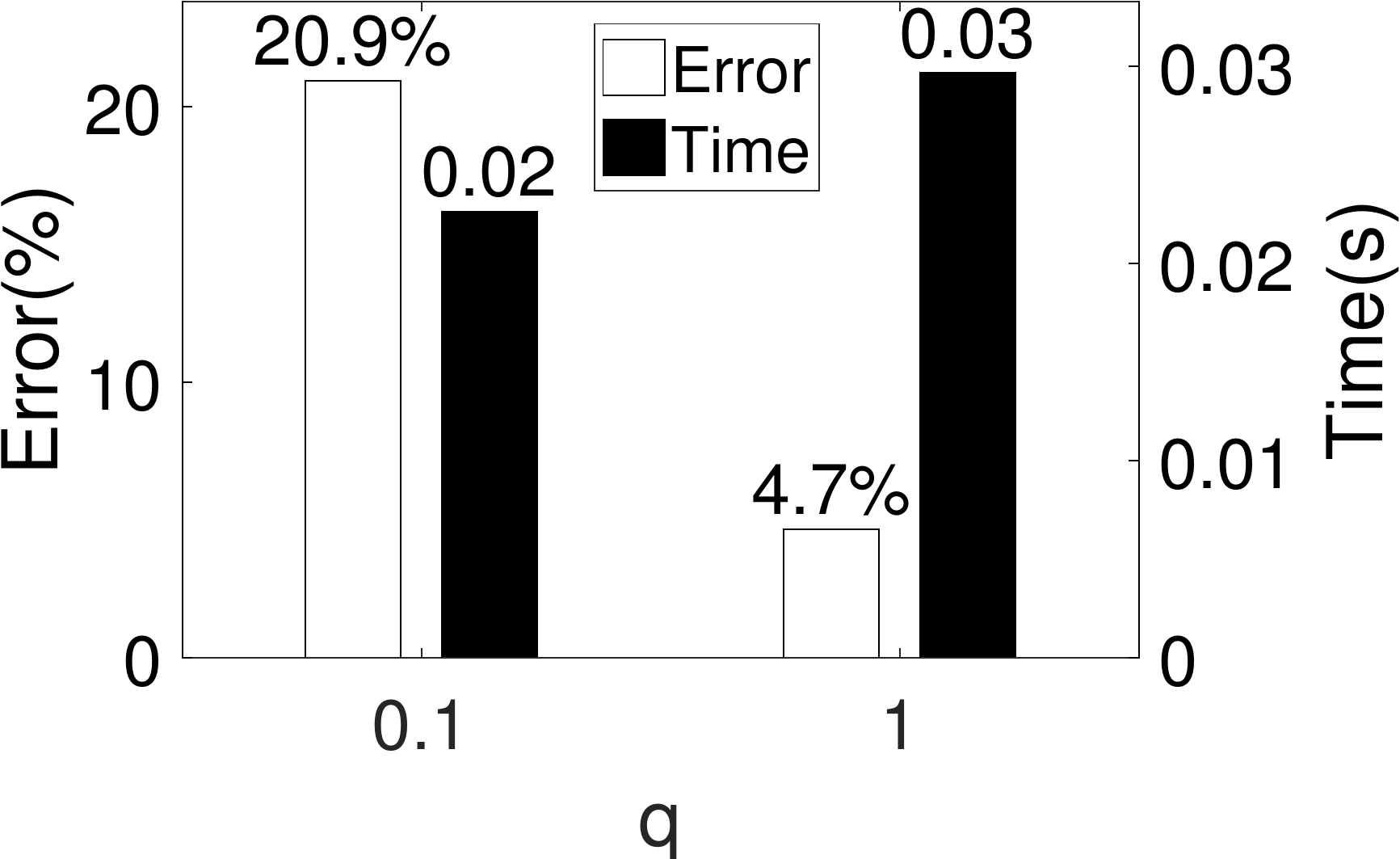}
  }
  \subfigure[Q2 on BC]{
    \label{subfig:q:b1}
    \includegraphics[height=0.9in]{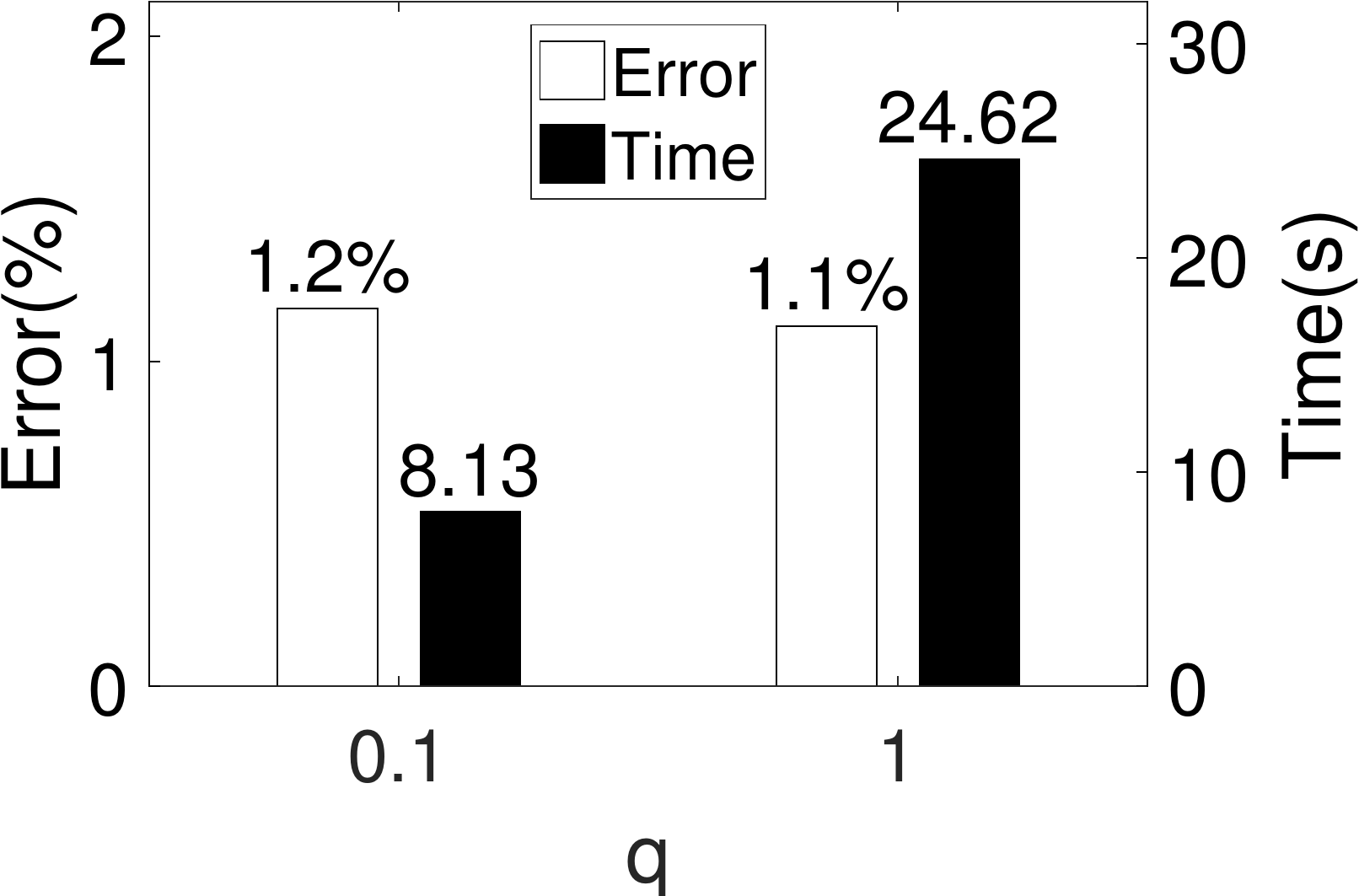}
  }
  \subfigure[Q3 on BC]{
    \label{subfig:q:b2}
    \includegraphics[height=0.9in]{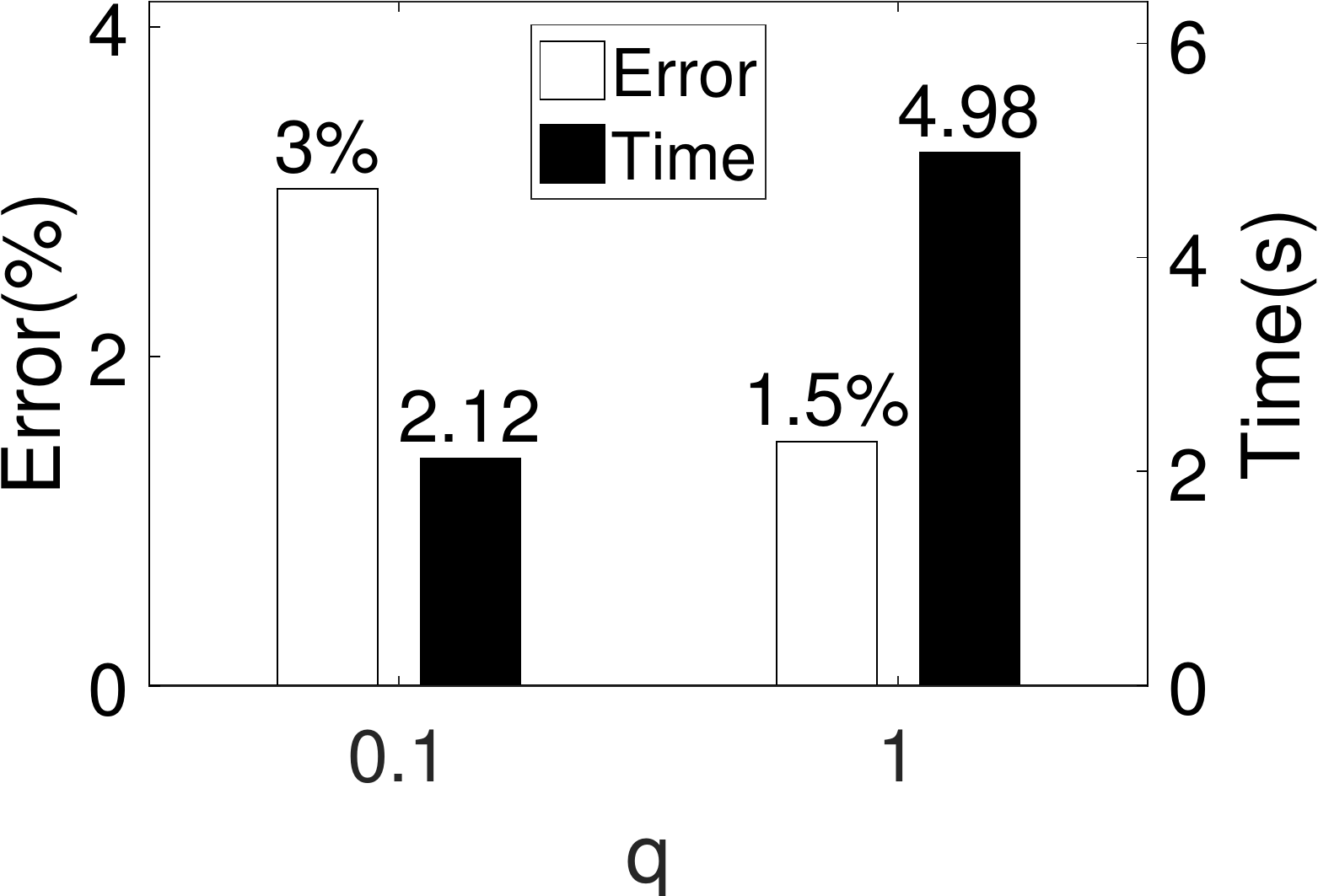}
  }
  \vspace{-1em}
  \caption{Comparison of the performance of EWS when $q=0.1$ and $1$}
  \label{fig:q}
\end{figure*}
\begin{figure*}
  \centering
  \hspace{1.35in}
  \subfigure{
    \includegraphics[height=0.12in]{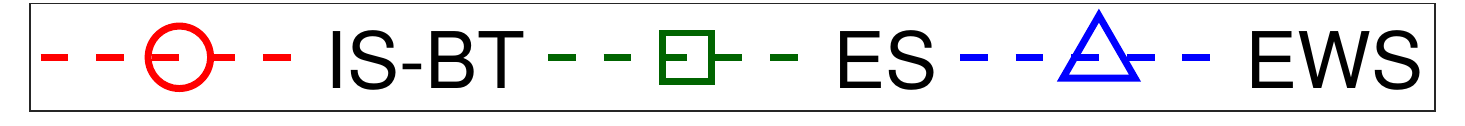}
  }
  \vspace{-1em}
  \newline
  \vspace{-1em}
  \subfigure[Q1 on AU]{
    \label{fig:p:m1:au}
    \setcounter{subfigure}{1} 
    \includegraphics[height=0.9in]{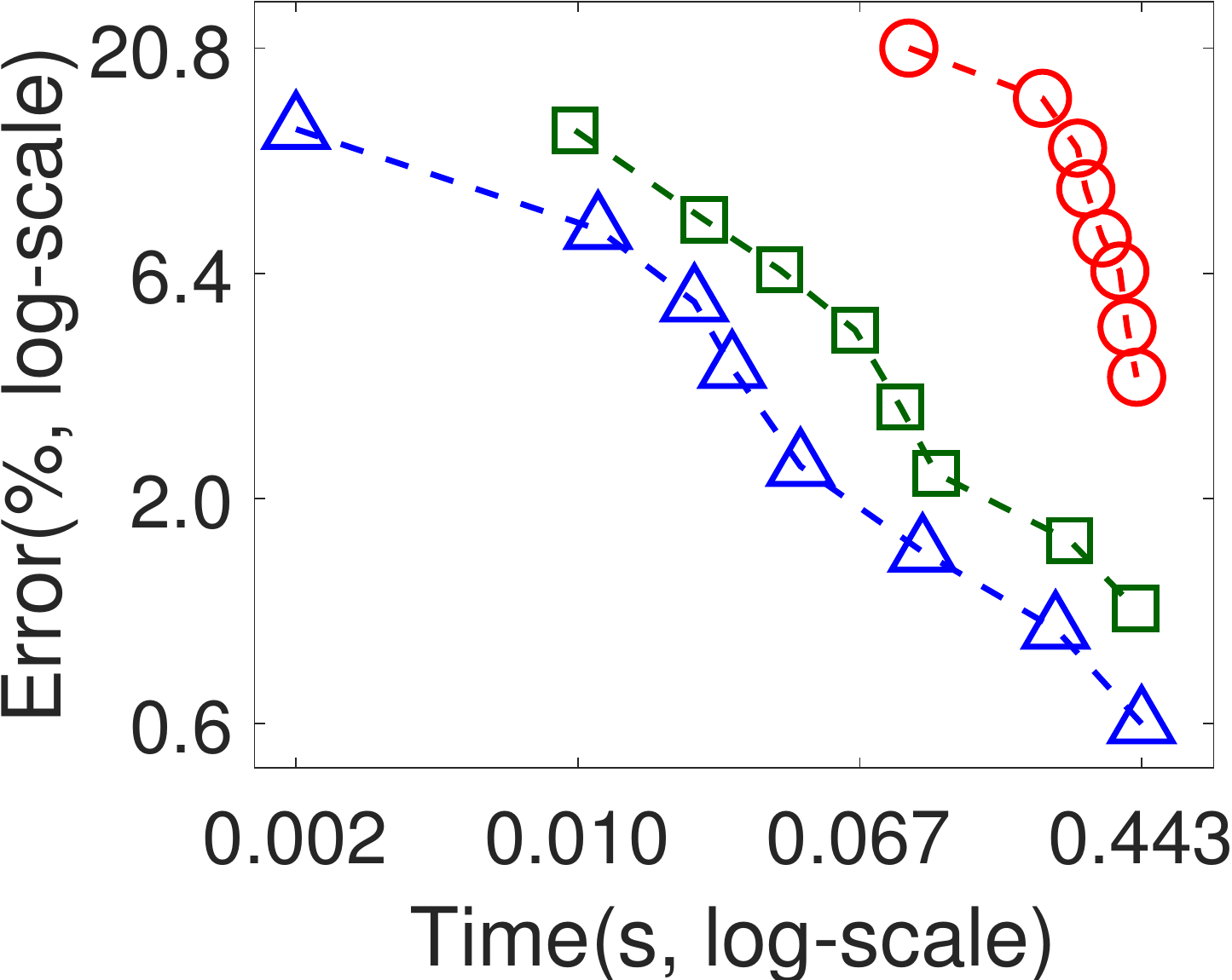}
  }
  \hfill
  \subfigure[Q2 on AU]{
    \label{fig:p:m2:au}
    \includegraphics[height=0.9in]{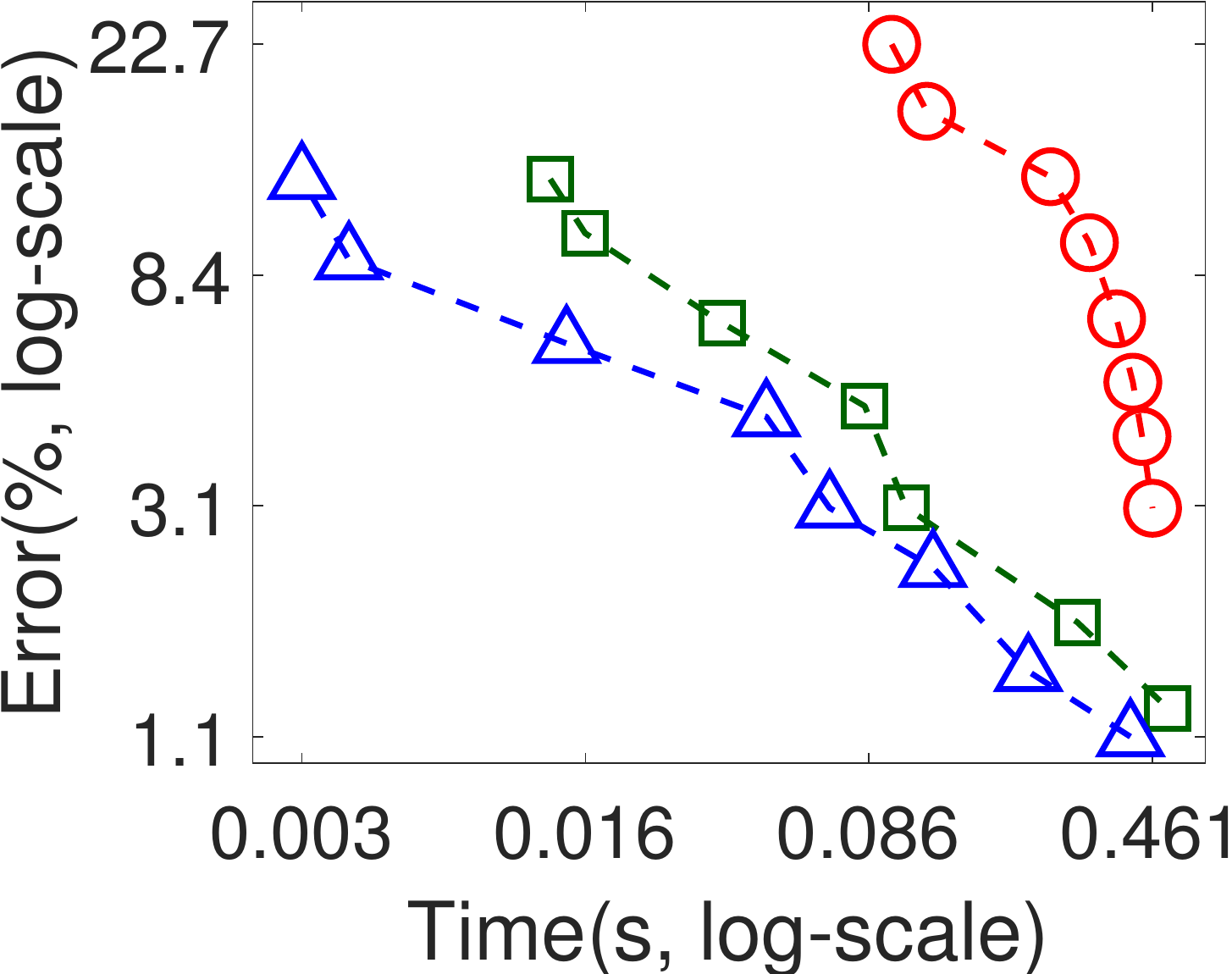}
  }
  \hfill
  \subfigure[Q3 on AU]{
    \label{fig:p:m3:au}
    \includegraphics[height=0.9in]{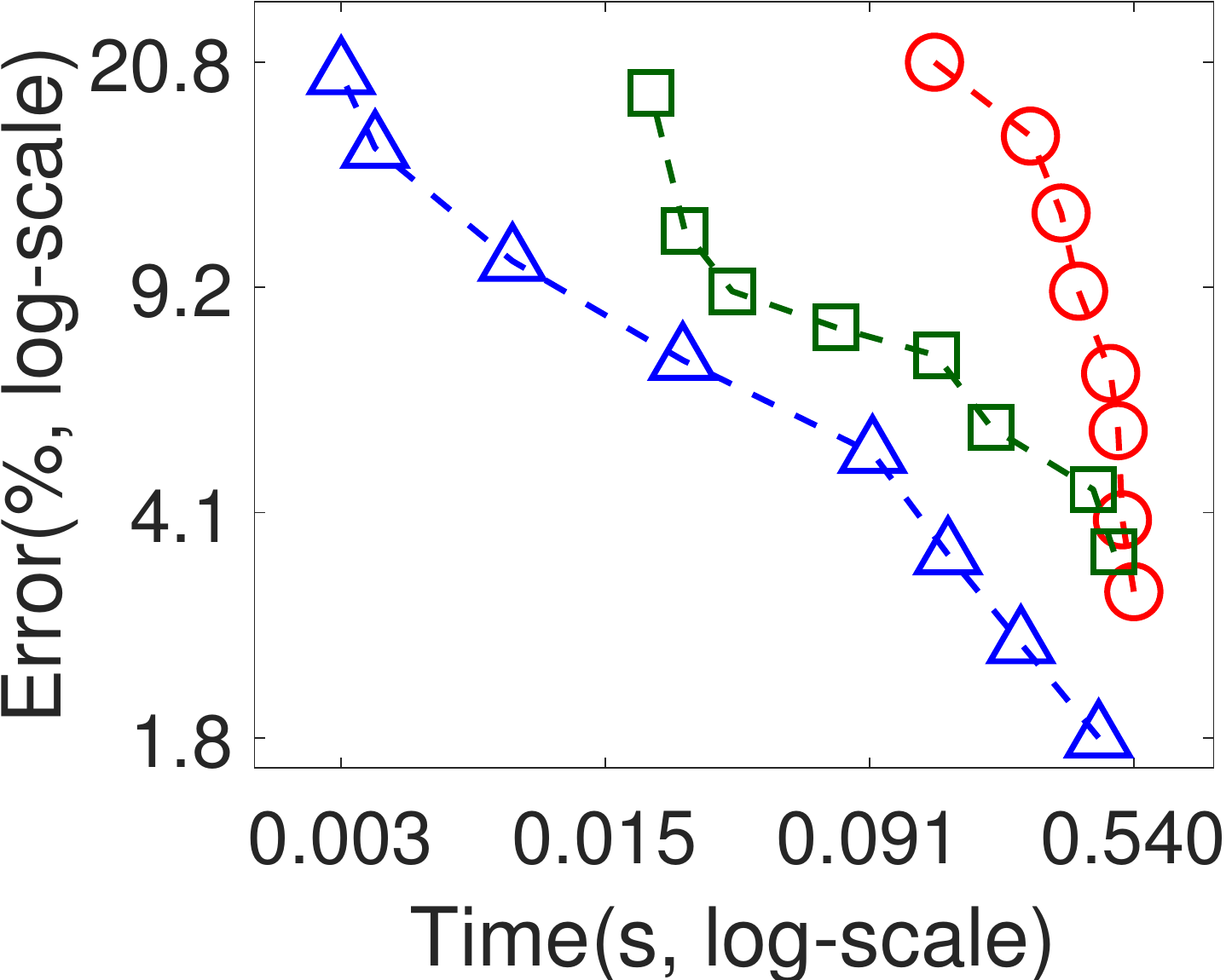}
  }
  \hfill
  \subfigure[Q4 on AU]{
    \label{fig:p:m4:au}
    \includegraphics[height=0.9in]{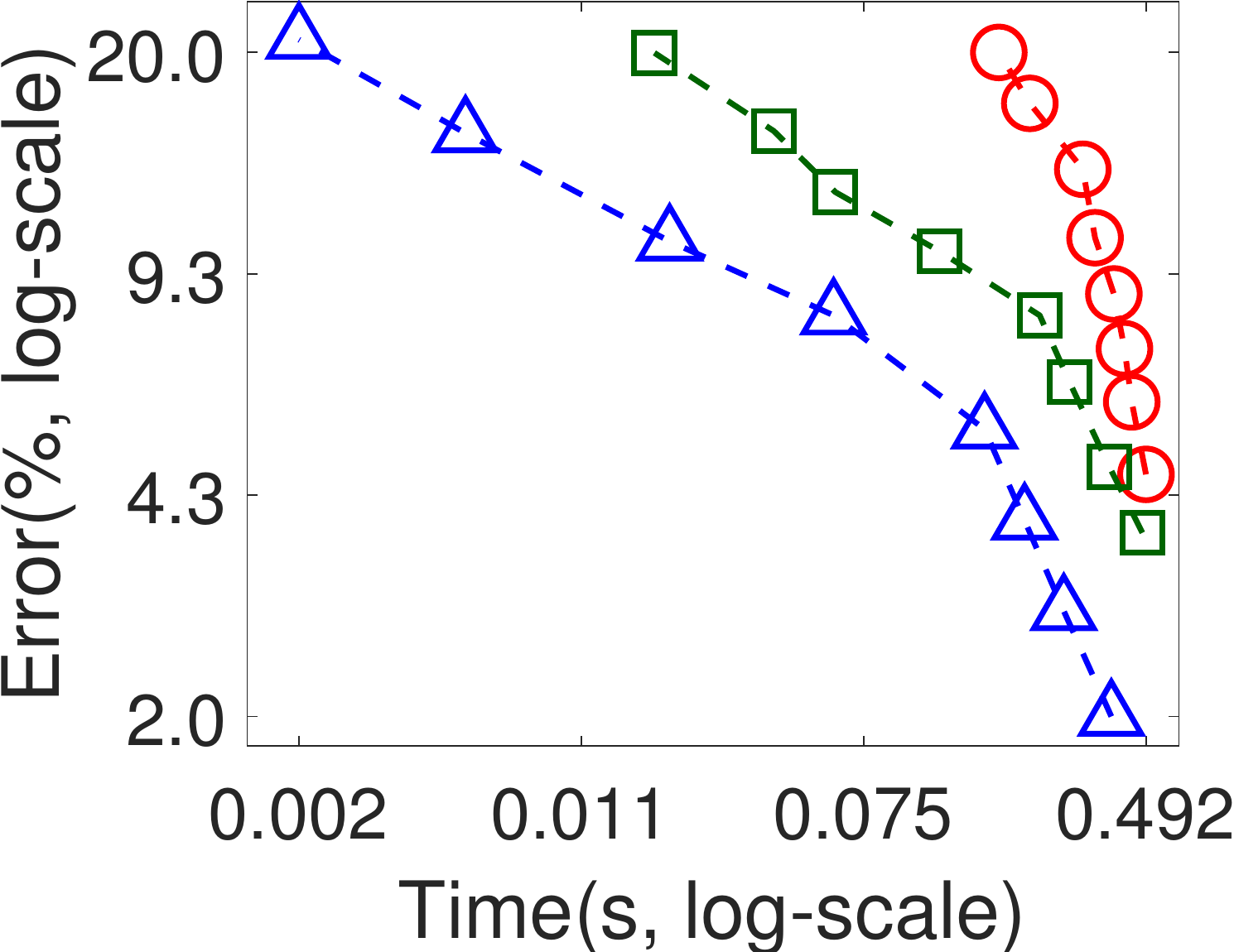}
  }
  \hfill
  \subfigure[Q5 on AU]{
    \label{fig:p:m5:au}
    \includegraphics[height=0.9in]{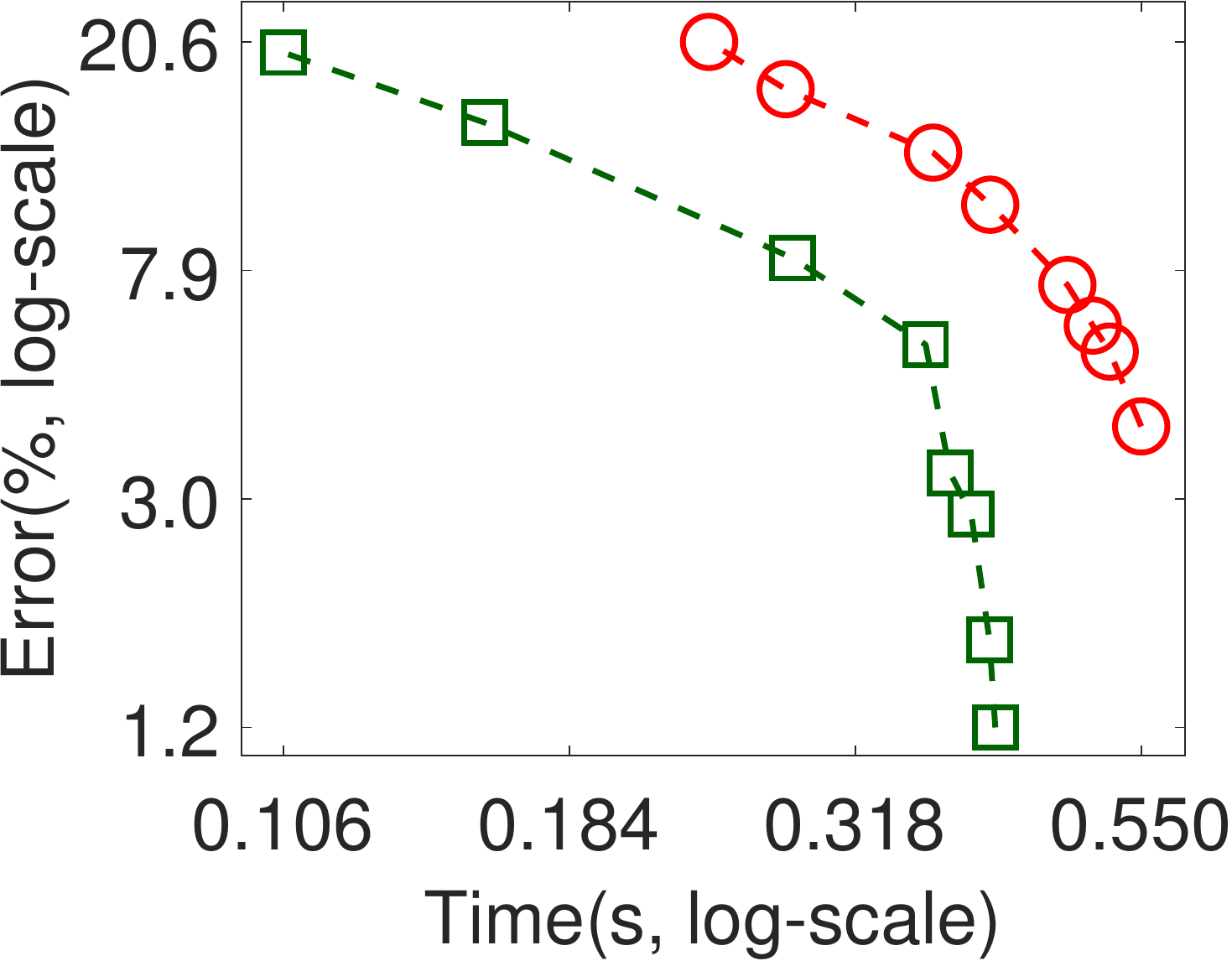}
  }
  \newline
  \vspace{-1em}
  \subfigure[Q1 on SU]{
    \label{fig:p:m1:su}
    \includegraphics[height=0.9in]{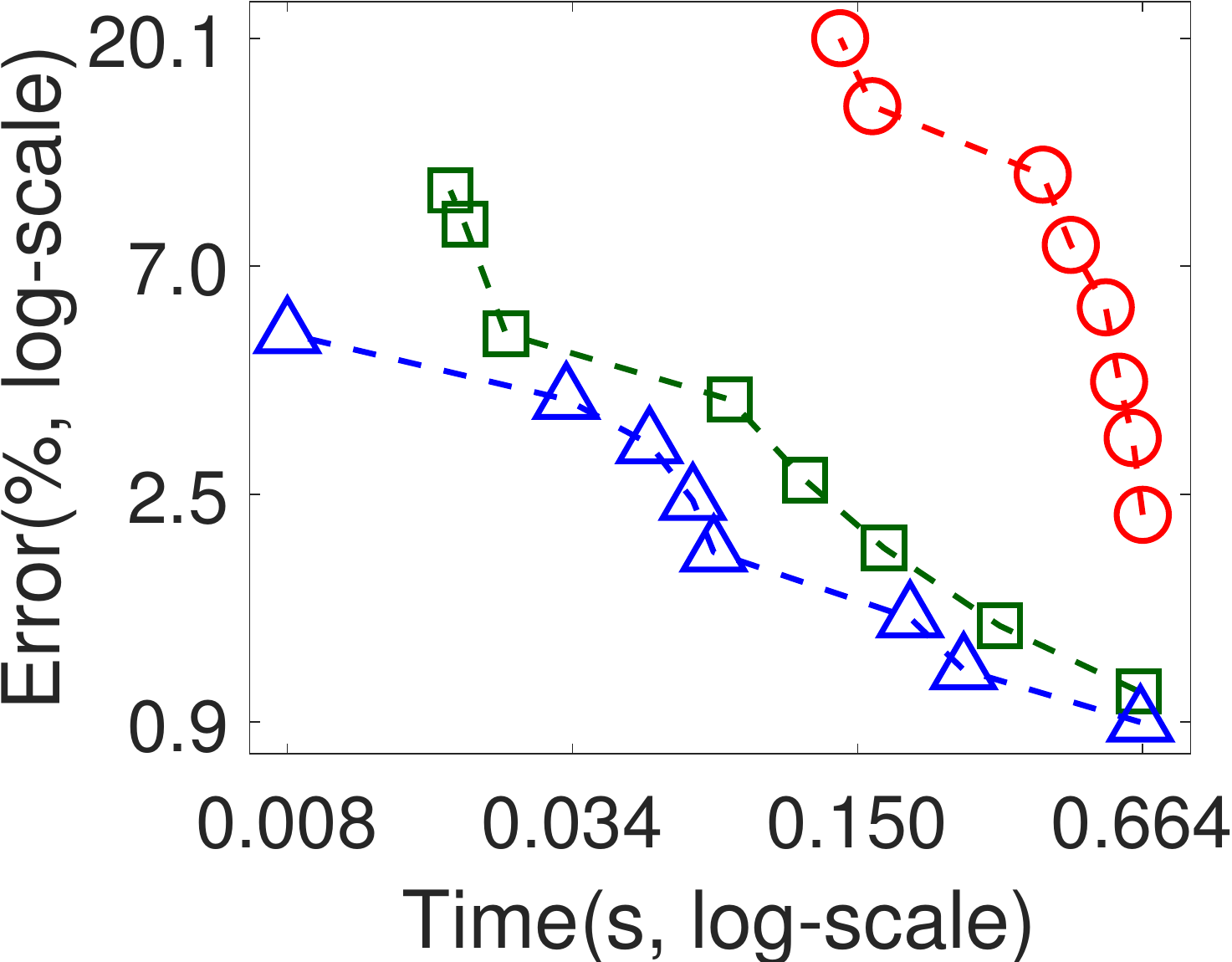}
  }
  \hfill
  \subfigure[Q2 on SU]{
    \label{fig:p:m2:su}
    \includegraphics[height=0.9in]{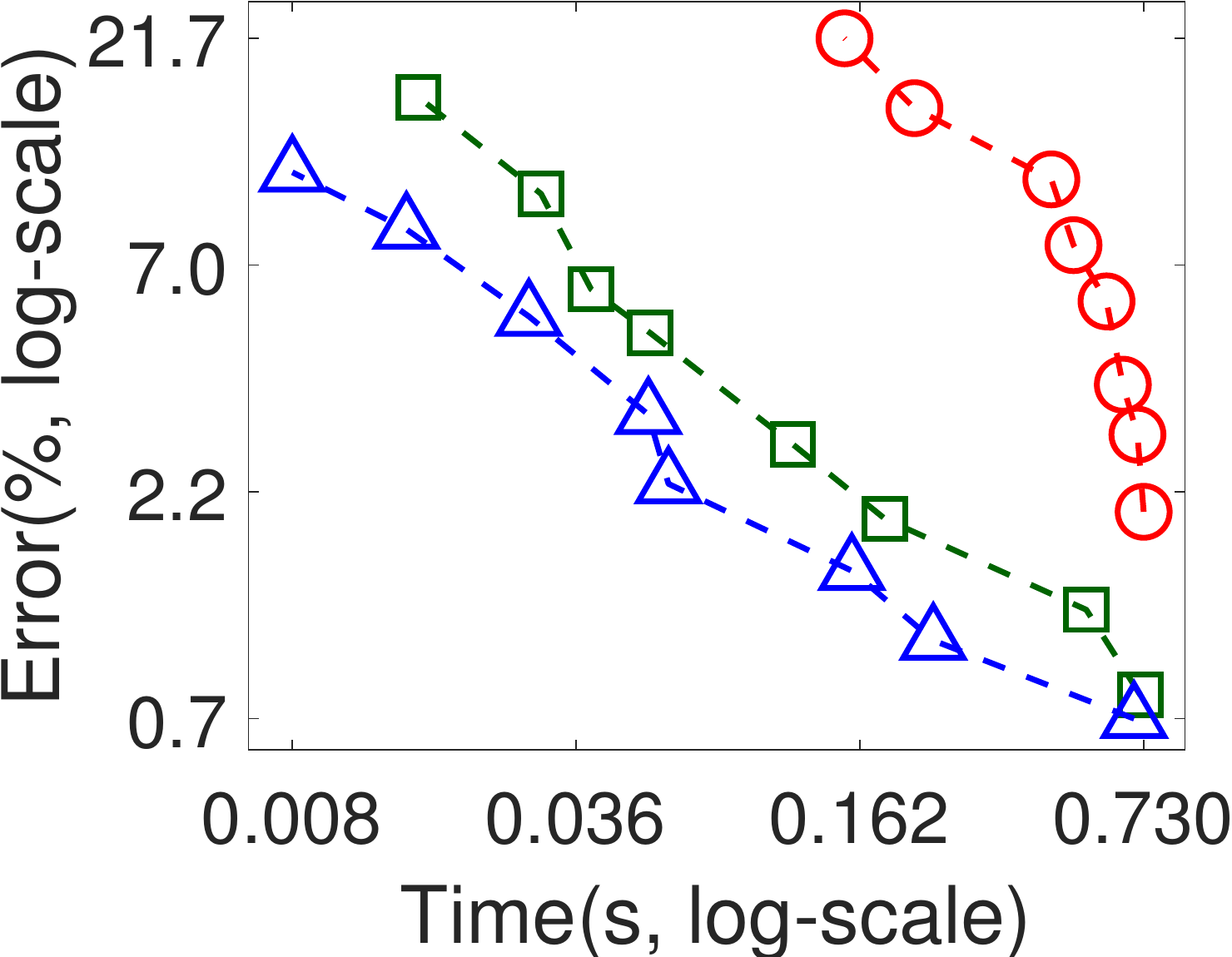}
  }
  \hfill
  \subfigure[Q3 on SU]{
    \label{fig:p:m3:su}
    \includegraphics[height=0.9in]{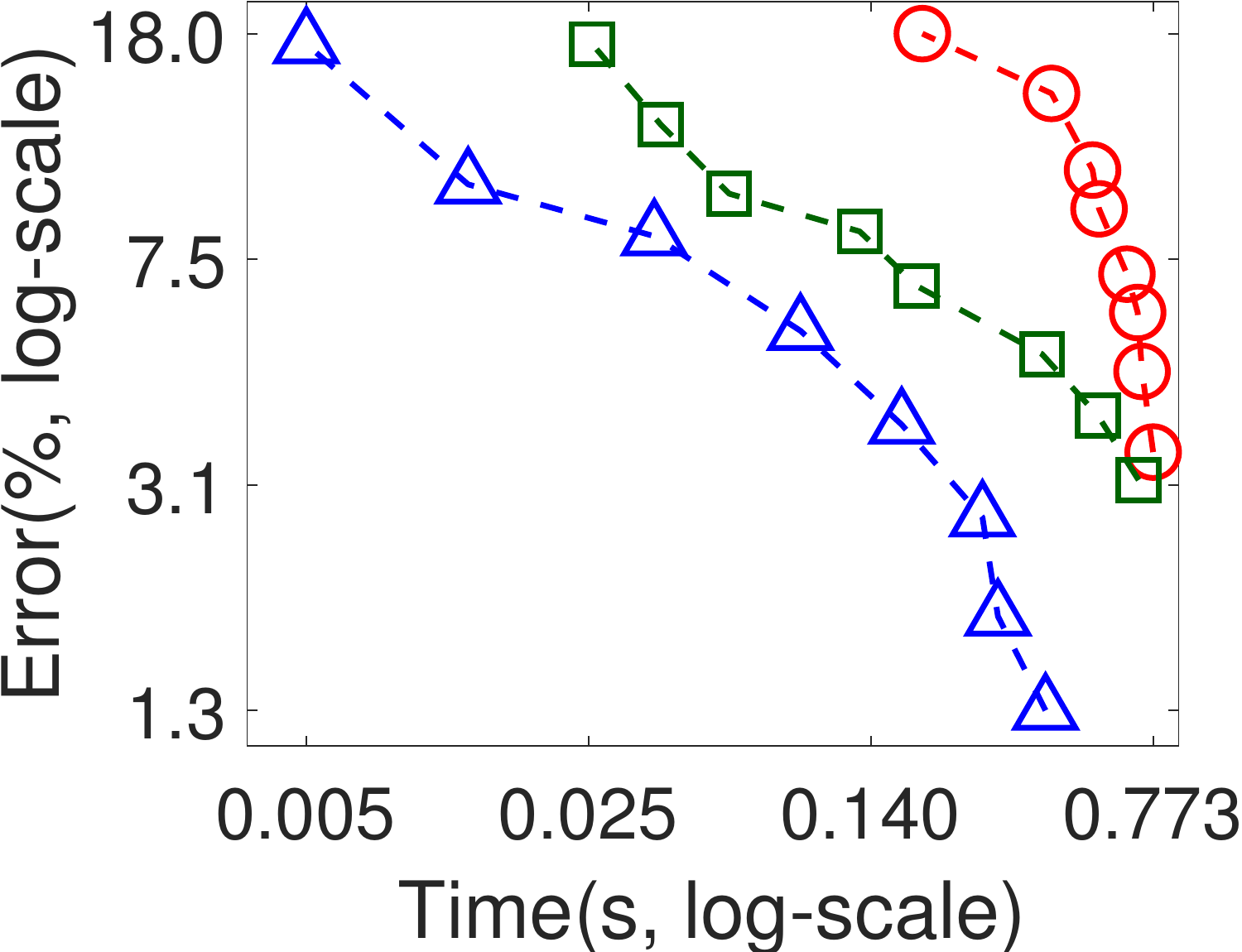}
  }
  \hfill
  \subfigure[Q4 on SU]{
    \label{fig:p:m4:su}
    \includegraphics[height=0.9in]{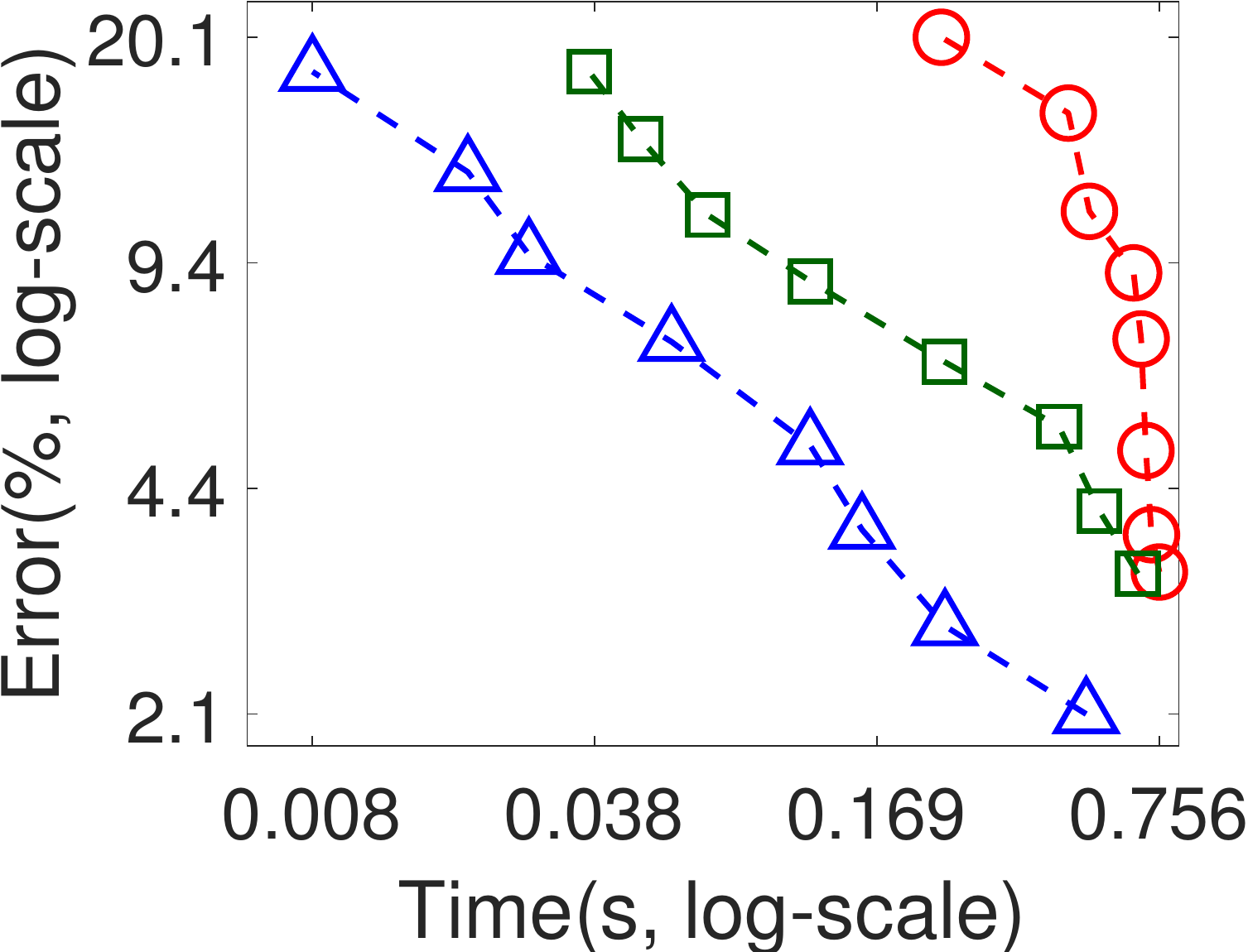}
  }
  \hfill
  \subfigure[Q5 on SU]{
    \label{fig:p:m5:su}
    \includegraphics[height=0.9in]{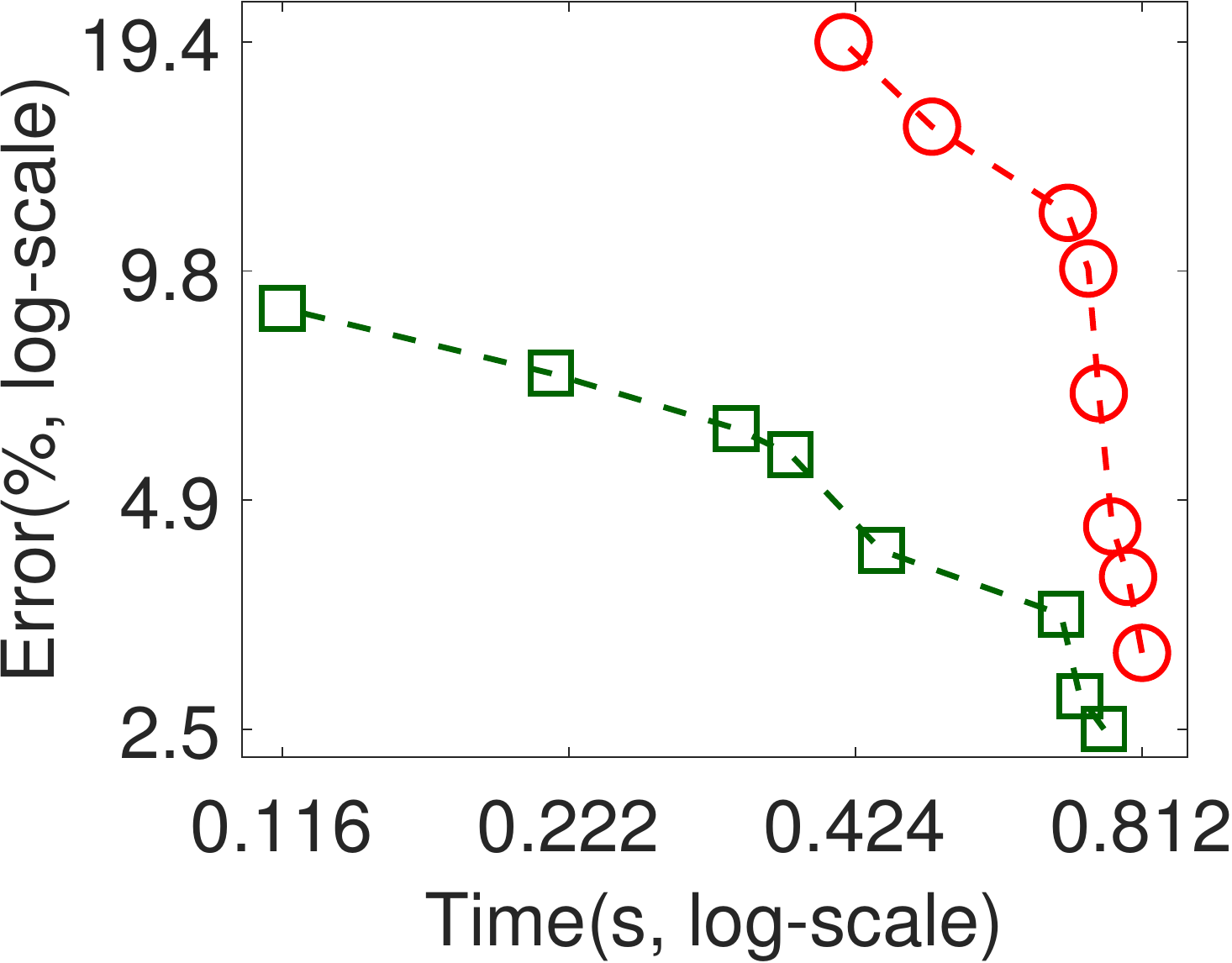}
  }
  \newline
  \vspace{-1em}
  \subfigure[Q1 on SO]{
    \label{fig:p:m1:so}
    \includegraphics[height=0.9in]{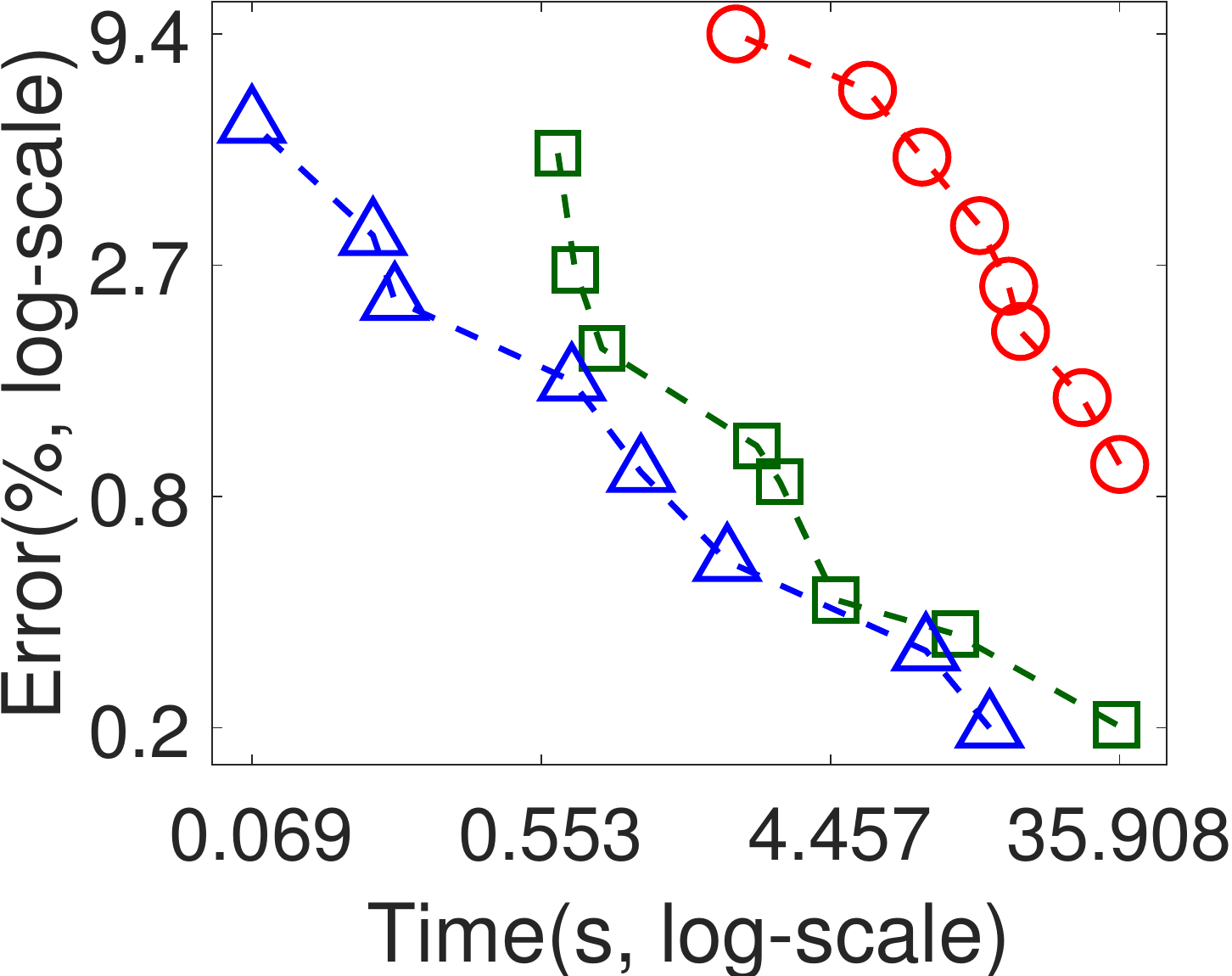}
  }
  \hfill
  \subfigure[Q2 on SO]{
    \label{fig:p:m2:so}
    \includegraphics[height=0.9in]{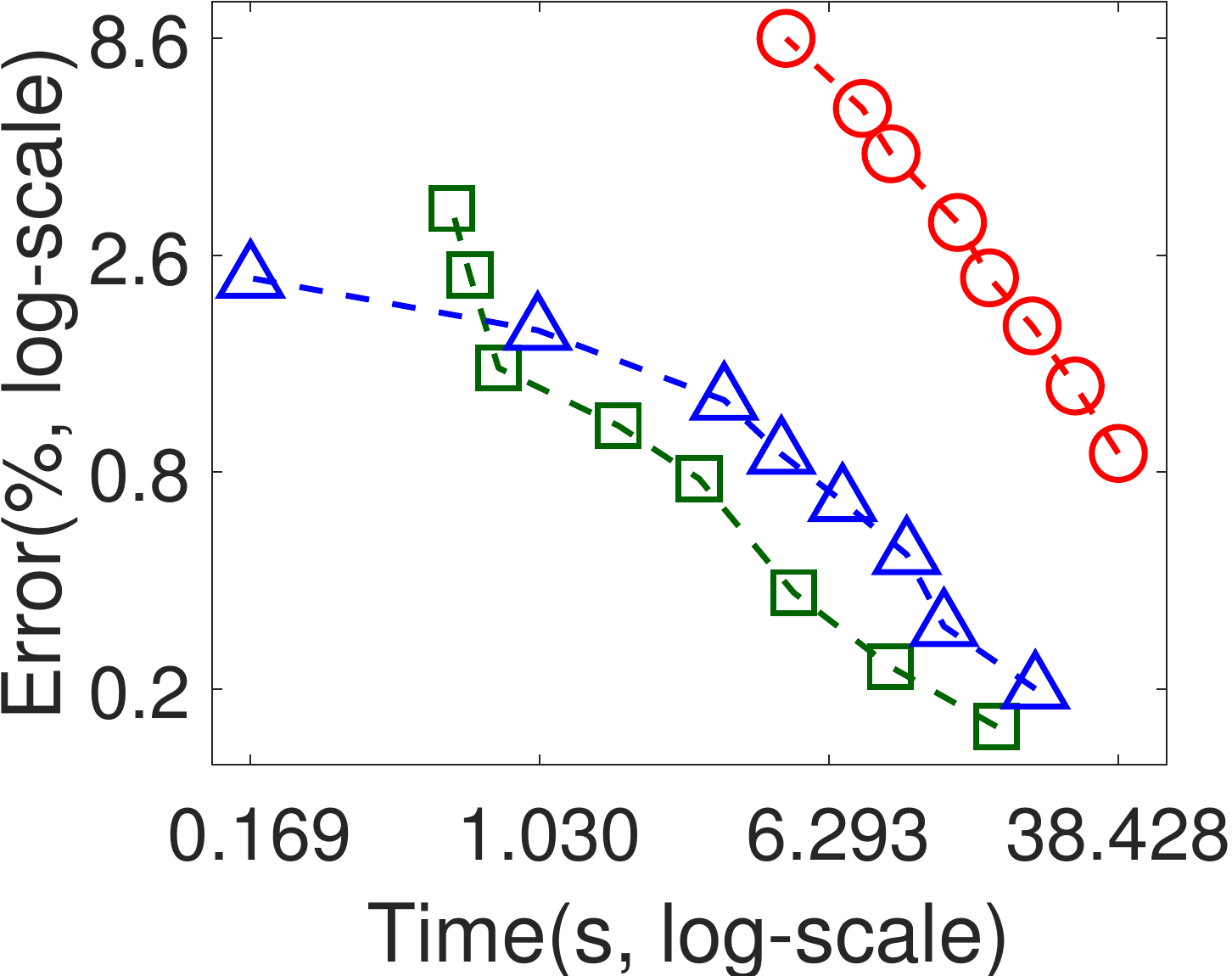}
  }
  \hfill
  \subfigure[Q3 on SO]{
    \label{fig:p:m3:so}
    \includegraphics[height=0.9in]{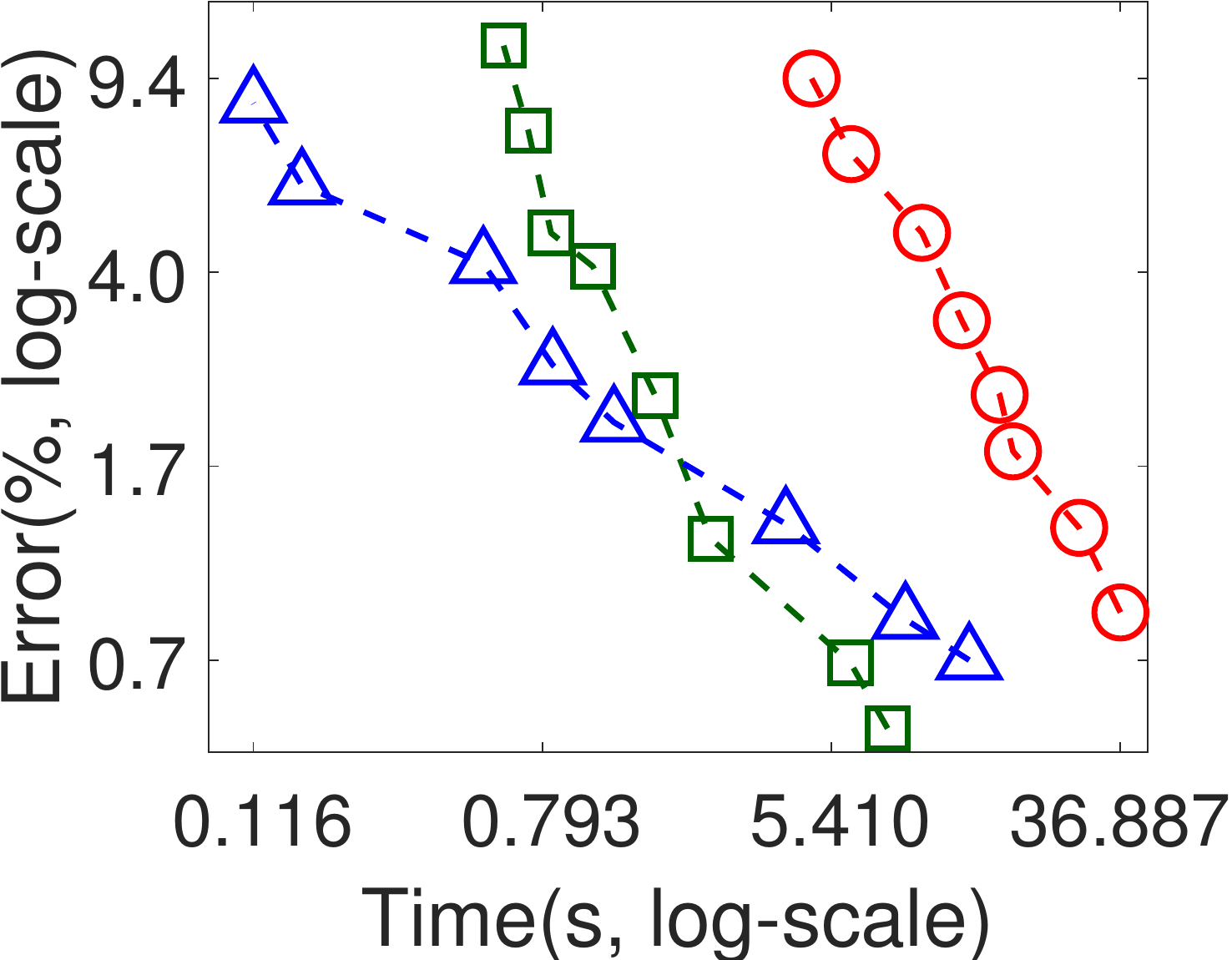}
  }
  \hfill
  \subfigure[Q4 on SO]{
    \label{fig:p:m4:so}
    \includegraphics[height=0.9in]{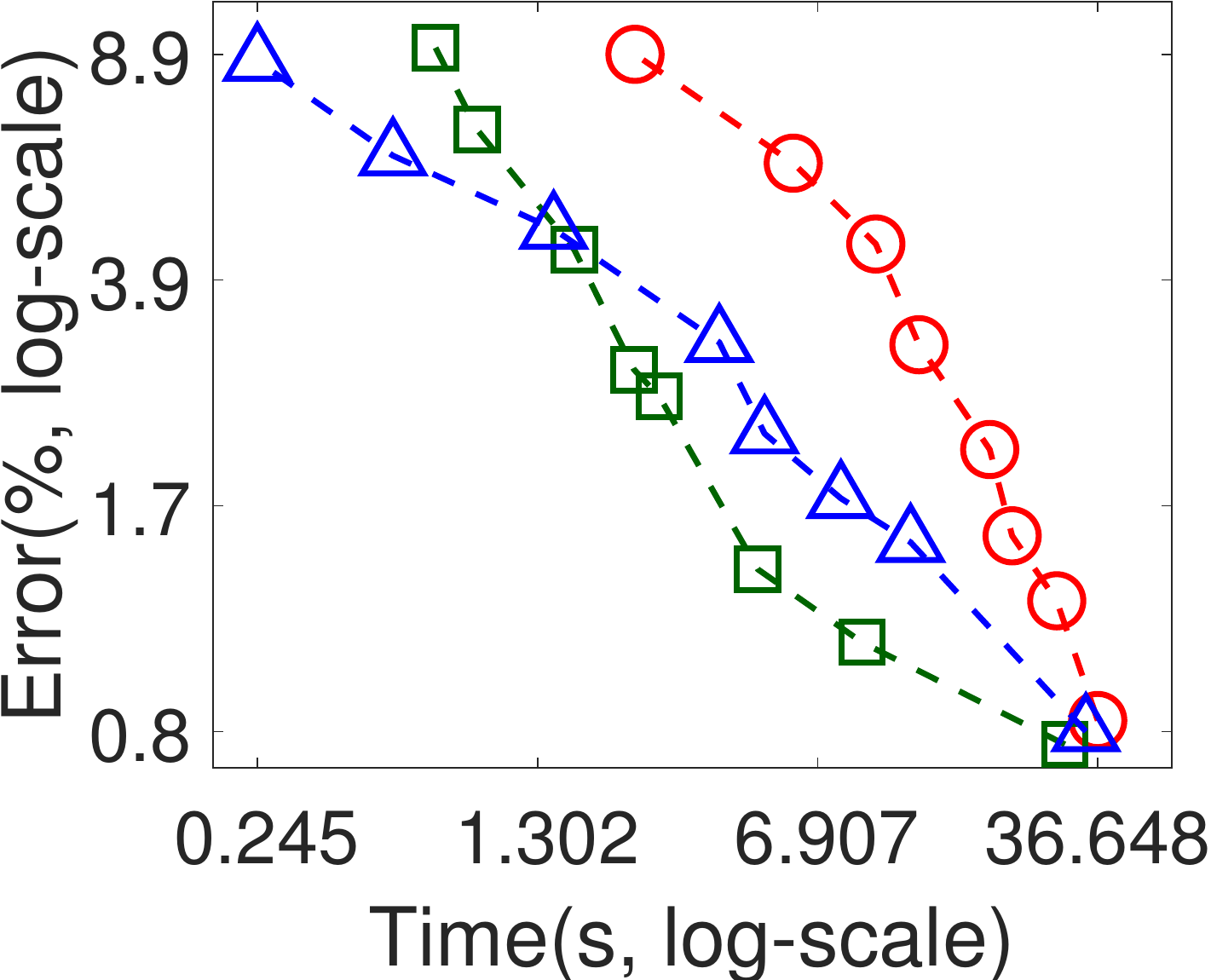}
  }
  \hfill
  \subfigure[Q5 on SO]{
    \label{fig:p:m5:so}
    \includegraphics[height=0.9in]{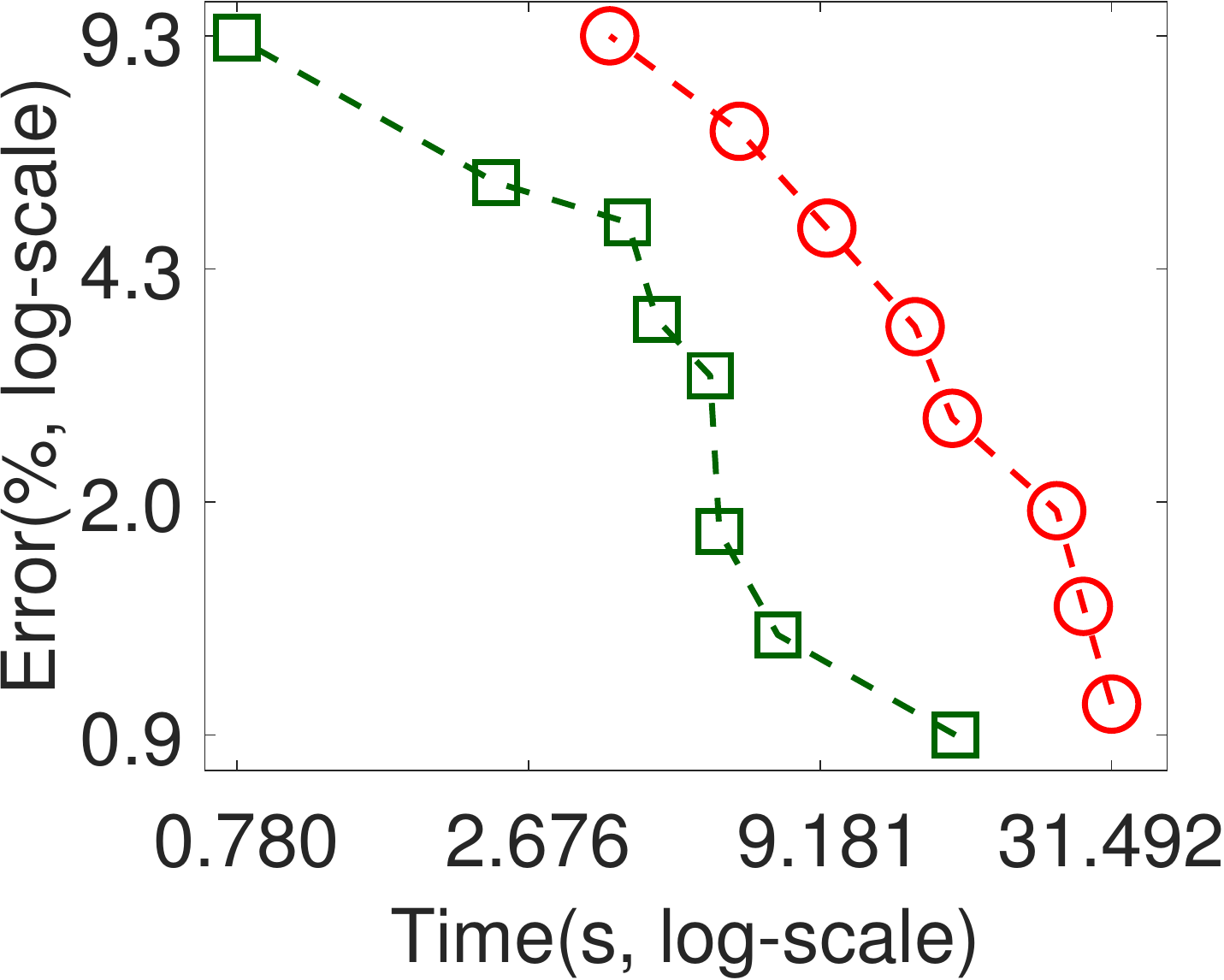}
  }
  \newline
  \vspace{-1em}
  \subfigure[Q1 on BC]{
    \label{fig:p:m1:bt}
    \includegraphics[height=0.9in]{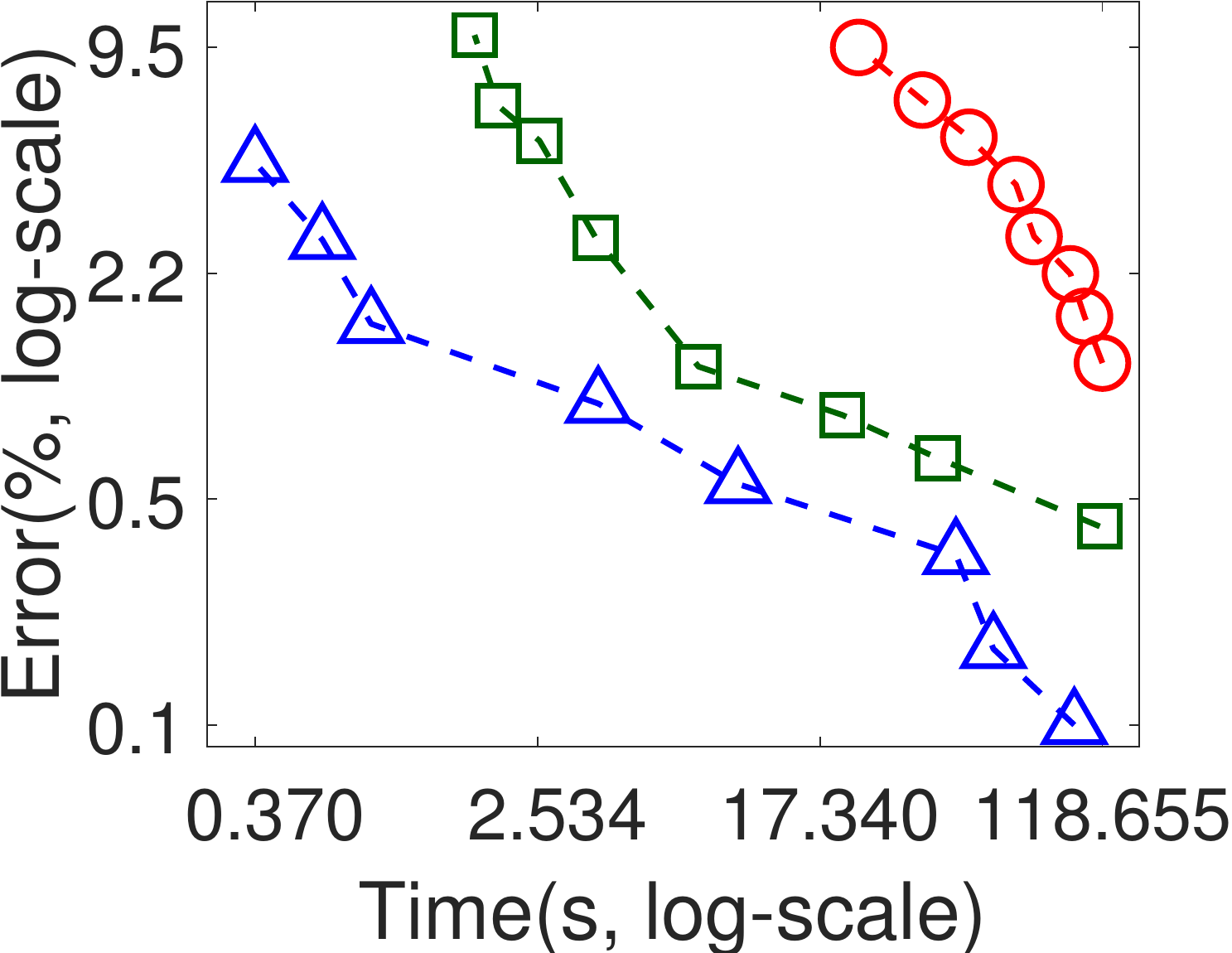}
  }
  \hfill
  \subfigure[Q2 on BC]{
    \label{fig:p:m2:bt}
    \includegraphics[height=0.9in]{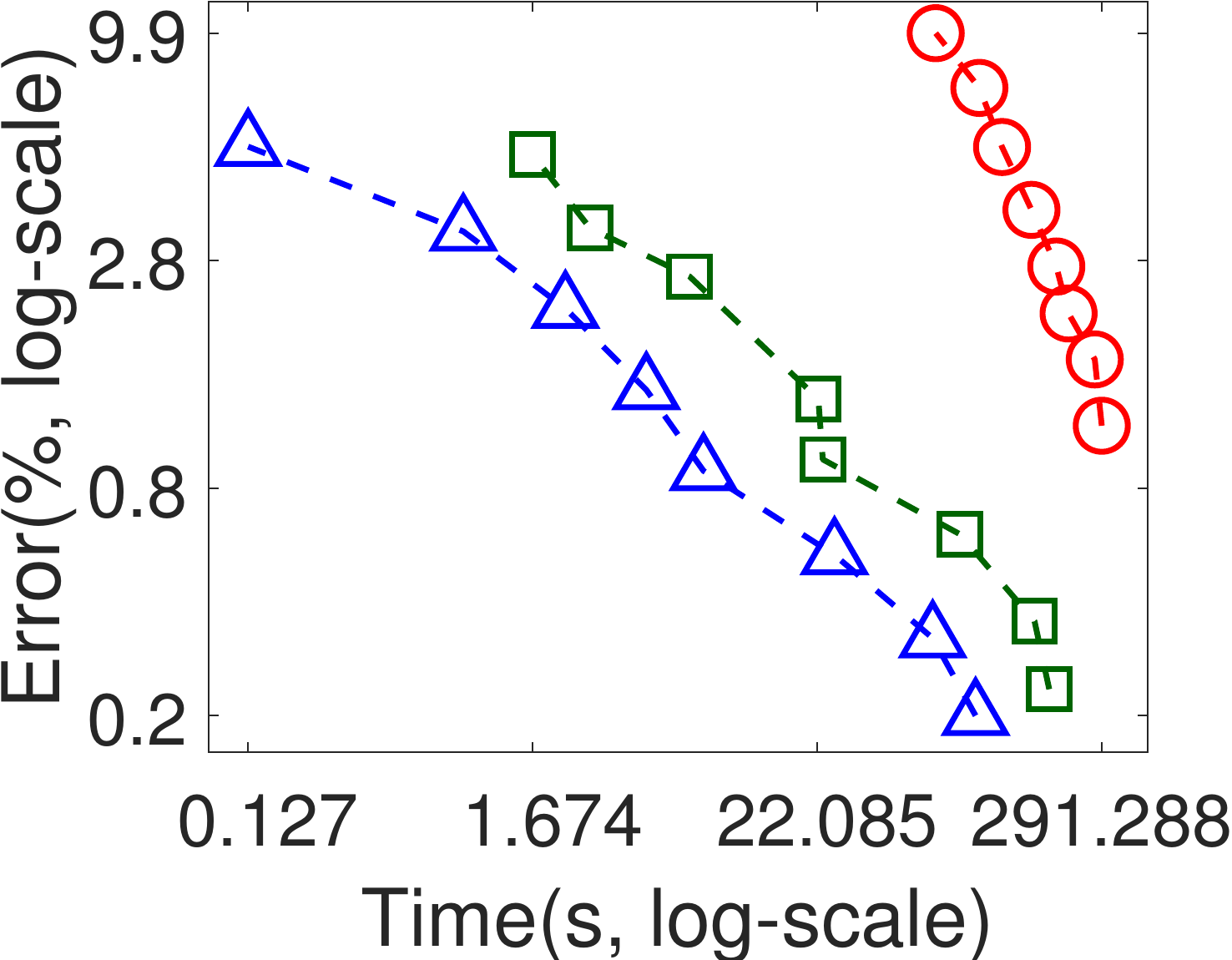}
  }
  \hfill
  \subfigure[Q3 on BC]{
    \label{fig:p:m3:bt}
    \includegraphics[height=0.9in]{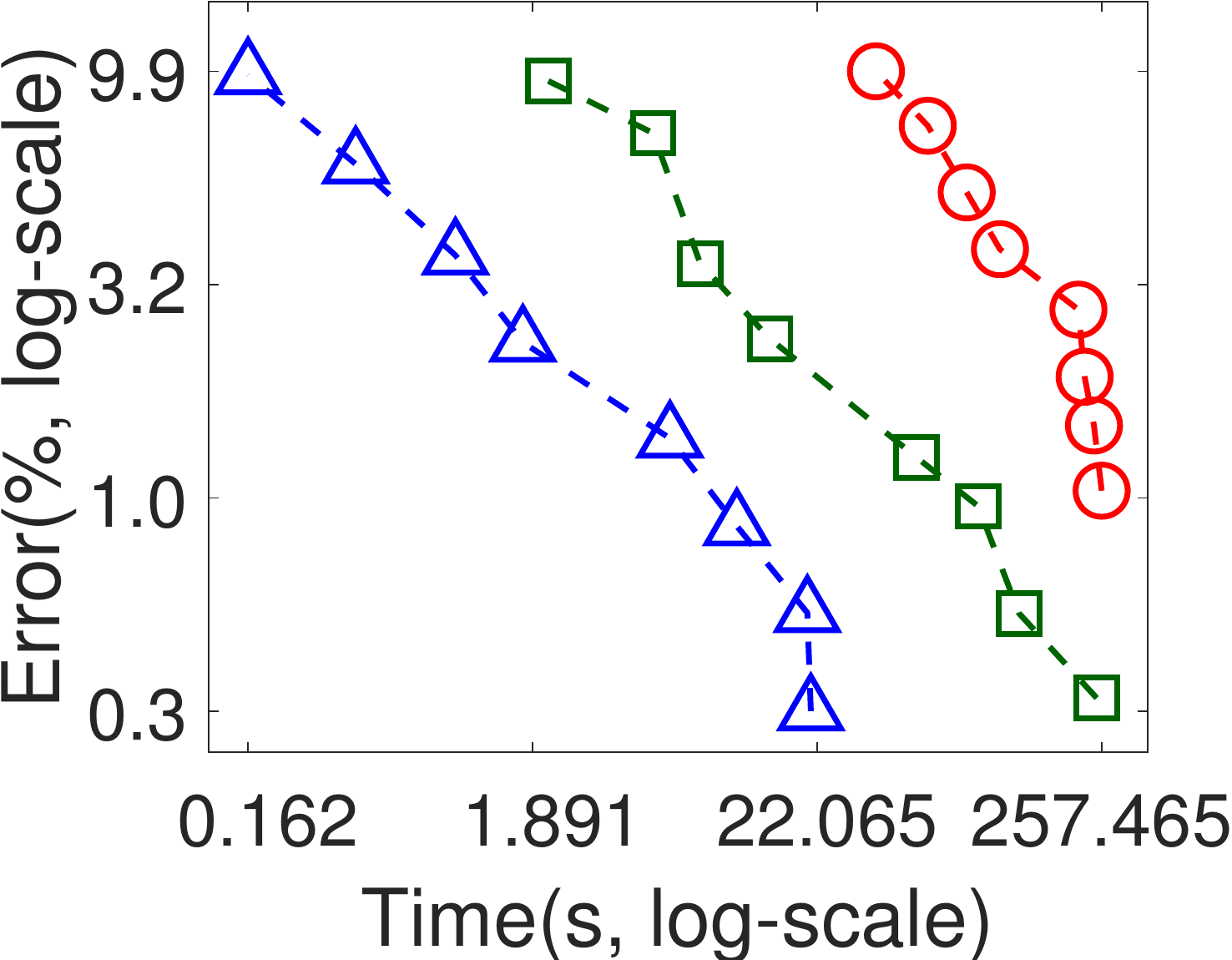}
  }
  \hfill
  \subfigure[Q4 on BC]{
    \label{fig:p:m4:bt}
    \includegraphics[height=0.9in]{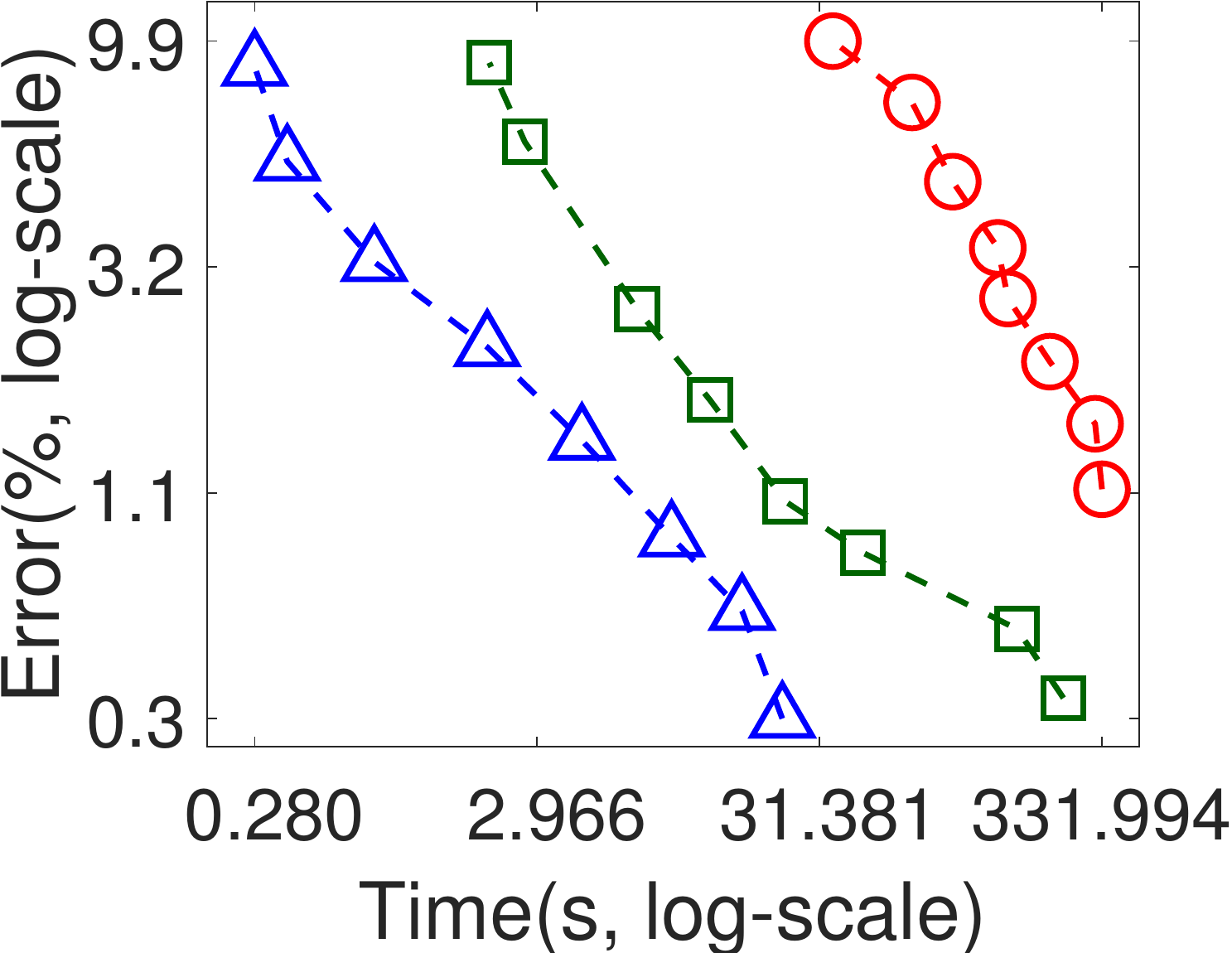}
  }
  \hfill
  \subfigure[Q5 on BC]{
    \label{fig:p:m5:bt}
    \includegraphics[height=0.9in]{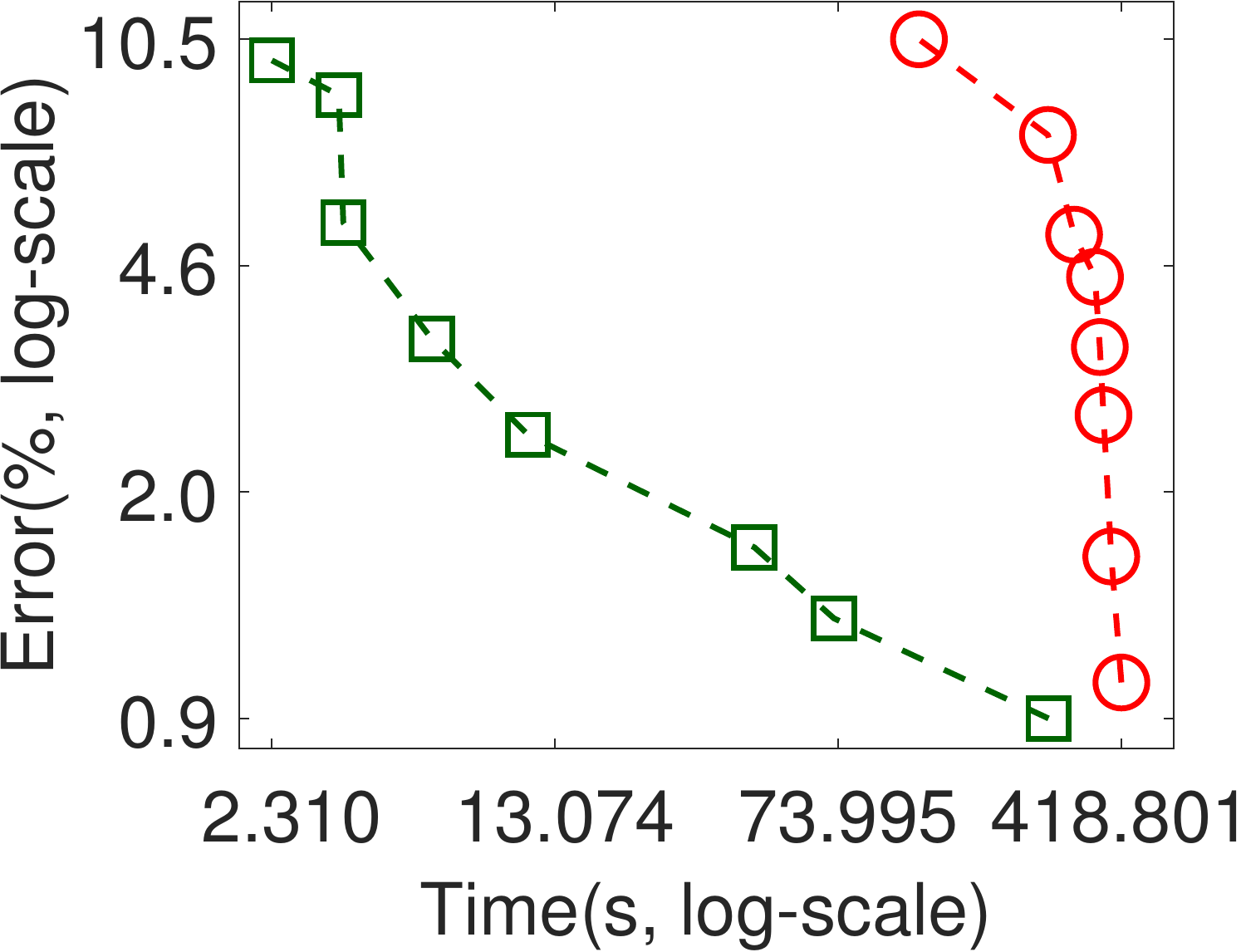}
  }
  \newline
  \subfigure[Q1 on RC]{
    \label{fig:p:m1:rc}
    \includegraphics[height=0.89in]{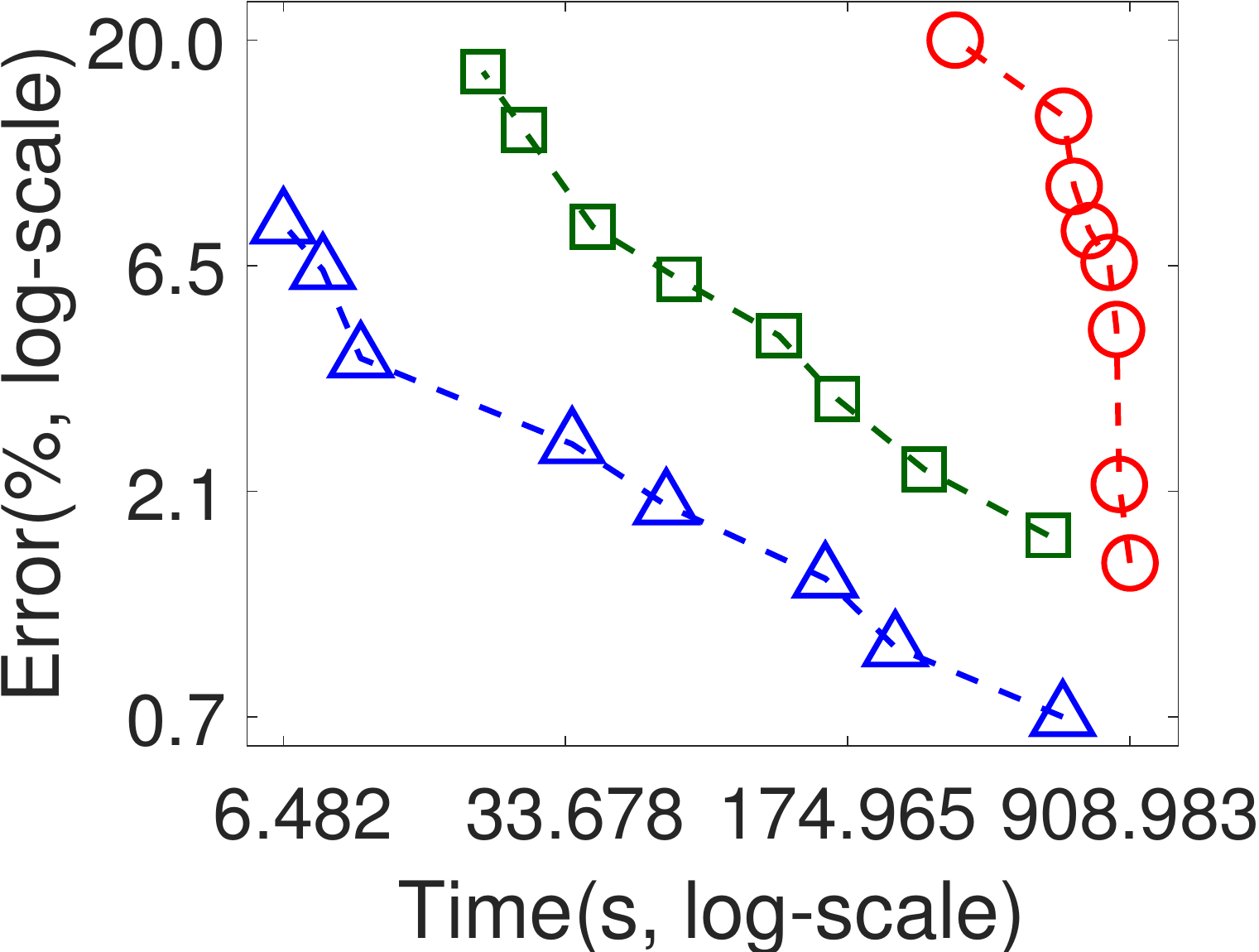}
  }
  \hfill
  \subfigure[Q2 on RC]{
    \label{fig:p:m2:rc}
    \includegraphics[height=0.89in]{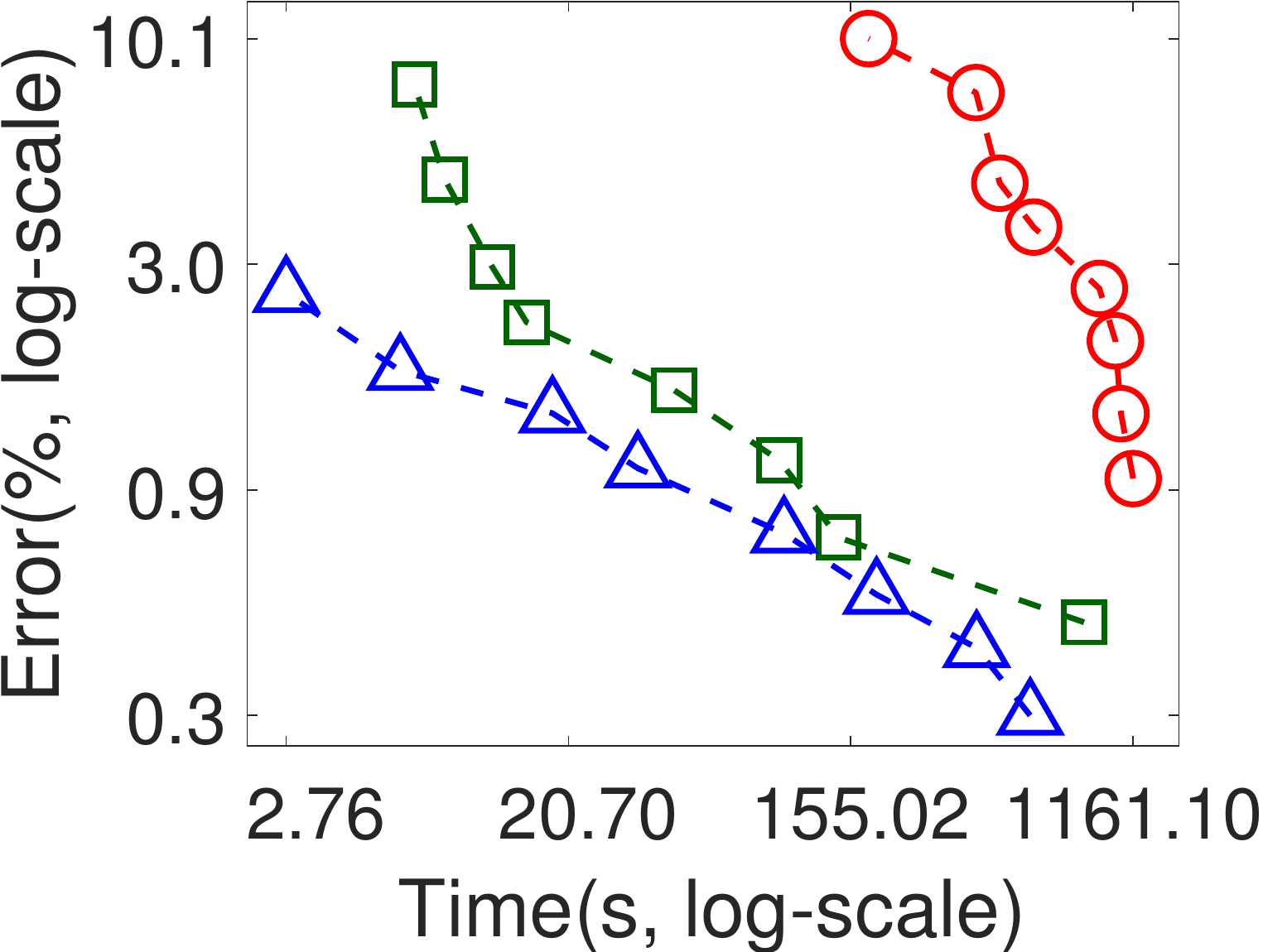}
  }
  \hfill
  \subfigure[Q3 on RC]{
    \label{fig:p:m3:rc}
    \includegraphics[height=0.89in]{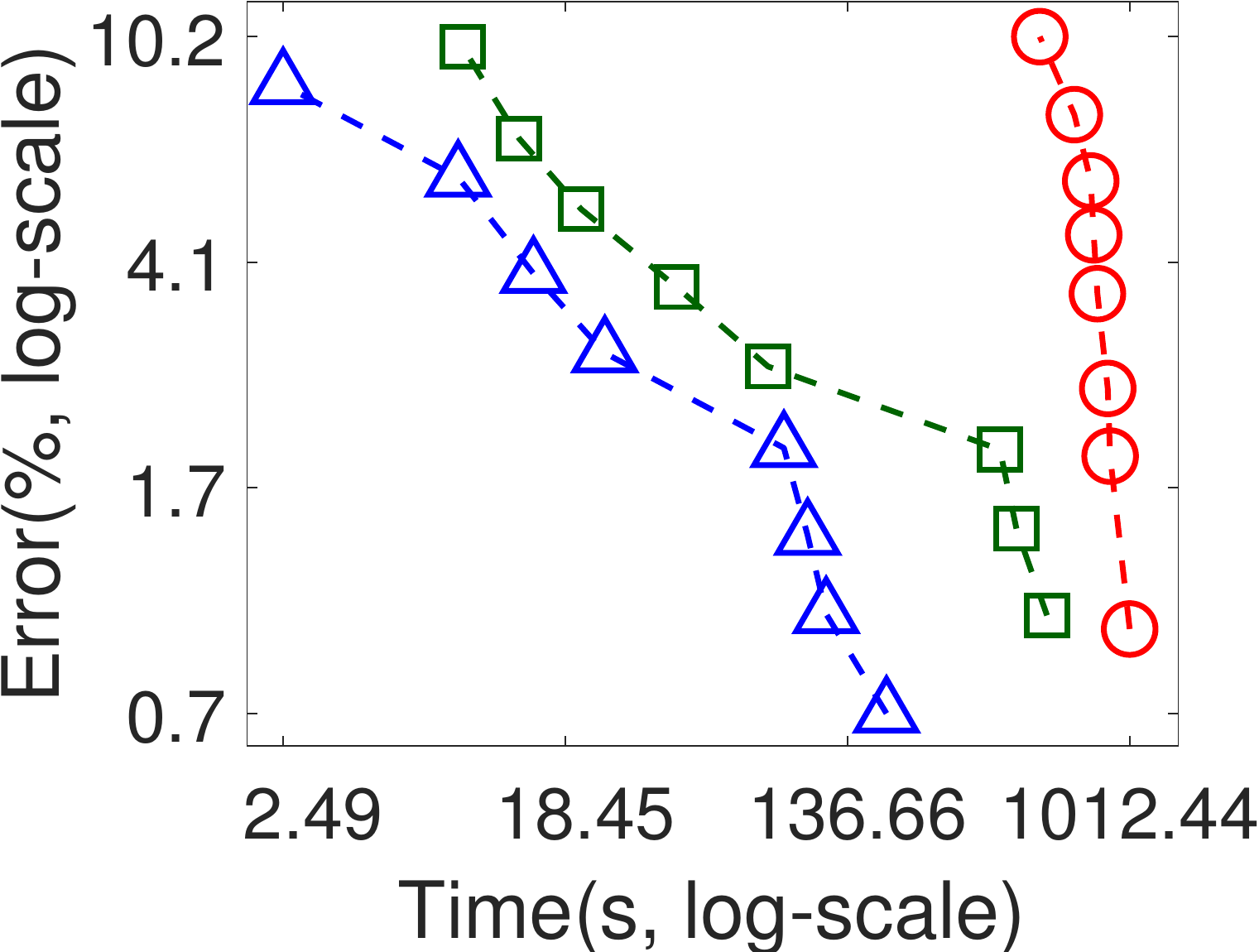}
  }
  \hfill
  \subfigure[Q4 on RC]{
    \label{fig:p:m4:rc}
    \includegraphics[height=0.89in]{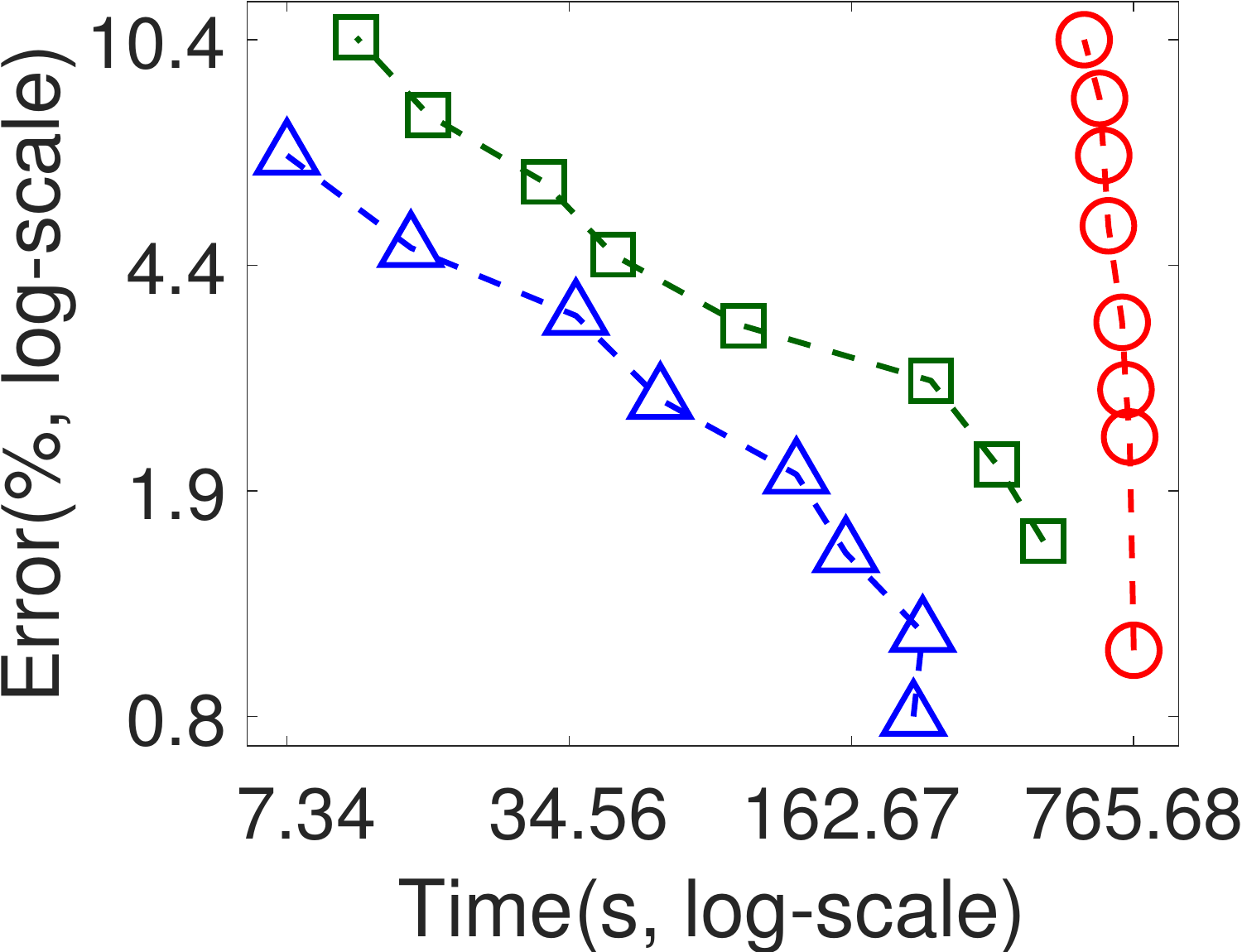}
  }
  \hfill
  \subfigure[Q5 on RC]{
    \label{fig:p:m5:rc}
    \includegraphics[height=0.89in]{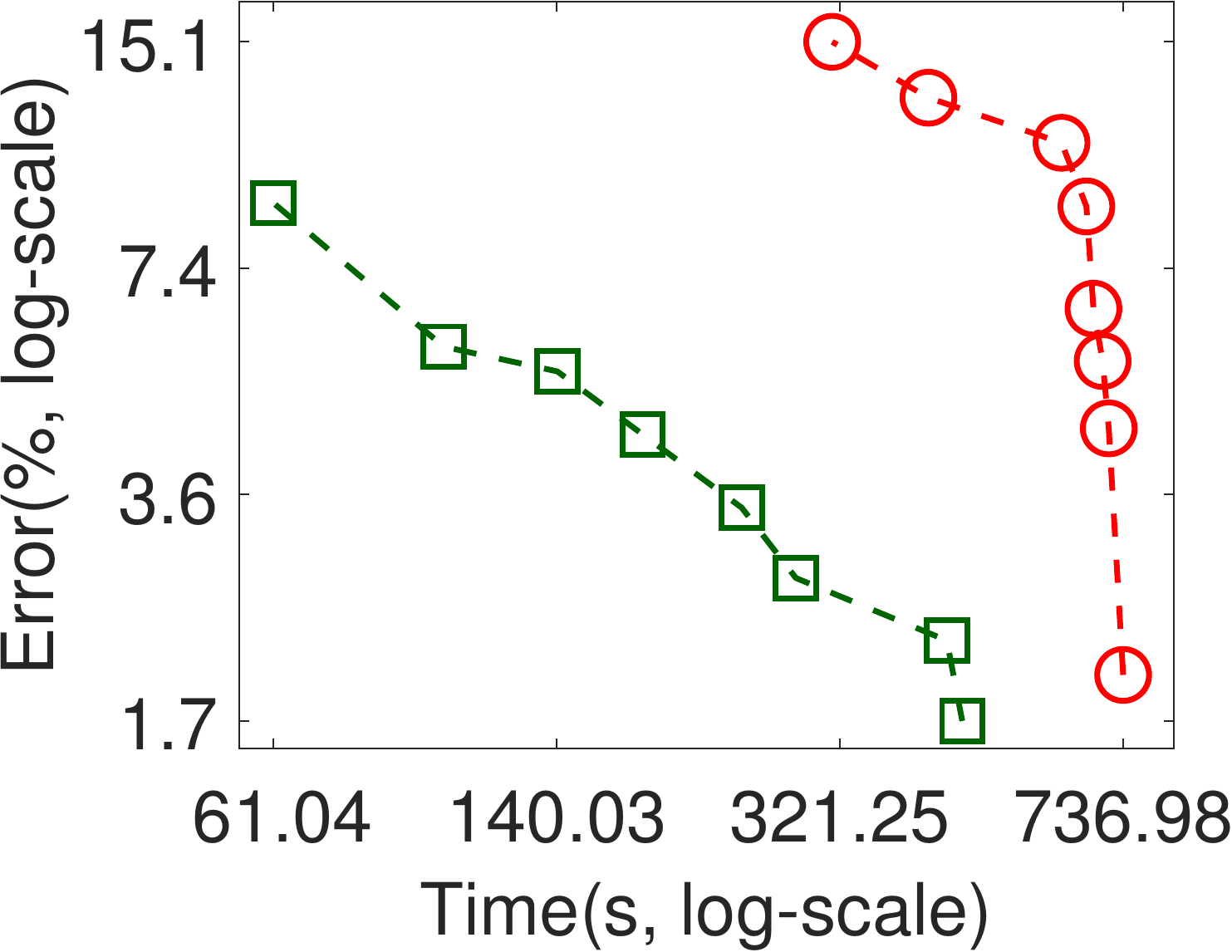}
  }
  \vspace{-1em}
  \caption{Relative error ($\%$) vs. running time (seconds) with varying sampling probability}
  \label{fig:p}
\end{figure*}
\begin{figure*}
  \centering
  \subfigure[Q3 on BC with varying $\delta$]{
    \label{subfig:delta}
    \includegraphics[height=0.9in]{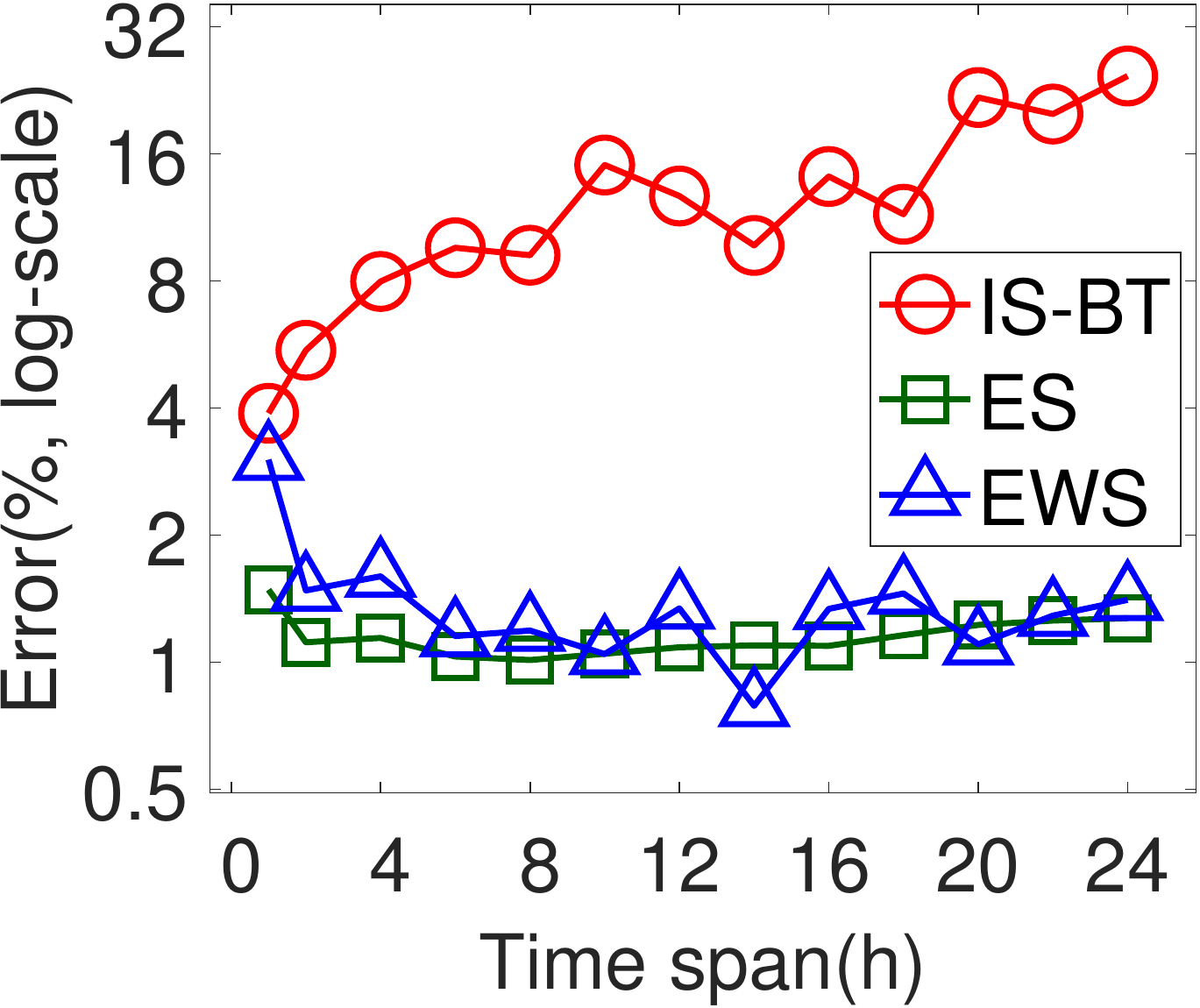}
    \includegraphics[height=0.9in]{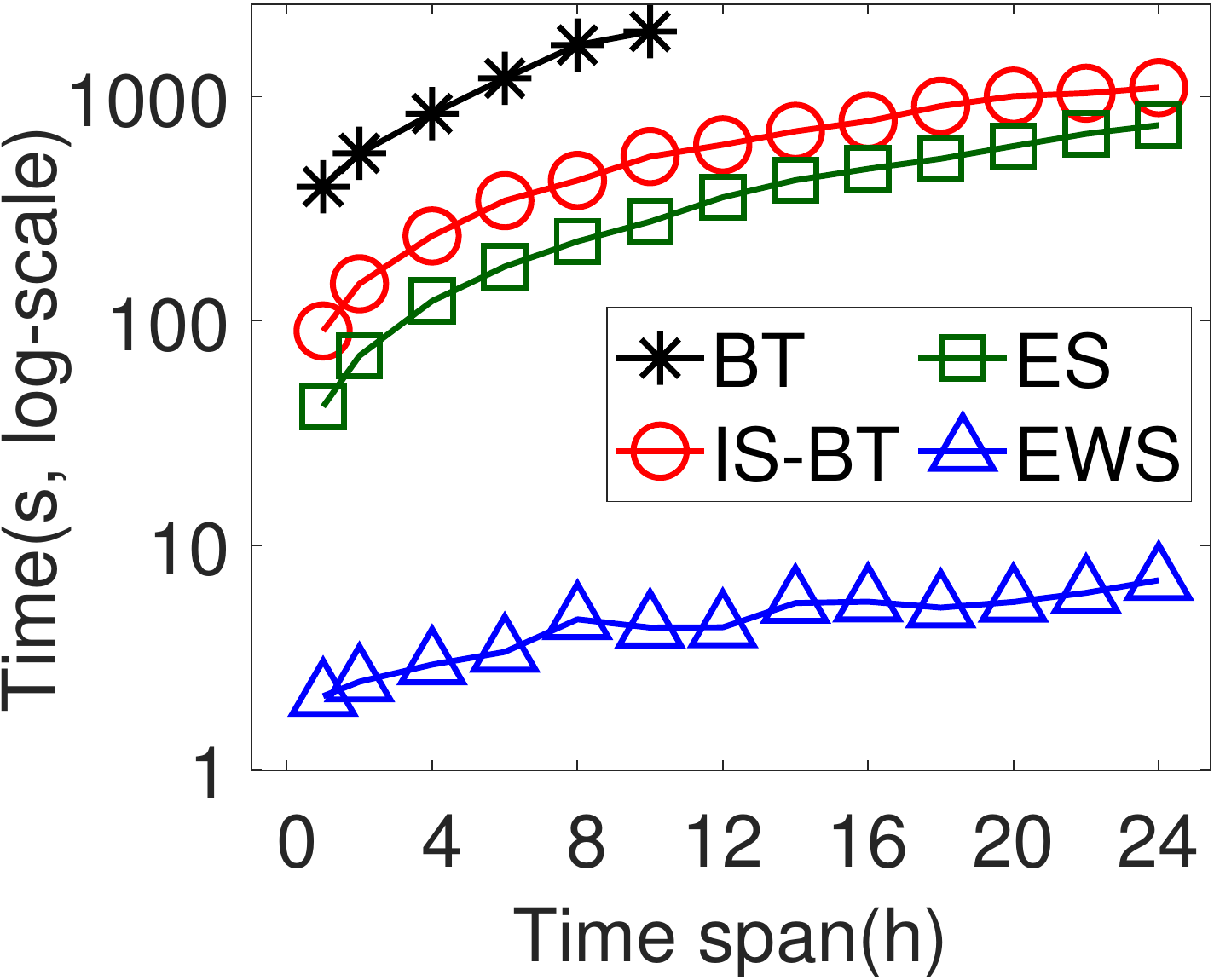}
  }
  \hspace{2em}
  \subfigure[Q2 on RC with varying $m$]{
    \label{subfig:size}
    \includegraphics[height=0.9in]{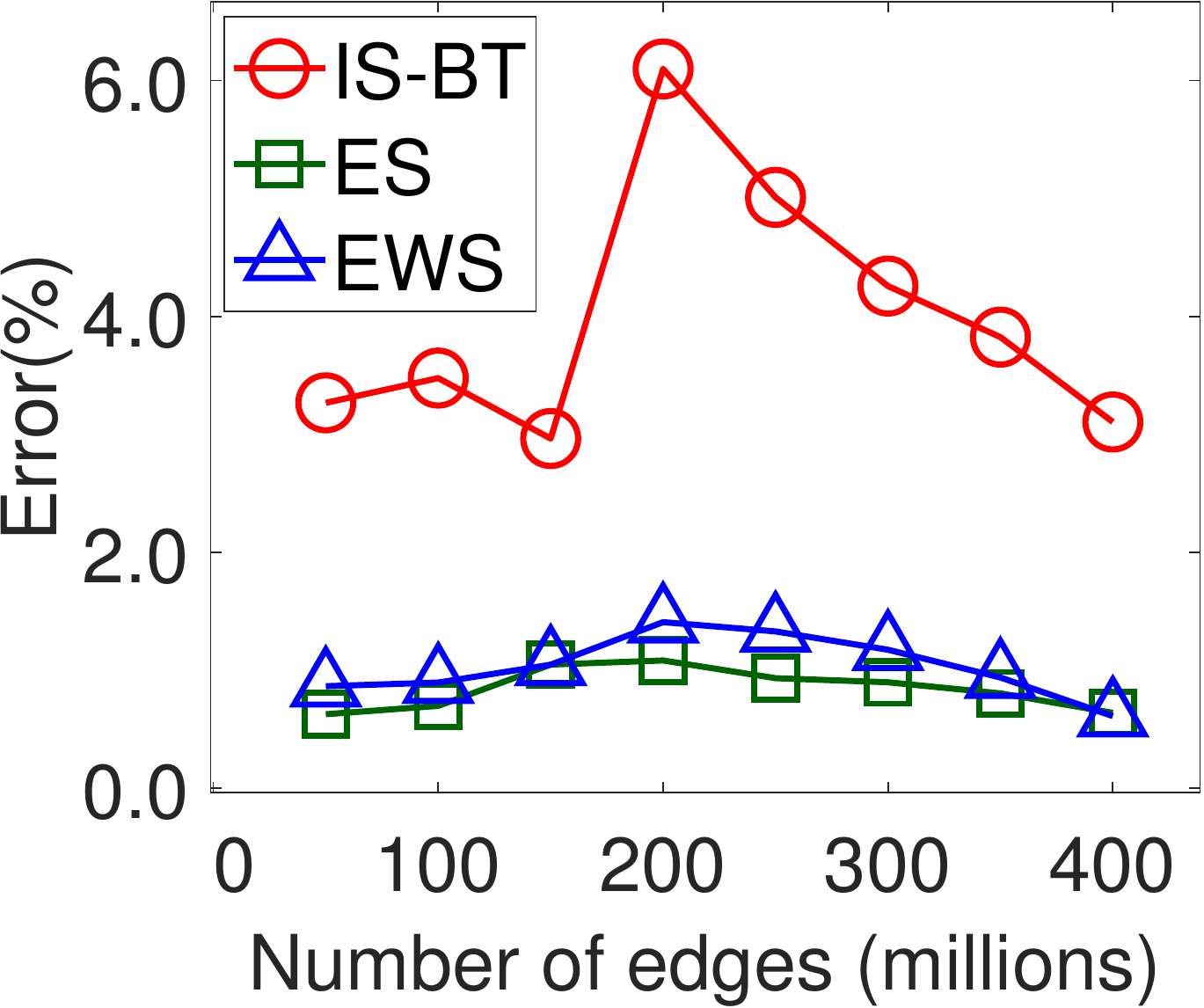}
    \includegraphics[height=0.9in]{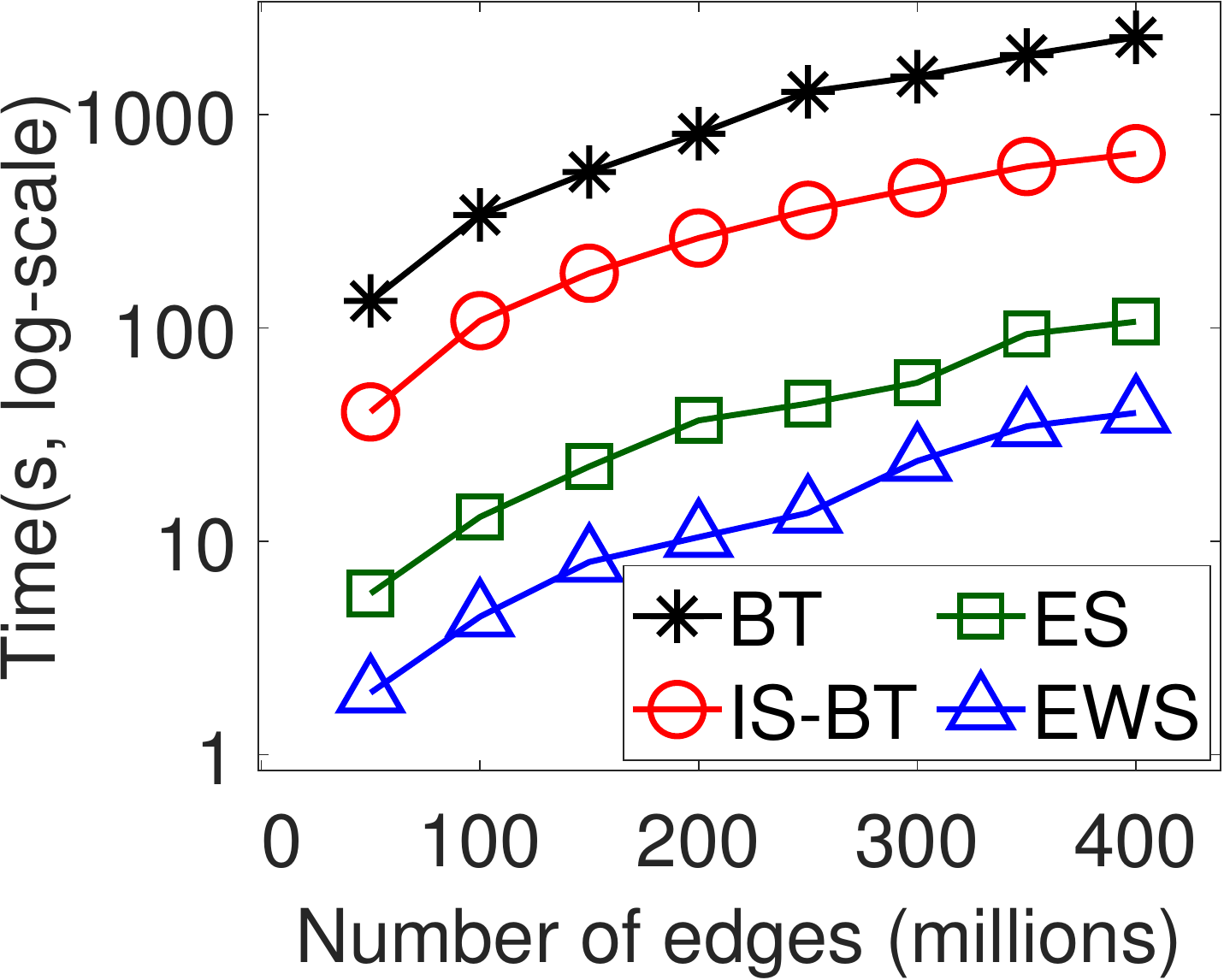}
  }
  \vspace{-1em}
  \caption{Scalability tests with varying time span $\delta$ and number of temporal edges $m$}
  \label{fig:scalability}
\end{figure*}

The overall performance of each algorithm is reported in Table~\ref{tab:results}.
Here, the time span $\delta$ is set to $86400$ seconds (i.e., one day) on AU and SU,
and $3600$ seconds (i.e., one hour) on SO, BC, and RC (Note that we use the same
values of $\delta$ across all experiments, unless specified).
For IS-BT, we report the results in the default setting
as indicated in~\cite{DBLP:conf/wsdm/LiuBC19}, i.e., we fix the interval length to $30\delta$
and present the result for the smallest interval sampling probability that
can guarantee the relative error is at most $5\%$.
For ES and EWS, we report the results when $p=0.01$ by default;
in a few cases when the numbers of motif instances are too small
or their distribution is highly skewed among edges,
we report the results when $p=0.1$ (marked with ``*'' in Table~\ref{tab:results})
because ES and EWS cannot provide accurate estimates when $p=0.01$.
In addition, we set $q$ to $1$ on AU and SU, and $0.1$ on SO, BC, and RC for EWS.

First of all, the efficiencies of EX and 2SCENT are lower than the other algorithms.
This is because they use an algorithm for subgraph isomorphism or cycle detection in static graphs
for candidate generation without considering temporal information.
As a result, a large number of redundant candidates are generated
and lead to the degradation in performance.
Second, on medium-sized datasets (i.e., AU and SU),
ES runs faster than IS-BT in most cases;
and meanwhile, their relative errors are close to each other.
On large datasets (i.e, SO, BC, and RC),
ES demonstrates both much higher efficiency (up to $10.3$x speedup)
and lower estimation errors ($2.42\%$ vs. $4.61\%$) than IS-BT.
Third, EWS runs $1.7$x--$19.6$x faster than ES
due to its lower computational cost per edge.
The relative errors of ES and EWS are the same on AU and SU because $q=1$.
When $q=0.1$,
EWS achieves further speedups at the expense of higher relative errors.
A more detailed analysis of the effect of $q$ is provided
in the following paragraph.

\vspace{1mm}
\noindent\textbf{Effect of $q$ for EWS:}
In Figure~\ref{fig:q}, we compare the relative errors and running time of EWS
for $q=1$ and $0.1$ when $p$ is fixed to $0.01$.
We observe different effects of $q$
on medium-sized (e.g., SU) and large (e.g., BC) datasets.
On the SU dataset, the benefit of smaller $q$ is marginal:
the running time decreases slightly but the errors become obviously higher.
But on the BC dataset, by setting $q=0.1$, EWS achieves $2$x--$3$x speedups
without affecting the accuracy seriously.
These results imply that \emph{temporal wedge sampling} is more effective on larger datasets.
Therefore, we set $q=1$ on AU and SU, and $q=0.1$ on SO, BC, and RC for EWS
in the remaining experiments.

\vspace{1mm}
\noindent\textbf{Accuracy vs. Efficiency:}
Figure~\ref{fig:p} demonstrates the trade-offs between \emph{relative error}
and \emph{running time} of three sampling algorithms, namely IS-BT, ES, and EWS.
For IS-BT, we fix the interval length to $30\delta$ and vary the interval sampling
probability from $0.01$ to $1$. 
For ES and EWS, we vary the edge sampling probability $p$
from $0.0001$ to $0.25$.
First of all, ES and EWS consistently achieve better trade-offs between accuracy
and efficiency than IS-BT in almost all experiments. Specifically, ES and EWS can
run up to $60$x and $330$x faster than IS-BT when the relative errors are at the same level.
Meanwhile, in the same elapsed time, ES and EWS are up to $10.4$x and $16.5$x more accurate
than IS-BT, respectively. Furthermore, EWS can outperform ES in all datasets except SO
because of lower computational overhead. But on the SO dataset, since the distribution
of motif instances is highly skewed among edges and thus the \emph{temporal wedge sampling}
leads to large errors in estimation, the performance of EWS
degrades significantly and is close to or even worse than that of ES.
Nevertheless, the effectiveness of \emph{temporal wedge sampling} for EWS can still be confirmed
by the results on the BC and RC datasets.

\vspace{1mm}
\noindent\textbf{Scalability:}
We evaluate the scalability of different algorithms with varying
the time span $\delta$ and dataset size $m$.
In both experiments, we use the same parameter settings as
used for the same motif on the same dataset in Table~\ref{tab:results}.
We test the effect of $\delta$ for $Q3$ on the BC dataset by varying $\delta$
from $1$h to $24$h. As shown in Figure~\ref{subfig:delta}, the running time
of all algorithms increases near-linearly w.r.t.~$\delta$. BT runs out of memory
when $\delta>10$h. The relative errors of ES and EWS keep steady for different $\delta$
but the accuracy of IS-BT degrades seriously when $\delta$ increases.
This is owing to the increase in cross-interval instances
and the skewness of instances among intervals.
Meanwhile, ES and EWS run up to $2.2$x and $180$x faster than IS-BT,
respectively, while always having smaller errors.
The results for $Q2$ on the RC dataset with varying $m$
are presented in Figure~\ref{subfig:size}.
Here, we vary $m$ from $50$M to near $400$M by extracting
the first $m$ temporal edges of the RC dataset.
The running time of all algorithms grows near-linearly w.r.t.~$m$.
The fluctuations of relative errors of IS-BT
explicate that it is sensitive to the skewness of instances among intervals. 
ES and EWS always significantly outperform IS-BT for different $m$:
they run much faster, have smaller relative errors,
and provide more stable estimates than IS-BT.

\section{Conclusion}\label{sec:conclusion}

In this paper, we studied the problem of approximately counting a temporal motif
in a temporal graph via random sampling. We first proposed a generic Edge Sampling
(ES) algorithm to estimate the number of any $k$-vertex $l$-edge temporal motif
in a temporal graph. Furthermore, we improved the ES algorithm
by combining edge sampling with wedge sampling
and devised the EWS algorithm for counting $3$-vertex $3$-edge temporal motifs.
We provided comprehensive theoretical analyses on the unbiasedness, variances,
and complexities of our algorithms. Extensive experiments on several real-world
temporal graphs demonstrated the accuracy, efficiency, and scalability of our algorithms.
Specifically, ES and EWS ran up to $10.3$x and $48.5$x faster
than the state-of-the-art sampling method while having lower estimation errors.

\bibliographystyle{abbrvnat}
\bibliography{references}

\end{document}